# Toward Refactoring of DMARF and GIPSY Case Studies – a Team 12 SOEN6471-S14 Project Report

Dipesh Walia[1], Pankaj Kumar Pant[2], Mahendra Neela [3], Naveen Kumar[4], Ram Babu Kunchala[5]
[1, 2,3,4,5] Computer Science & Software Engineering Department, Concordia University, Montreal, Quebec, Canada

[1]dipeshwalia2013@gmail.com
[2]itp.pankaj@gmail.com,
[3]venkataneela.usapp@gmail.com,
[4]naveenkumar.bojedla@gmail.com,
[5]ramkunchala702@gmail.com

**ABSTRACT**

The main significance of this document is two source systems namely GIPSY and DMARF. Intensional languages are required like GIPSY for absoluteness and forward practical investigations on the subject.DMARF mainly focuses on software arechetictual design and implementation on Distributed Audio recognition and its applications such as speaker identification which can run distributively on web services architecture. This mainly highlights security aspects in a distributed system, the Java data security framework (JDSF) in DMARF. ASSL (Autonomic System Specification Language) frame work is used to integrate a self-optimizing property for DMARF. GIPSY mainly depends on Higher-Order Intensional Logic (HOIL) and reflects three main goals Generality, Adaptability and Efficiency

## I. INTRODUCTION

In this chapter we will learn about two frame works MARF and GIPSY.GIPSY (General Intentional Programming System) is a multi-intentional programming system which the programs are written in lucid programming languages. MARF (Modular Audio Recognition Framework) is a java based platform acts as a library in its applications. In DMARF we specify ASSL (Autonomic System Specification Language) a number of autonomic properties such as self-healing, self-optimization and self-protection. In GIPSY design the complier frame work (GIPC), the eduction engine (GEE), and programming environment (RIPE) are the compliers which produces the binary format which is compiled by GIPSY program as a binary output.

## II. BACKGROUND

The following paper emphasis on various research papers based on open source systems namely DMARF [11] and GIPSY [39]. Below is the summary of the research papers that were divided among the team, each had chosen one paper for GIPSY, and one for DMARF (Mapping of list of the individual to the case studies read can be found in Appendix A).

### *A. OSS Case Study*

#### *1) DMARF*

DMARF is based on classical MARF in which pipelines stages were made into distributed nodes. MARF mainly consists of pipeline stages that communicate with each other to get data in a chained manner. Pipelines are of four grouping similar kinds of algorithms.1. Sample loading, 2. Pre-processing. 3. Feature extraction and 4. Training/ classification. "MARF is mainly concerned with collection of voice, sound, speech, text and natural language processing (NLP) algorithms"[31] that are being used in SpeakerIdentApp is an MARF application which can identify the people using their voice, and it consists of a database of speakers.

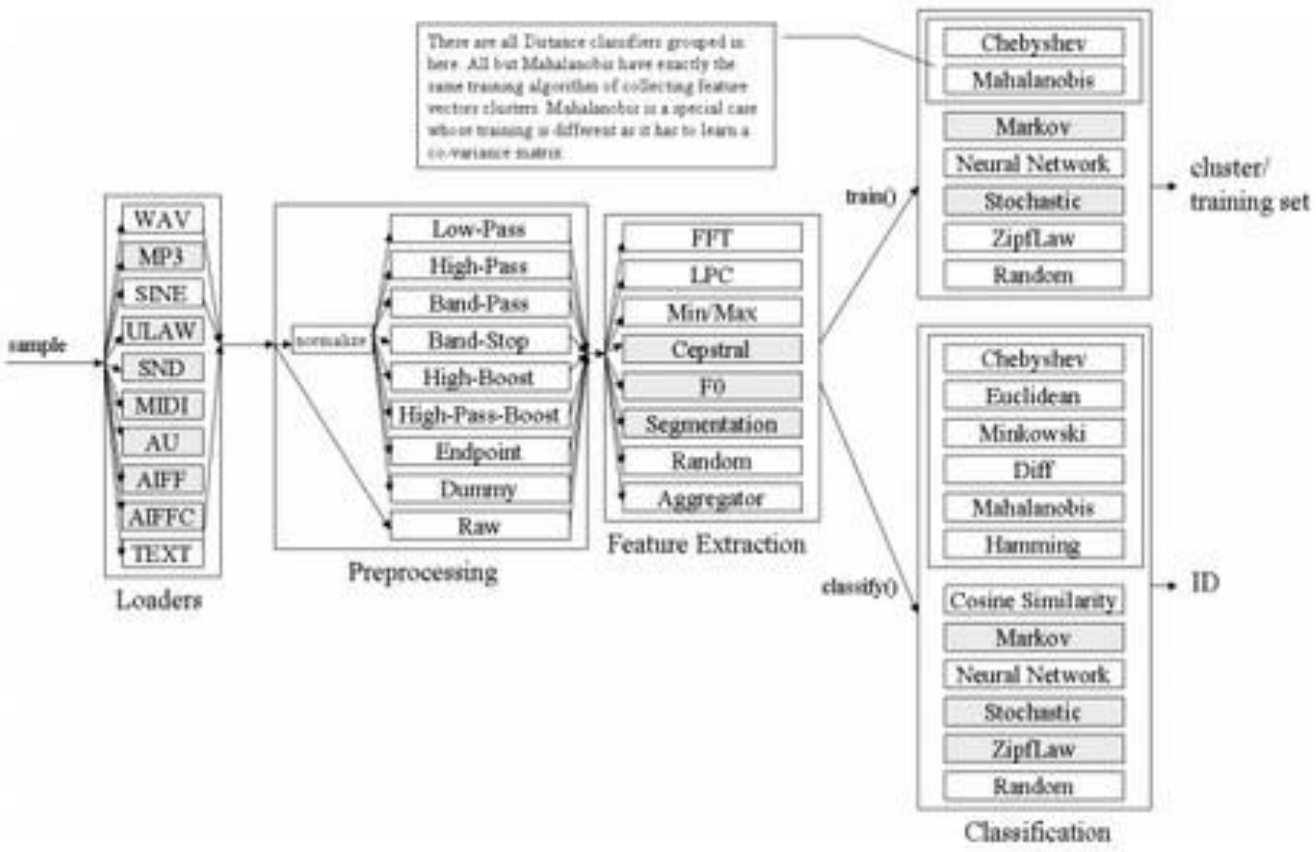

*Figure 1 Core MARF Pipeline Data Flow [4]*

Though MARF has several applications, most of them are dependent on recognition pipeline which is main part in MARF. In traditional MARF as shown in figure, if there are bulks of audio files or texts it is difficult to identify at once, the process is done sequentially which is time consuming process but in DMARF, the pipeline is distributed and runs on a cluster. The main scope of DMARF is to manage the SNMP protocol. It helps in producing SNMP agents and managing the protocols. DMARF components doesn't understand SNMP, so one-proxy

SNMP is located on one-management service. By using SNMP, proxy talks with the manager and agents. Proxy agents produced are not fully instrumented. Usually agents are produced by the MIB Compiler which is a part of Advent Net also produces two agents namely master agents and sub-agents which in term helps in feature extraction. By using SNMP along with DMARF, can produce good results as they are good at managing and can manage the stress which uses networks.

In general, DMARF is collection of signal processing and natural processing (NLP) algorithms written in java.

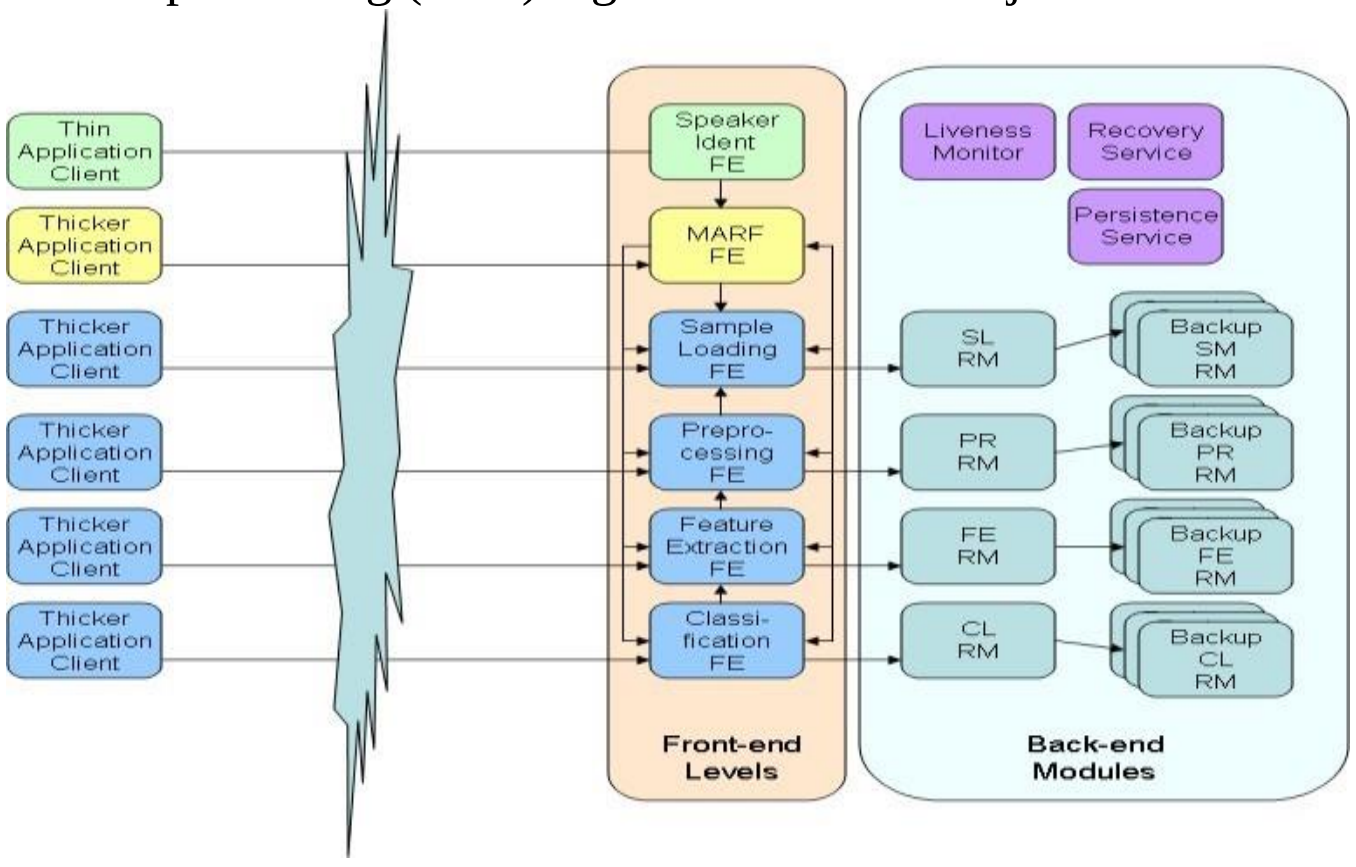

*Figure 2 Distributed MARF Pipeline [4]*

The objective is to have an autonomic figuring layer spreading DMARF by determining autonomic properties at each of the example distinguishing phases of the same. Autonomic computing [12] promises reduction of the workload needed to maintain complex systems by transforming them into self-managing autonomic systems. Security autonomic computing principles were used to solve the security problem. Let's discuss about Autonomic computing.

The main principle is to apply complex hiding so that it emphasizes on reducing the workload needed to maintain complex systems by transforming managing autonomic systems. To develop the artifacts to evaluate policies framework has developed by IBM, which is Policy Management for Autonomic Computing (PMAC).

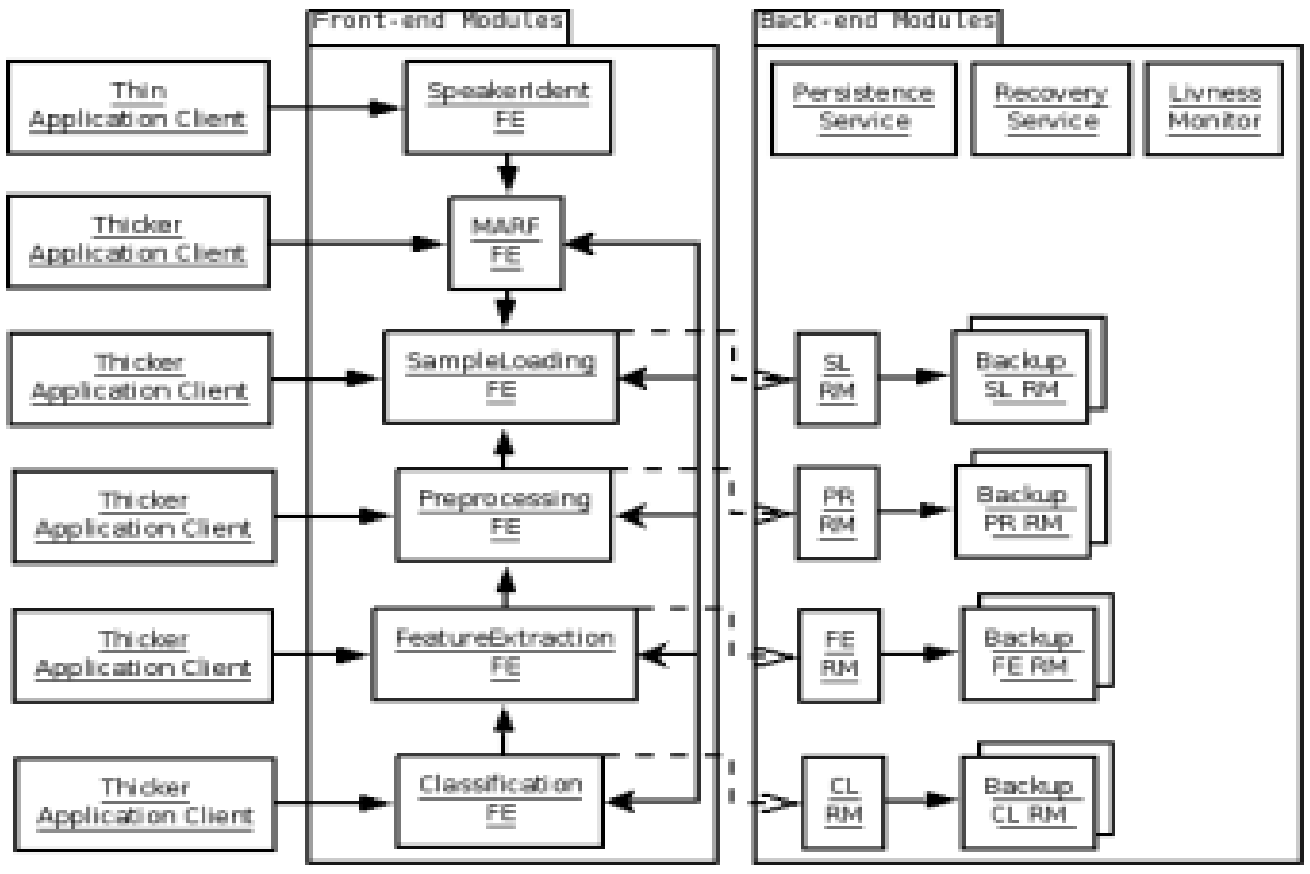

*Figure 3 Distributed MARF Pipeline [5]*

ASSL[13] [14] mainly focuses on the problem of formal specification and code generation of autonomic systems (Ass) within the framework. From any ASSL valid specification. All in all, ASSL considers autonomic frameworks (Ass) as made out of autonomic components (AEs) conveying over collaboration conventions. To detail those, ASSL is characterized through formalization of levels. Over these levels, ASSL expert videos a multi-level particular model that is intended to be adaptable and uncovered a wise choice and setup of framework components and instruments required by an AS.

- AS tier – It presents a general and global AS perspective, where author will define general autonomic system rules in terms of service-level objectives (SLO).
- AS interaction protocol (ASIP) tier – Forms a communication protocol which define the means of communication between AEs.
- AE tier – AEs with their own behavior forms a unit-level perspective.

Next the self-optimization, as mentioned above ASSL (Autonomic System Specification Language) provide DMARF an autonomic layer for autonomous purpose, and DMARF as an autonomic system fundamentally focused on distributed pattern recognition and related optimization. Two major operative conditions for optimization purpose includes training set classification data replication [1] which implies data replication in order to optimize data shift among stage nodes and dynamic communication protocol selection [1] which is to enable DMARF automatic adoption of effective protocol available. ASSL contains autonomic (ASs) inside framework and hub of the structure AS is something called as autonomic elements (AEs), which form the primary unit and to define the functioning which includes 3 major abstractions constituted of sub-tires, i.e., AS tier, AS Interaction Protocol (ASIP) tier and AE tier.

The idea behind ADMARF (autonomic DMARF) is explained below:

- “Self-optimization takes place when ADMARF enters in the classification stage
- The Classification stage itself forces the stage nodes synchronize their latest cached results.
- Here each node is asked to get the results of the other nodes. Before starting with the real computation, each stage node strives to adapt to the most efficient currently, available communication protocol." [1]

AS tier specification and AE tier specification are fundamentally responsible for executing above mention algorithm.

For property of self-forensics [15] [16] could be a new thought introduced to cover and formally apply to or be enclosed in the design of not solely involuntary hardware and/or software package systems, that area unit inherently advanced and laborious to research in general once incidents happen, however additionally as associate degree no mandatory demand for smaller projects. Self-forensics in a very shell includes an avid module or modules observant the regular modules in a way, logging the observations and events that led to default structure applicable. The modules will optionally have a capability of mechanically diagnose themselves supported the collected proof and build additional advanced and complete selections once the analysis than ad-hoc binary selections. In a sense, the self-forensic modules are often seen as sensible “black boxes” like in planes, however are often enclosed in orbiter, road vehicles, massive and little software package systems that assist with the incident analysis. Human specialists are often

Also trained in investigation techniques supported the rhetorical information sets collected throughout totally different incidents.

The specifications area unit there to be designed into the system for the aim of tracing and understanding advanced relationships and events among some parts of the aforesaid systems, particularly the distributed ones during this work, we have a tendency to moderately narrowly focus on “exporting” the states of the systems as information structures in written account encoded order as Forensic Lucid contexts using its syntax in accordance with the synchronic linguistics for collection the cases. Additionally proceed to explain the background and also the connected work, followed by the specification of the core information structures and information flows in rhetorical Lucid of the case studies, like the Distributed standard Audio Recognition Framework (DMARF), General connotative Programming System (GIPSY), Java information Security Framework (JDSF), and Crypto lysis –an automated cryptanalytics framework for classical ciphers. Forensic Computing: It mainly identifies, recovering, analyzing and presenting the facts and opinions with in computer crime investigations .For example in self-forensics, a state-tracing Linux kernel [17] exhibits by its own state for the analysis. Many of web-based applications are written in java, which provides a good basics for a self-forensics.

Self-forensics [18, 19, 20] is additional property with forensic logging with Forensic Lucid. Forensic lucid was initially proposed for specification, automatic deduction and event reconstruction in cybercrime based on digital forensic [21]. Which is proposed to extend its working on other domain including vehicle investigation in various vehicle crash investigations, and autonomous software and hardware systems. Forensic Lucid primary experiment platform is GIPSY. Forensic Lucid includes primarily Compiler (General Intensional Programming Compiler (GIPC) member). Moreover GIPSY run time system, General Eduction Engine (GEE) is flexible enough for various modes of execution. For achieving SFAP (Self-Forensics Autonomic Property) property, the property of SELF_FORENSIC policy is added to AS and autonomic element (AE).

- Adding syntax and semantic support for lexical analyzer, parser and semantic checker of ASSL.
- Adding proper code generator for JOOIP and forensic lucid to translate events.

ASSL management element (ME) is responsible to encode any module or subsystem to increase or decrease forensic evidence amount as forensic lucid events depending on criticality of faults. To allow mixing of Java and lucid codes (by placing lucid fragments anywhere in Java classes), the team came up with hybrid Intentional Object Oriented language, JOOIP [46, 47]. JOOIP + Forensic Lucid Code evaluation is performed by GEE. (The Forensic Lucid Compilation and Evaluation Flow in GIPSY is shown in following figure)

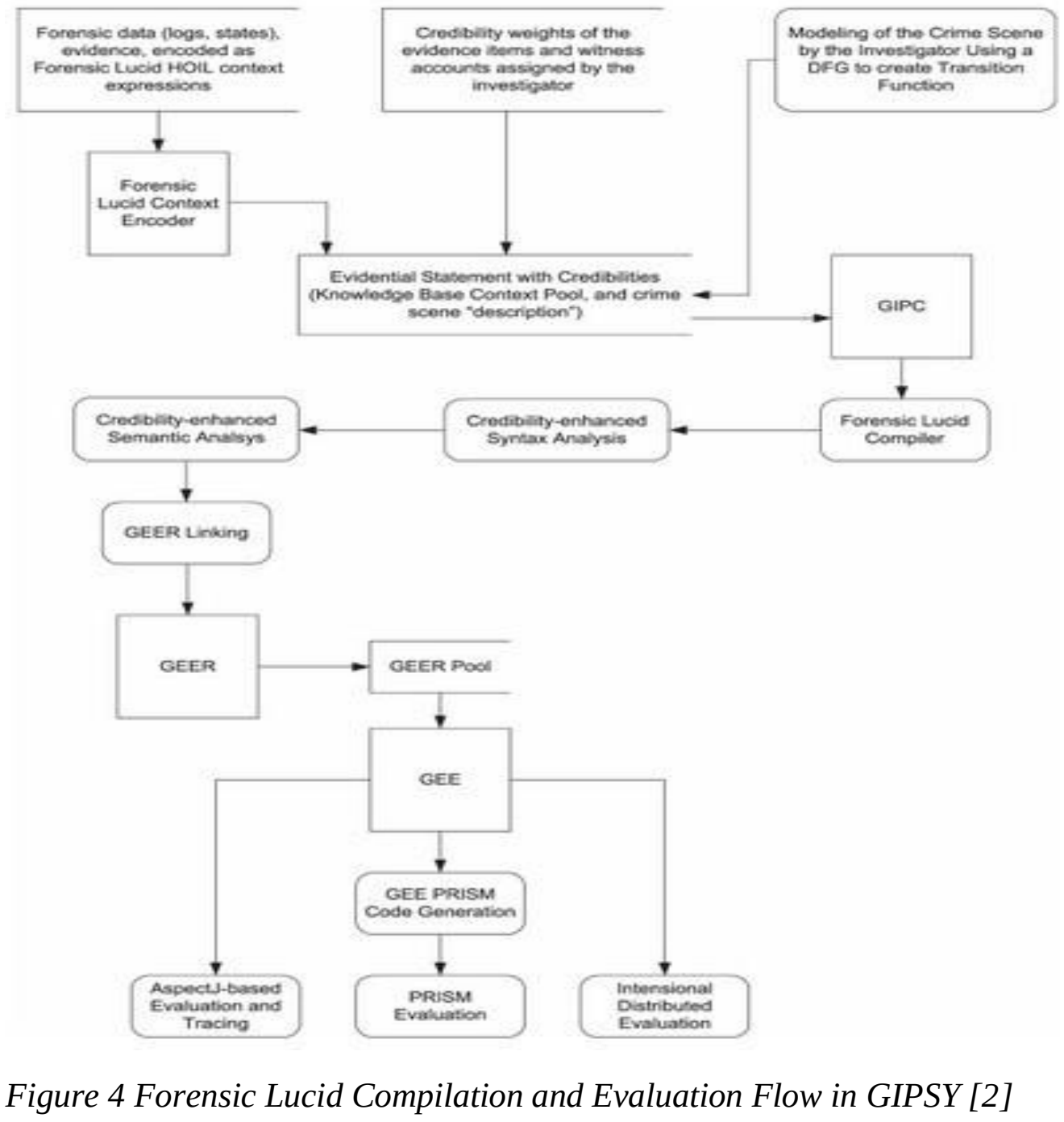

*Figure 4 Forensic Lucid Compilation and Evaluation Flow in GIPSY [2]*

The INITIATED_BY and TERMINATED_BY clauses would correspond to beginning and end of data stream Lucid Operators. Finally, the JOOIP code with Lucid fragments generated by ASSL toolkit is passed to hybrid GIPSY compiler, GIPC, to properly compile and link them together in an executable code inside the GEE engine resources (GEER), this enable to have three choices of evaluation, the traditional eduction model of GEE, AspectJ-based eduction model, and probabilistic model checking with the PRISM backend.

Ground work for implementing self-forensic autonomous property within ASSL toolset was analyzed for implementation. Future work will be to complete implementation and export to other target software system of ADMARF, AGIPSY [23], and others described conceptually in [22].

### 2) *GIPSY*

The GIPSY model is a gathering of replaceable Java segments [24] they are orchestrated in three bundles

- Collection of compilers for programming dialect (GIPC).
- Runtime Programming Environment (RIPE).
- General Eduction Engine (GEE).

In Earlier days, clear projects were composed by the designers which are basic, clear, characteristic and utilized for particular useful reckoning [24]. Wanderer gives a schema to executing projects composed in clear programming dialect [25]. These projects does not help much I/O frameworks and totally rely on upon existing records (like libraries). In Later stages, GIPSY structural planning was proposed so as to fabricate mixture projects joining Java with these clear projects [24].

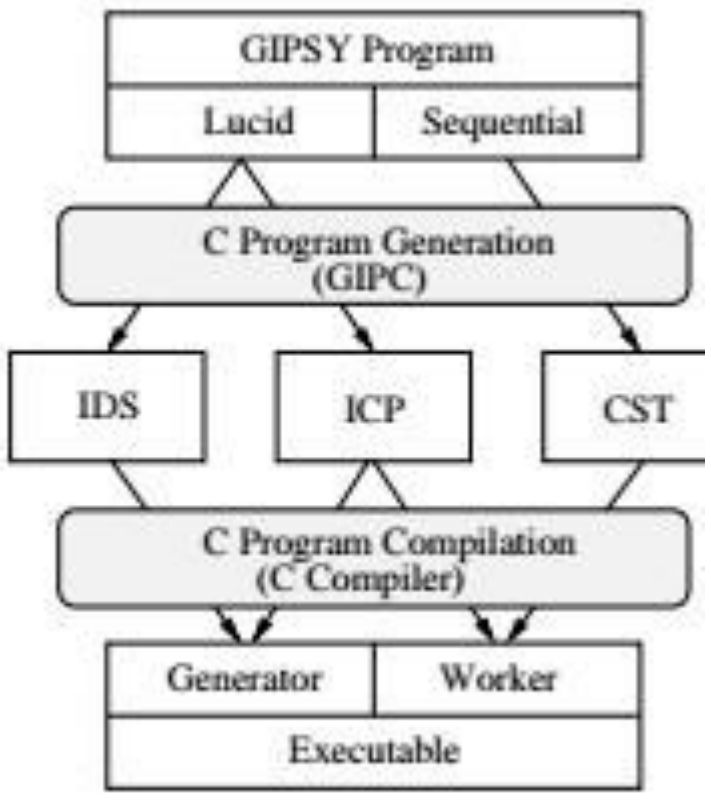

*Figure 5 GIPSY Program Compilation Process [10]*

The source code comprises of two parts: the Lucid part and the granular part.

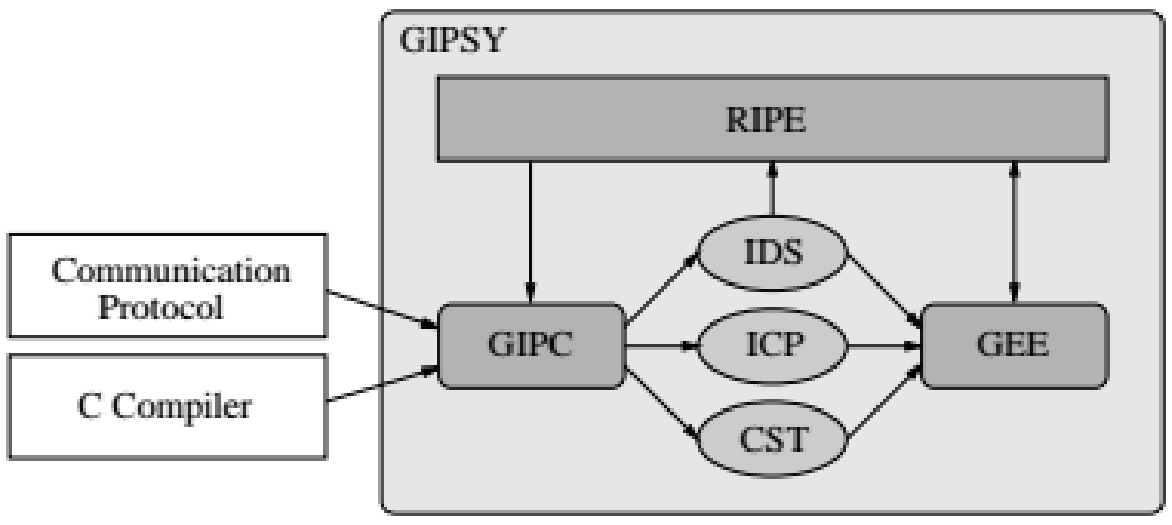

*Figure 6 GIPSY Software Architecture [10]*

The essential objective of the GIPSY architecture is the dialect freedom. Generic education eduction engine resources (GEER) are a middle presentation produced by the compiler to accomplish dialect autonomy. Gipsy projects a stage for researching the deliberate and half and half purposeful and basic programming. Beginning configuration postpones the execution of the run-time framework, which is General Eduction Engine (GEE). GEE primarily concentrates on the re- planning the building design and backing the new components in the complex situations [26]. GIPSY has a multi-level construction modeling, every level is a theoretical substance speaking to a free unit and collaborate with different units to execute a project [25]. Keeping in mind the end goal to have self-versatile ability to oversee itself on developing complex nature's domain, Autonomic General Intentional Programming System (AGIPSY) was presented which takes a shot at autonomic registering for making the work of the complex workstation frameworks in a simple and more astute way. The primary point of the AGIPSY framework is to develop GN into autonomic components [27]. GIPSY framework is chiefly composed with a plan to help a typical run-time environment and concurrence of the Intensional programming.

AGIPSY Architecture: GIPSY is a multi-layered construction modeling. The GIPSY hubs (GN) can transform nearby information and productively speak with one another to process a specific appeal in spite of the fact that coordination of these GN is a real issue in AGIPSY.

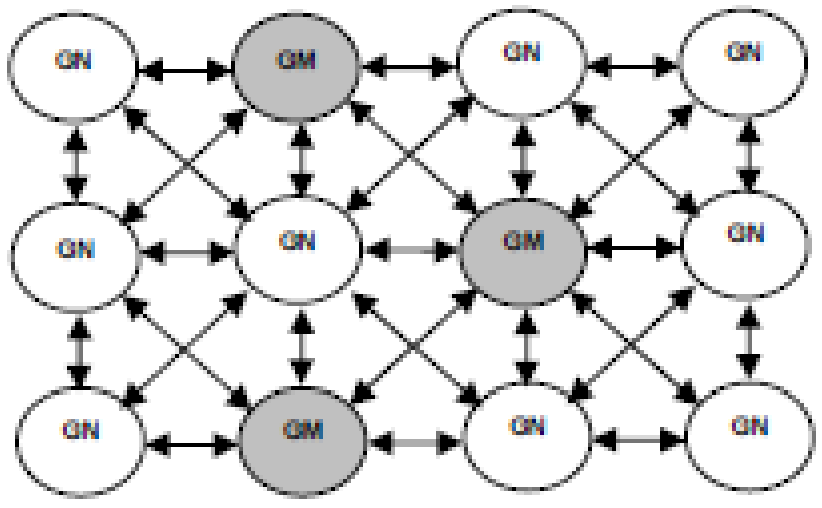

*Figure 7 AGIPSY ARCHITECTURE [10]*

AGIPSY comprises autonomous GNs (GIPSY Nodes). Each GNs are autonomous and has control over its actions and state. Control provided by node Manager (NM). The GNs running a GMT instance are global autonomic managers, which monitor and manage the work of the entire system. These GNs are called GIPSY managers (GMs).

Some of the autonomic features are also described as sub section stated as:

- Fault Tolerance & Recovery: GNs can recover from many types if failures.
- Self-Maintenance: With Node Managers (NMs) having full control of GNs, the self-maintenance can be achieved and defined by: 1) Self-Configuration; 2) Self-Optimization; 3) Self-Healing; 4) Self-Protection

GIPSY AE: By its working a GIPSY AE is GIPSY node and autonomic properties is achieved using node Manager (NM). There are four distinct control-loop components- Monitor, Simulator, Decision Maker and Executor. In addition there are two controllers: channel controller and sensor controller. Channels are means of communication among NMs and channel Controller is responsible for sending and receiving messages over the channels. Sensors are used to measure parameter of GIPSY tier instance and Effectors are kind of "manageability interface" used by executor to control managed GIPSY tier instance. Sensor Controller is responsible for controlling the sensors. Moreover ASSL Knowledge is full

ASSL specification of the GIPSY AE. There are different trade-offs as:

- Performance Trade-off: To monitor and control is performance overhead.
- Scalability-Complexity Trade-off: Scalability is also important issue while designing AGIPSY architecture.

Then with the advances in software engineering design, using the Demand Migration Framework (DMF) for implementing multi-tier runtime system for GIPSY in order to execute the Intensional-imperative programs which are using java [28]. JINI and JMS which are a part of DMF are emerged with distributed computation for GIPSY. The development of ongoing architecture deals with constructing a group of wrapper classes for each tier, specifically to detail how the demands are propagated within the GIPSY node [29]. This project mainly deals with demand propagation and their values, for each and every demand a unique syntax is defined to identify the existence within the GIPSY network.

The goal of this is to provide a feasible solution for the existing intentional imperative programs to achieve high scalability and flexibility. The framework which is produced can achieve some qualities such as maintainability and extensibility. As extensibility is possible the classes and interfaces for implementation of a program are kept under a single package called gipsy.GEE.multitier with sub packages with corresponding wrappers.

```
import gipsy.Configuration;
public interface IMultiTierWrapper extends Runnable
{
. . .
startTier(); stopTier();
setConfiguration(Configuration);
ConfigurationgetConfiguration();
. . .
}
```

Primary API of the IMultiTierWrapper Interface [30]

The multitier architecture which is implemented is more flexible and can adapt to the changes quickly. Additional layers such as

MultiTierWrapper, ITransportWrapper, ITransportAgent and class tool GEERPool are included in the architecture in order to synchronize any furthermore changes in the design of the multitier runtime system to achieve complete flexibility and extensibility.

Moving on to GIPSY network automation, to simplify tiresome GIPSY network management there is development of a new element which enable the user to interact, create a network for example graph. Also, it allows visually to master the network parameters at runtime. The overall objective is to provide more control to the user at runtime, below are the requirement and design specification of the above concept. To start with GIPSY GMT plays the crucial role specifically in network related tasks hence is used for originating and terminating nodes, as well as the allocating and de-allocation tiers at run time. GMT graphical user interface was executed in Java JFC/Swing library also configuration the file of GIPSY was used to store various configuration information. JUNG, an open source library, was also used in addition to GMT to remove manual complexity. As shown in figure below from mouse click user can start or stop GIPSY nodes which uses GMT in back also below is the architectural level representation of JUNG classes.

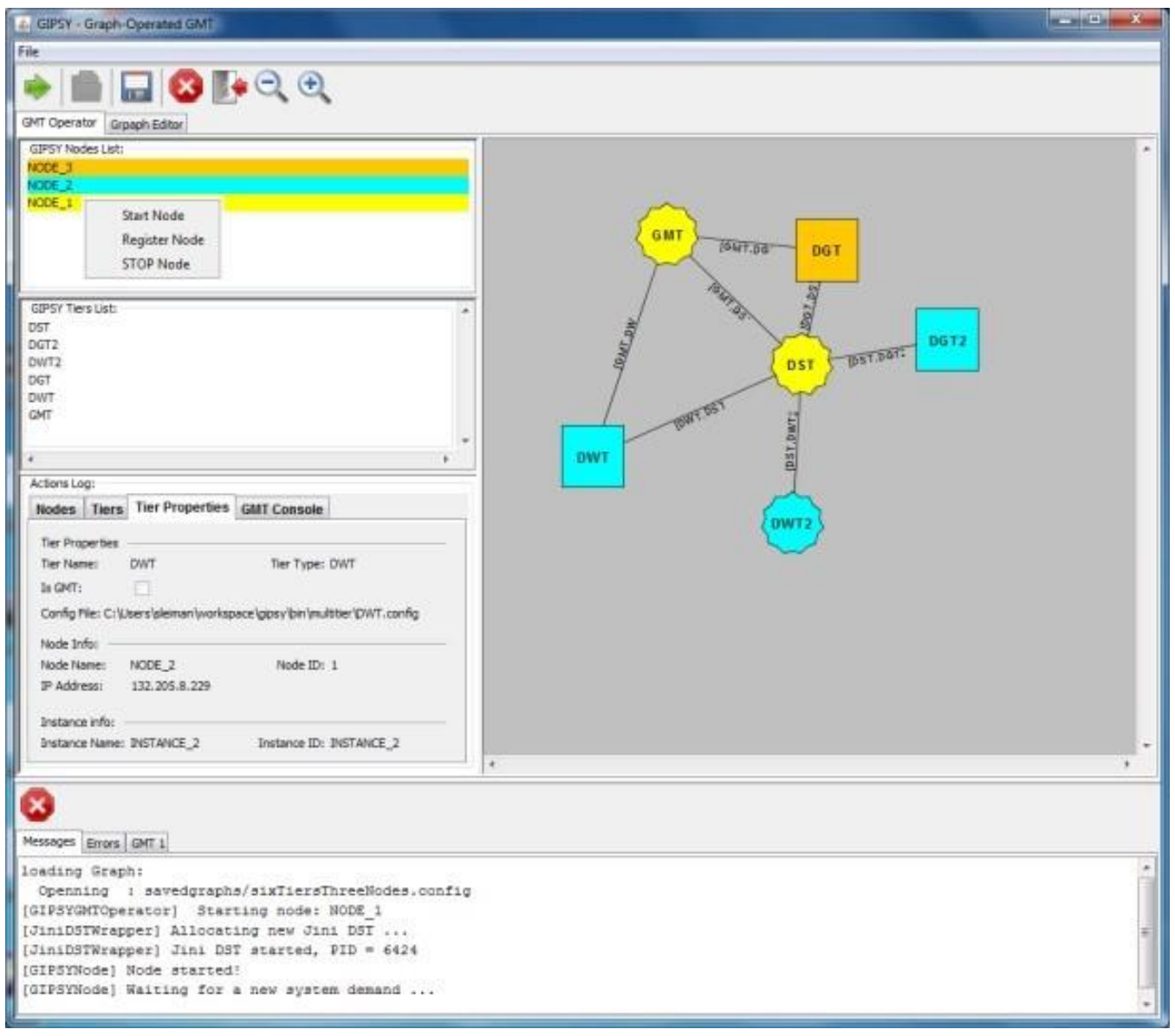

*Figure 8 GMT OPERATOR VIEW*

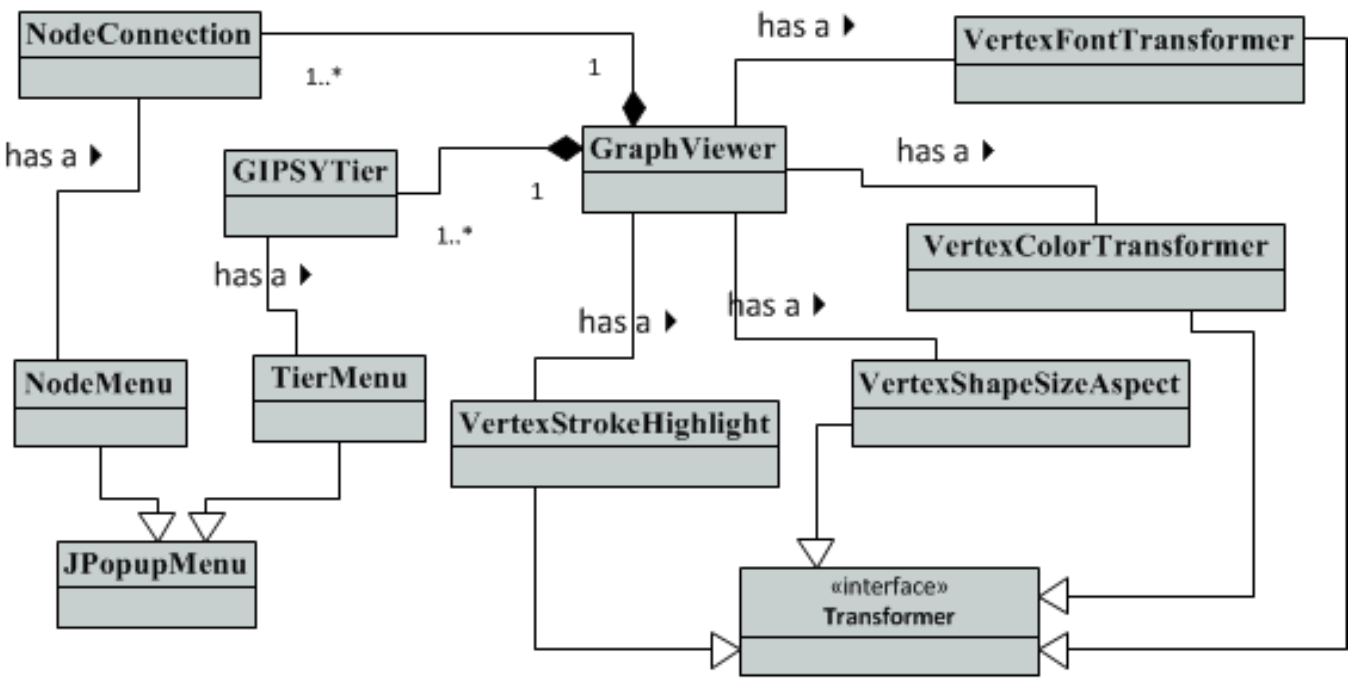

*Figure 9 Visualization Classes[6]*

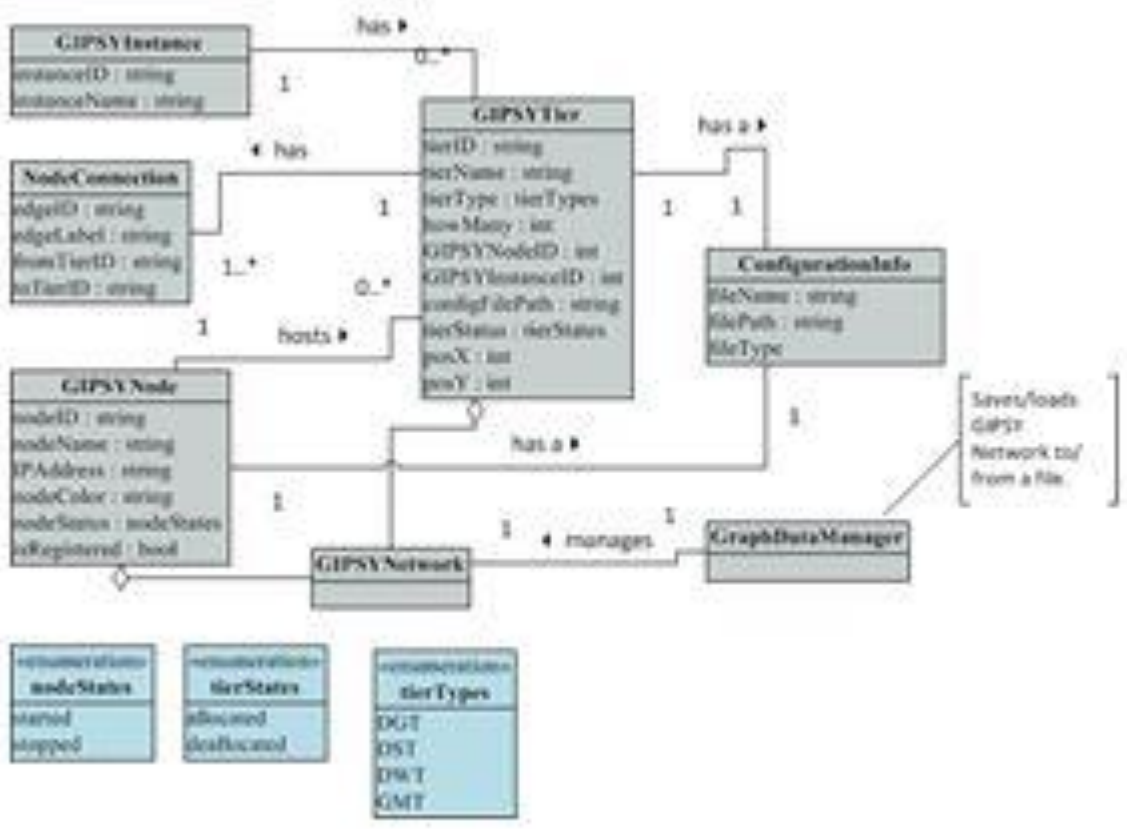

*Figure 10 Graph-Related Data Structures [6]*

GIPSY scales very well, but when interacting many GNs its complexity grows rapidly. AGIPSY is autonomic version of GIPSY based on multiple Interacting autonomic GNs. At the GN level, every GN have embedded NM (node manager) which monitors and manages the GIPSY tiers deployed on that GN.

To conclude for GIPSY, JLucid, JOOIP and cybercrime forensic act as a domain model for GIPSY Architecture. Mainly in cybercrime digital forensic act as an event reconstruction, Forensic Lucid is mainly used for automatic deduction in cybercrime domain.

And for Distributed MARF, the classical MARF was extended [33], [34], [35] to run on distributed environment comprising distributed nodes as well as front-end. The medium of communication over the network is Java RMI [36], CORBA [37] and finally, XML-RPC [38]. The detailed Pipeline is shown below using replica manager and Backup facility.

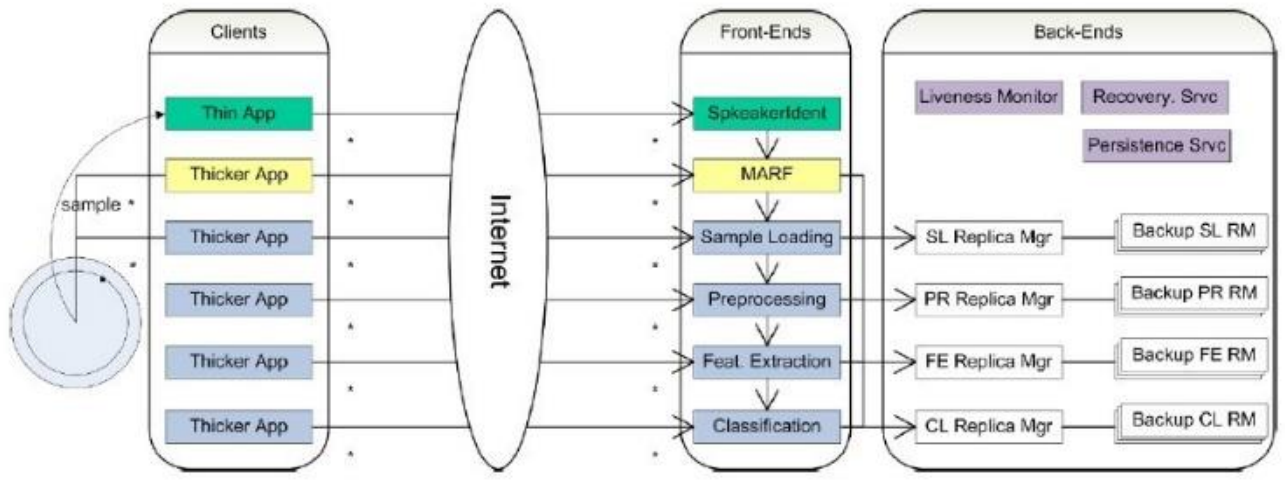

*Figure 11 Distributed MARF Pipeline [32]*

### *3) SUMMERY*

Software measurement aspects like LOC, NUMBER OF FILES, NUMBER OF METHODS and NUMBER OF CLASSES is calculated below. For this purpose we chose different tools to verify the results, as the definition of software measurement properties varies with respect to tools we got different results.

- Line of code and Java Files (Linux System) Procedure followed to determine Line of code in an OSS study:

To calculate Line of code in a particular project, some key steps were taken into consideration. First, the source code of the project was downloaded into a Linux system. Then with a build in command of Linux Operating system as defined below, the line of code was found.

The command used is: find –name "*.java" | xargs wc –l

This command finds all the java files in the current and all the sub folders along with Line of code status, the only disadvantage is that any file containing blank line will counted as Line of code. Moreover, the above command without xargs argument will display the number of files with "*.java" extension.

*Table 1 DMARF GIPSY LOC and Java File Calculation (Linux Command)*

| | LOC | JAVA FILES |
|---|---|---|
| GIPSY | 139632 | 602 |
| DMARF | 131706 | 1024 |

- Number of Classes and Number of Methods calculation Method (using eclipse)

Procedure followed to determine Number of Classes and Methods in an OSS study using Eclipse. While determining the number of classes and Methods procedure the was to search for the regular expression with options as "all files with type as declaration for counting number of classes" and "all files with Methods as declaration for counting number of Methods", respectively.

*Table 2 DMARF GIPSY Number of methods and number of classes using declaration type in Eclipse*

| | Number of Methods | Number of Classes |
|---|---|---|
| GIPSY | 5,680 | 666 |
| DMARF | 6,305 | 1054 |

CodePro Eclipse plugin Results (LOC, Number of methods)

*Table 3 DMARF GIPSY LOC and Number of Methods using CodePro Eclipse plugin*

| | Number of Methods | Line of code |
|---|---|---|
| GIPSY | 5680 | 104,073 |
| DMARF | 6,305 | 77,297 |

InCode[24] Results

*Table 4 DMARF GIPSY Using inCode*

| | Number of Methods | Number of Classes |
|---|---|---|
| GIPSY | 6,468 | 702 |
| DMARF | 7,554 | 1058 |

Snapshot of the calculation are attached in Appendix B.

## III. Requirements and Design Specifications

### *A. Personas, Actors, and Stakeholders*

#### *1) DMARF*

DMARF is an extension to MARF over distributed network. MARF (Modular Audio Recognition Framework) is basically collection of different algorithms for analysis for audio (sound, speech and voice) and natural languages, moreover it recognize with sample application (NLP, etc.) for its use and implementation in Java.

MARF is composed of different Algorithms namely SpeakerIdentApp (used for identifying the speech or voice of the Speaker), LangIdentApp (used to identify the language in voice or speech sample). While processing the speech and natural languages, the use of Zipf's Law is considered, which states that "in any given natural language, frequency of any word is inversely proportional to its rank in frequency table. Therefore most frequency word will occur approx. twice as often as second most frequent word and similarly three times as often as third most frequent word."[40]

These applications inside MARF framework are used for different purposes. For instance, SpeakerIdentApp for Identification of Speaker (location) based on its voice sample. In this system you train the system based on your own voice and then test any sound wave for result.

This framework also used for Natural language parsing using probabilistic method. "A natural language parser is a program that specifies the grammatical structure of sentences, for example, which groups of words in Natural Language go together (as "phrases") and which words are the subject or object of a verb. Probabilistic parsers uses knowledge of language gained from hand-parsed sentences for testing to produce the most likely analysis of new sentences."[41]

This type of framework can be used by some of the professionals in following areas

- Security based areas like securing a lock with speech, voice
- Speech recognition software developers
- End users using different applications based on voice control, e.g. Siri, google now, etc.
- Automated identification of User
- Using Natural Language Parser to help understand basic syntax and structure of any language
- People with disabilities can be benefited by making voice commands
- Using speech recognition for automating control of a device from any where

Based on different domains where the framework can be used, the primary and secondary actors can be defined as below:

**Primary Actor**: Audio recognition software developer

**Secondary Actors**: Authentication System

The **stakeholders** for specific domain will be:

- Audio recognition software developer
- Maintainer of MARF (i.e. Serguei Mokhov)

**Primary Actor:**

Audio recognition software developer: Audio recognition software developer is a person who wants to develop a software for security devices, so that an authenticated person is allowed or permitted. He uses MARF framework for training the security system with voice of authenticated personnel and then checking using MARF while permitting the personnel.

**Secondary Actors:**

Maintainer of MARF: This is the person responsible, if problem arises in the part of MARF while testing and making the software. The relevant information regarding MARF and its working can be discussed and worked together for producing an efficient software system

**Primary Persona:**

Name: David N. Armenta
Photograph:

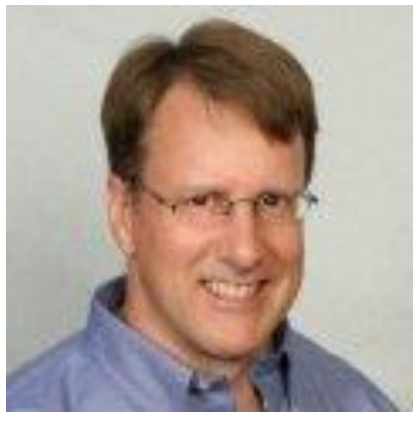

Personal: 27 years old, Single, pursuing Master degree from Concordia University. David is interested in developing Audio recognition software.

Background information: David grew up in Montreal, he was passionate about computers in his school days, so he followed to pursue his bachelor from Concordia University. He is very much energetic regarding developing different software and his interested field is Audio recognition.

Key Goal: Moving further for his passion he developed couple of audio recognition software, but he got struck while developing using network as a medium. He tried a couple of software for audio recognition and is looking forward to develop audio recognition software based on real time. He has heard about MARF and after knowing the fact that DMARF is extension of MARF in distributed systems will help him to develop real time based audio recognition software.

2) *GIPSY*

General Intensional Programming System (GIPSY): GIPSY is an intensional programming language which is mainly used for investigating intensional programming language of LUCID family. GIPSY is an open source platform which uses JAVA. Previously all intensional programming languages use C, but in GIPSY all the background work is done using JAVA. GIPSY is a language independent runtime system. It is a collection of three compilers General Intensional Program Compiler (GIPC), General Education Engine (GEE) and Generic Education Engine Resources (GEER).

GIPSY for reasoning of intensional expressions uses Higher Order Intension Logic (HOIL). GIPSY can be used in application of cyber forensic. Considering cyber forensic as a domain and retrieving the actors, stakeholders for this domain. Class, as these class exists independently but are part of the super

Cyber Forensic is also known as computer forensic or digital forensic uses digital media for identifying, preserving, recovering and analyzing digital information. Cyber forensics mainly deals with retrieving the digital data with a legal audit trail. In the recent days this system is mainly used for investigating crimes related to fraud, murder and cyberstalking. Cyber Forensic is also known as computer forensic or digital forensic uses digital media for identifying, preserving, recovering and analyzing digital information. Cyber forensics mainly deals with retrieving the digital data with a legal audit trail. In the recent days this system is mainly used for investigating crimes related to fraud, murder and cyber stalking.

To overcome difficulties and benefits we apply intentional logic to automated cyber forensics and compares with previously automata approach. Intentional logic makes regular mathematical and logical expressions context-aware i.e. first class value can be controlled by logical expression via context operators. Forensic lucid language shows directions for evidential statement to a particular crime scene.

**Actors and Stakeholders:**
**Actors: Primary Actor (**Physics Experts**)**

Physics Experts is a primary actor for this system. He uses system very often to retrieve the information and for analyzing the data related to any crime by identifying the digital information. System is mainly operated by data analyst and has the right to access the information and modify the data under certain circumstances. Analyst should follow certain legal procedures for investigating the crime if not the case may not be valid.

**Secondary Actor: System engineer**
System engineer is a secondary actor who rarely uses the system. As there will be large data to be managed and dealing with large data may lead to risk management so here comes the system engineer role by reducing the risk management and by maintaining the large complex data.

**Stakeholder:**
There are few stakeholders who are related to this system. Stakeholder is a person or organization who are affected by the result of the system actions. For this cyber forensics system there are few stakeholders they are

**Physics Expert, Forensic Scientist and Expert**

System uses GIPSY technology for finding out the significance contextual analysis by inputting source code written in international programming.

Primary Persona:
Name: Richard Shurtz
Photograph:

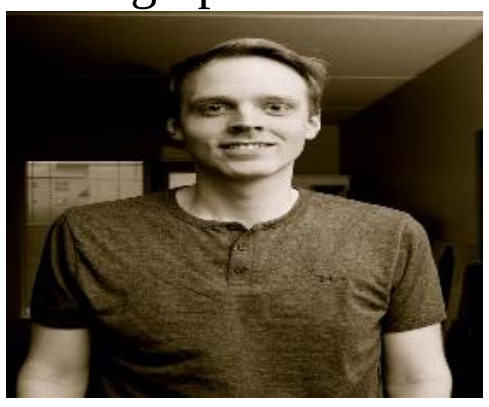

Personal: Richard Shurtz is working as a software engineer. He like to dig deep in security related software.

Background Information: Richard Shurtz is a software engineer at LUCID SOFTWARE. He persuaded his M.s degree from BYU University in 2014. He speaks English and Thai very good. He has done his masters in Computer Science and is good at developing programs. Previously he done his under graduation degree in Brigham Young University in the same stream as of his masters.

Key Goal: Richard Shurtz works in a software company which is using LUCID as a programming language. He develops applications using this language, as lucid is good for developing and can easily understood by any non-technical user. Recently he is facing many problems with operators and mathematical equations using this lucid programming language, he can't change his domain to another language as he is in a company where developing is based on LUCID. So he is looking for an extension in lucid which makes him easy with the previous difficulties and can be easily understood by any user, which can support multidimensional reasoning of expressions. Days, so he followed to pursue his bachelor from Concordia University

*B. Use Cases*

*1) DMARF*

**Use Case Brief**

UC: Get Authenticated

Actor: Audio recognition software developer (User)

The User trained the MARF framework with authenticated person audio sample. Then user inputs sample that is to be tested for comparison in the application using MARF framework. The Sample is loaded using Sample loading procedure, then preprocessing takes place to remove unwanted noise. Then the feature is extracted using algorithms in MARF to match the trained sample. Resulting in sample classification based on matching to the relevant source. The result is displayed and accessed by the User.

**Use Case Fully Dressed**

| Get Authenticated | |
|---|---|
| Use Case UC1: Get Authenticated | |
| Primary Actor: Audio recognition software developer (User) | |
| Stakeholders and Interests: | User: wants accurate and fast result |
| | Maintainer: wants software to be famous |
| Preconditions: | Application is Configured and trained for specific user |
| Success Guarantee (Post Condition): | Application compares and matches accurately |
| Main Success Scenario ( basic flow): | |
| | 1. **User** inputs **sample** to **application,** uses **MARF** |
| | 2. The Sample is **loaded** |
| | 3. The Sample is **preprocessed** for noise removal |
| | 4. The Sample **feature is extracted**. |
| | 5. The **sample is classified** and generated based on match. |
| | 6. The **result** is displayed and relevant action is performed. |
| Extensions (alternate flow): | |
| | 2a. The sample could not be loaded |
| | 3a. The sample can not pre processed. |
| | 4a. Features can not be extracted. |
| Special Requirements | legal and other formalities must be completed. |
| Open Issues: | Different mode of authentication to be discussed? |

*Figure 12 FullyDressed Use case*

2) *GIPSY*

**Use Case Brief**

UC: Get Result

Actor: Physics Experts

The Analyst inputs the specimen to GIPSY based software using GIPC Compiler, the input specimen is converted by GIPC into C program. The converted program is then compiled using C compiler and then the program is executed. The relevant result related to specimen is generated and after wards analyzed by Data Analyst.

**Fully Dressed Use Case**

| | |
|---|---|
| UseCase ID | UCGIPSY01 |
| Name | Generate Contextual Graph |
| Level | System Level |
| Primary Actor | Physics Experts |
| Secondary Actor | GEE |
| Stakeholders and Interests | Developer, Researcher |
| Success Guarantee | Contextual Graph is generated on input of lucid code. |
| Main Success Situation<br><br>1. **User** authenticates itself<br>2. The System opens the application and user provides the Specimen<br>3. Specimen is converted by **GIPC** to C program<br>4.C Program compiles with help of **C compiler**<br>5. System generates **result**<br>6. result analyzed by **User** | |
| Technology and Data Variations List | Lucid, C, Java |
| Special Requirements | GIPC (GIPSY compiler) |
| Frequency of Occurrence | High |

*Figure 13 Fully Dressed Usecase*

*C. Domain Model UML Diagrams*

*1) DMARF*

The conceptual classes are defined below taken while describing the domain of DMARF. It consists of classes used while developing an application based on DMARF for authentication of authorized personnel.

**Conceptual Classes and their Relationships:**

**User:** The user is application developer who takes the input of voice sample used to compare with the pre trained sample of DMARF for authentication. One user is capable of accessing one or many applications which uses MARF framework.

**Application:** This class is used for main interaction between User and DMARF. The application will make use of MARF framework for comparing the sample. There might be one or many applications which will use only one DMARF framework. There are three other classes which are a part of application class as they are controlled by the Application class whenever there is any use with these classes it uses them, they are SpeakerIdentApp, LangIdentApp and ProbabilisticParsLangApp. These three classes can exists independently and they are a part of application class.

**DMARF:** This is the main framework used for fulfillment of the comparison of voice samples. As applications are passed to DMARF the process is done and the result is displayed. One DMARF framework can provide only one result. DMARF will take the inputs from the specimen which consists of audio and voice samples and are used for comparison purpose. There are 4 other classes which can exists independently and are a part of DMARF class. They are

1) Loading Sample: This class corresponds to the inner component of DMARF for loading the sample. One Loading Sample class follows Sample Preprocess class.

2) Sample Preprocess: This is also inner component of DMARF, used for removing unwanted components like noise, etc. One Sample Preprocess class extracts from one Sample Feature Extract class.

3) Sample feature extract: This class is responsible for extracting the main features based on algorithms stored in the DMARF Framework. One sample feature extract class generates one Sample Classification class. Here sample feature extract class uses another class named Algorithm class for the sake of name. Only one sample feature extract class can have one or many Algorithm classes. Algorithm**s:** This inner Class constitute to algorithms used for feature extraction.

4) Sample Classification: This class provides the comparison and classification of both the sample (trained and testing sample)

**Specimen:** This class specifies the audio or voice sample used for comparing with pre-loaded trained voice sample for authentication. Here two other classes are inherited i.e.; Voice and Natural Language classes.

**Result:** This class specifies the final result, which is used by the User for further processing.

The inner classes or components are shown as aggregation relation to the parent class, as these class exists independently but are part of the super class. [Domain Diagram is on the next page]

2) *GIPSY*

**Conceptual Classes and their Relationships:**The Domain Model of Gipsy can be explained by using the class diagram. We are using different type of classes namely:

1) User: It is a class many number of user uses only one RIPE class to start the process.

2) RIPE (Run time Programming Environment): In this class many users interacts with one RIPE to run the program and RIPE uses the GIPC to compile the program.

3) GIPC (General Intensional Programming Compiler): It is a class where one GIPC uses the one C Compiler to compile the program file.

4) Source Directory: This a class which contains the source code. The file in this class is compile by the GIPC class. Lucid class and sequential classes are inherited by the program file.

5) GEE (General Eduction Engine):This is a class which is generated by the GIPC classes. This class consist of generator and worker classes. RIPE uses the worker classes to give the result of a particular input.
C Compiler: When GIPC uses the C Compiler to compile the program this class gets call.

3) Fused DMARF-Over-GIPSY Run-time Architecture (DoGRTA)

GIPSY'S circulated nature is to a great extent relying on its runtime structural engineering, where Lucid Language variations are assessed by a specialist generator pair utilizing build up level as a range. These demonstration as levels in a hub and every level can exist over a few hubs over the system, hence making the assessment offbeat. Though, the DMARF's pipeline method makes its assessment of the source record seem synchronous making headway in stages. Yet, its appropriated structural planning is made absolute through the front-end modules (preprocessing, Feature Extraction, Characterization) figuring on different hubs.

The main goal to wire DMARF over GIPSY, a dialect vigorously impacted from lucid family, MARFL [42], which is in charge of scripting the message to be performed is not yet processed. To use the GIPSY's multi-level architecture, we make an issue particular generator and laborer levels (PS-DGT and PS-DWT individually) for the MARFCAT application. The generator(s) produce requests of what needs to be processed as a document to be assessed and store such requests into a store look after by the interest store level (DST) as pending. Specialist's pickup pending requests from the store, also them transform then utilizing a conventional MARFCAT occurrence. When the result is figured, the PS-DWT store it go into the store with the status set to registered. The generator "harvests" all registered results and produces the last report for an experiments. Various experiments might be assessed all the while then again a particular case might be assessed distributive. This methodology serves to adapt to substantial sums of information and decline from computing warnings that have as of now been figured and reserved in the DST. The said PS-DGT produces the requests and stores in Demand store to be got by PS-DWT which executes it on disseminated hubs.

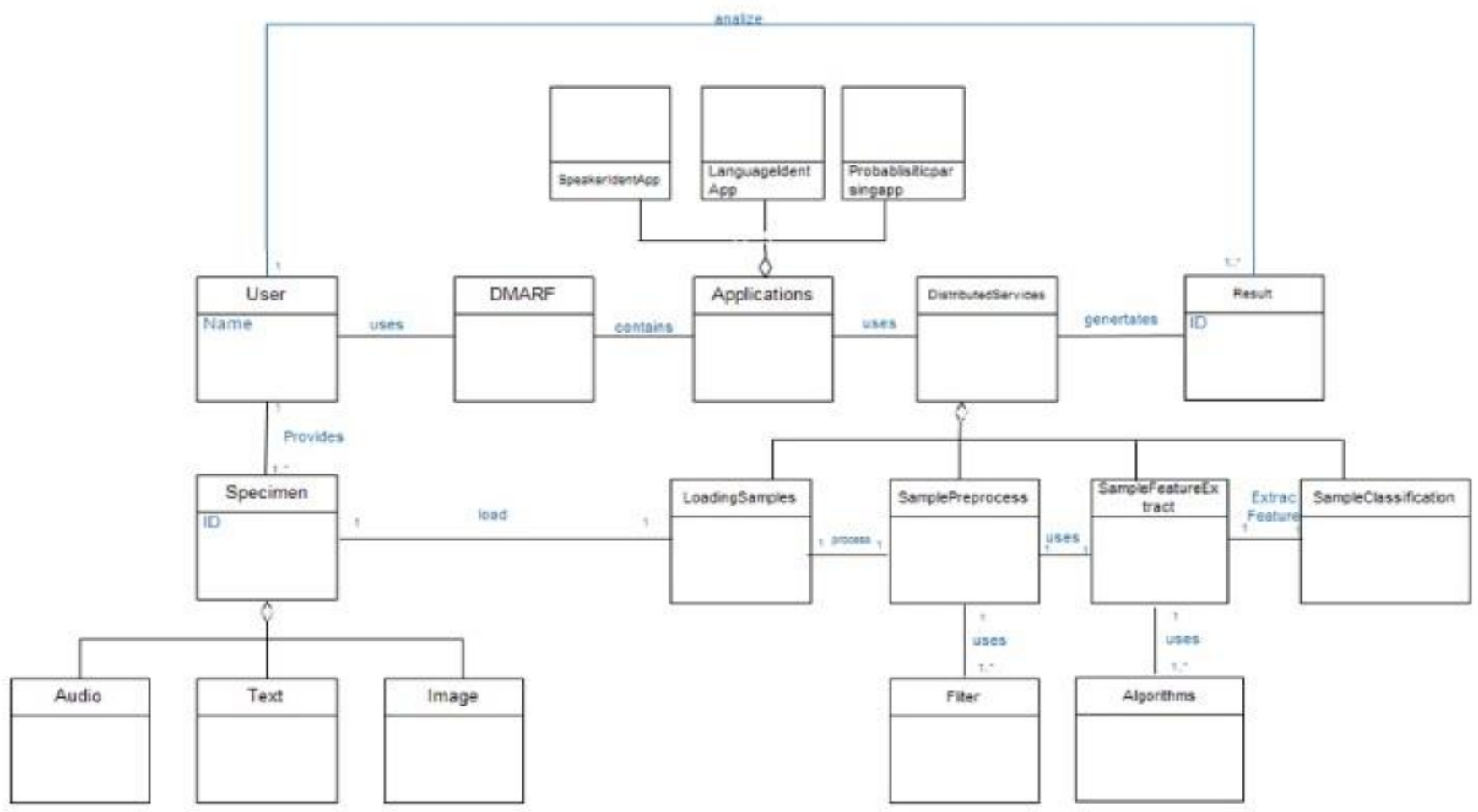

*Figure 14 Domain Model DMARF*

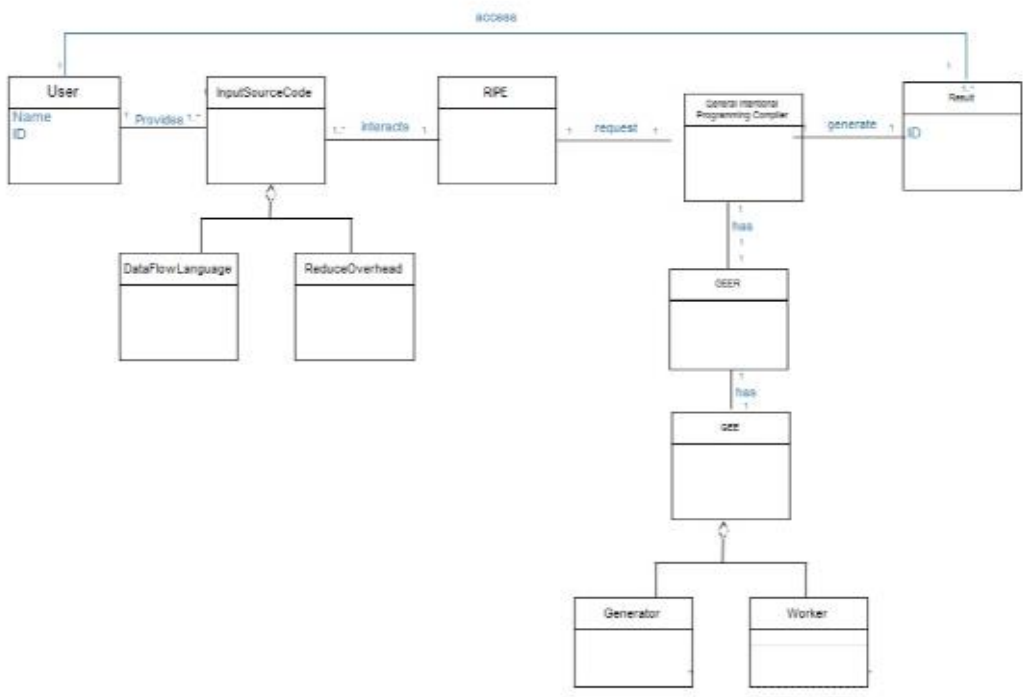

*Figure 15 Domain Model GIPSY*

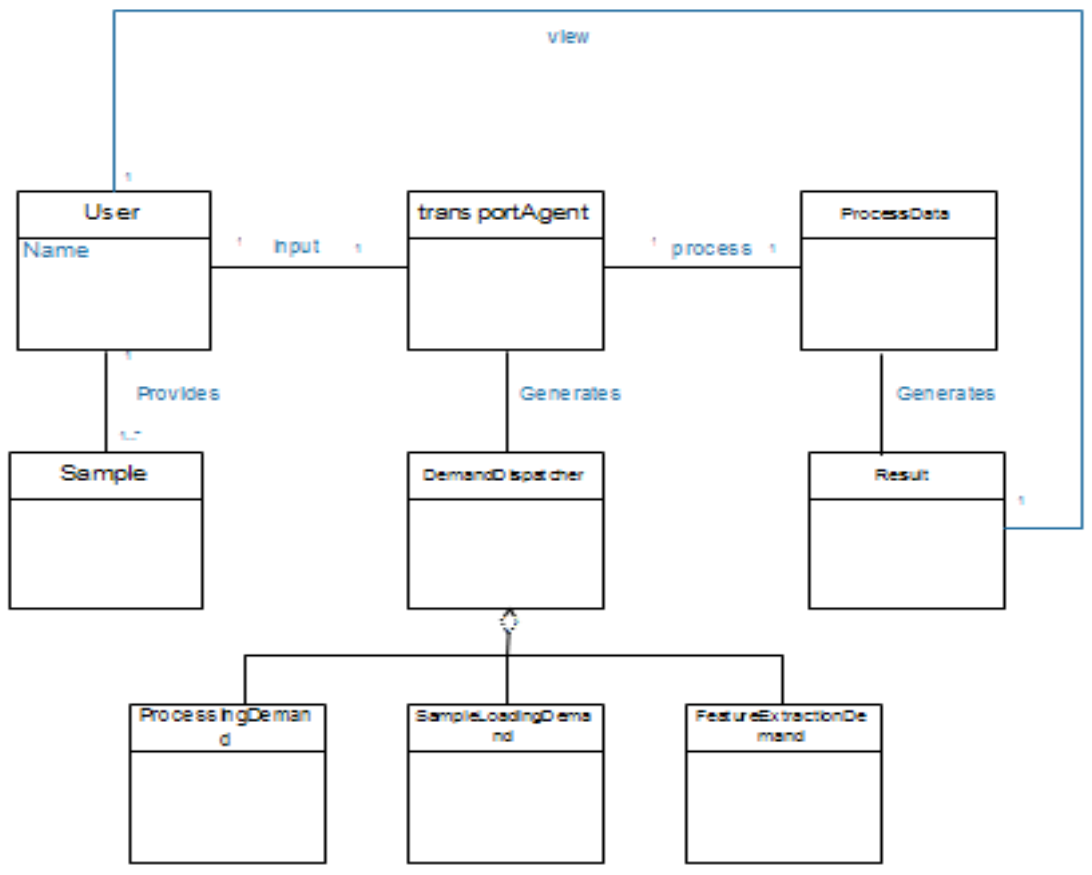

*Figure 16 DMARF over GIPSY run-time architecture*

## D. Actual Architecture UML Diagrams

### 1) DMARF

**Actual classes in DMARF**

Feature extraction:- In DMARF extract features () is used which is taken from IPreprocessing .GetMARFsource code revision() is used in feature extraction class.

Ipreprocessing : Removenoise() is used in this class and it removes unwanted noises.

Classifications :- In classification class train() , dump() ,restore () are inherited to get the result from distance stochastic , radom classification , cosine similarity measure classes.

MARF :- In MARF configuration() , sample format(), sample file () are used in executing the file whereas NLP checks the language and engram checks whether it is unigram or bigram ,EStatistical estimators are used to add one or group for executing the file.

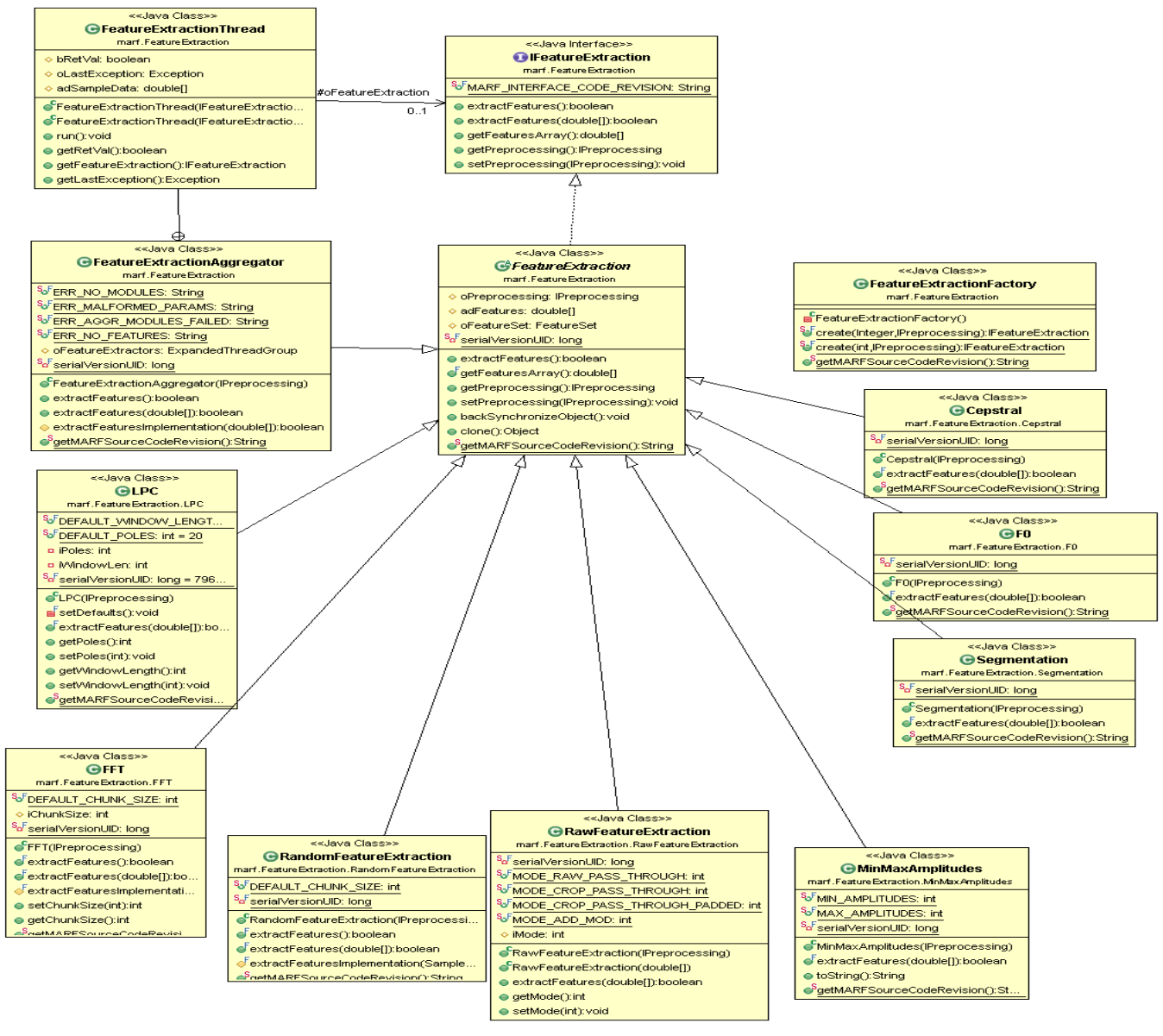

*Figure 17 FeatureExtraction class diagram*

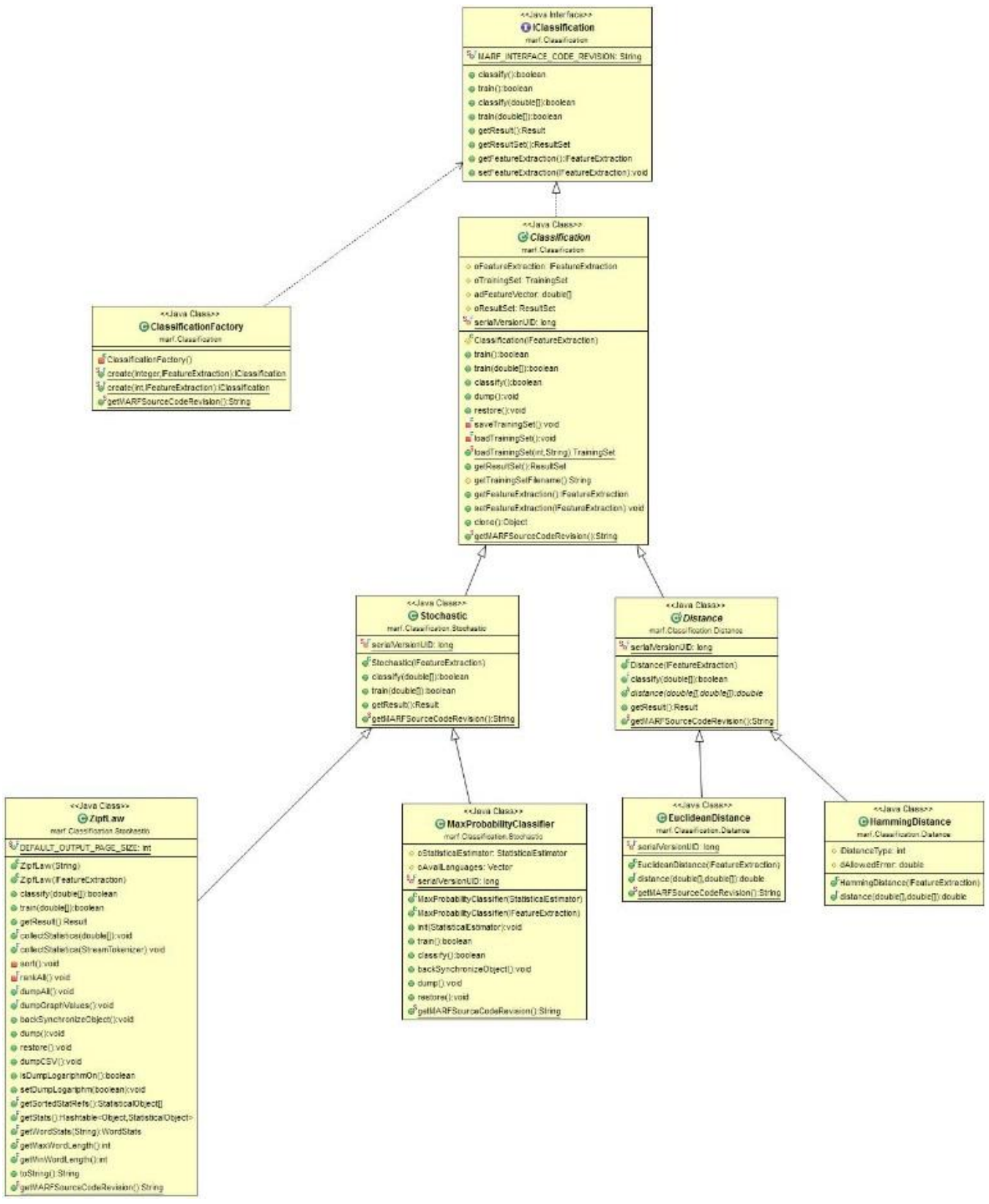

*Figure 18 Classification class diagram*

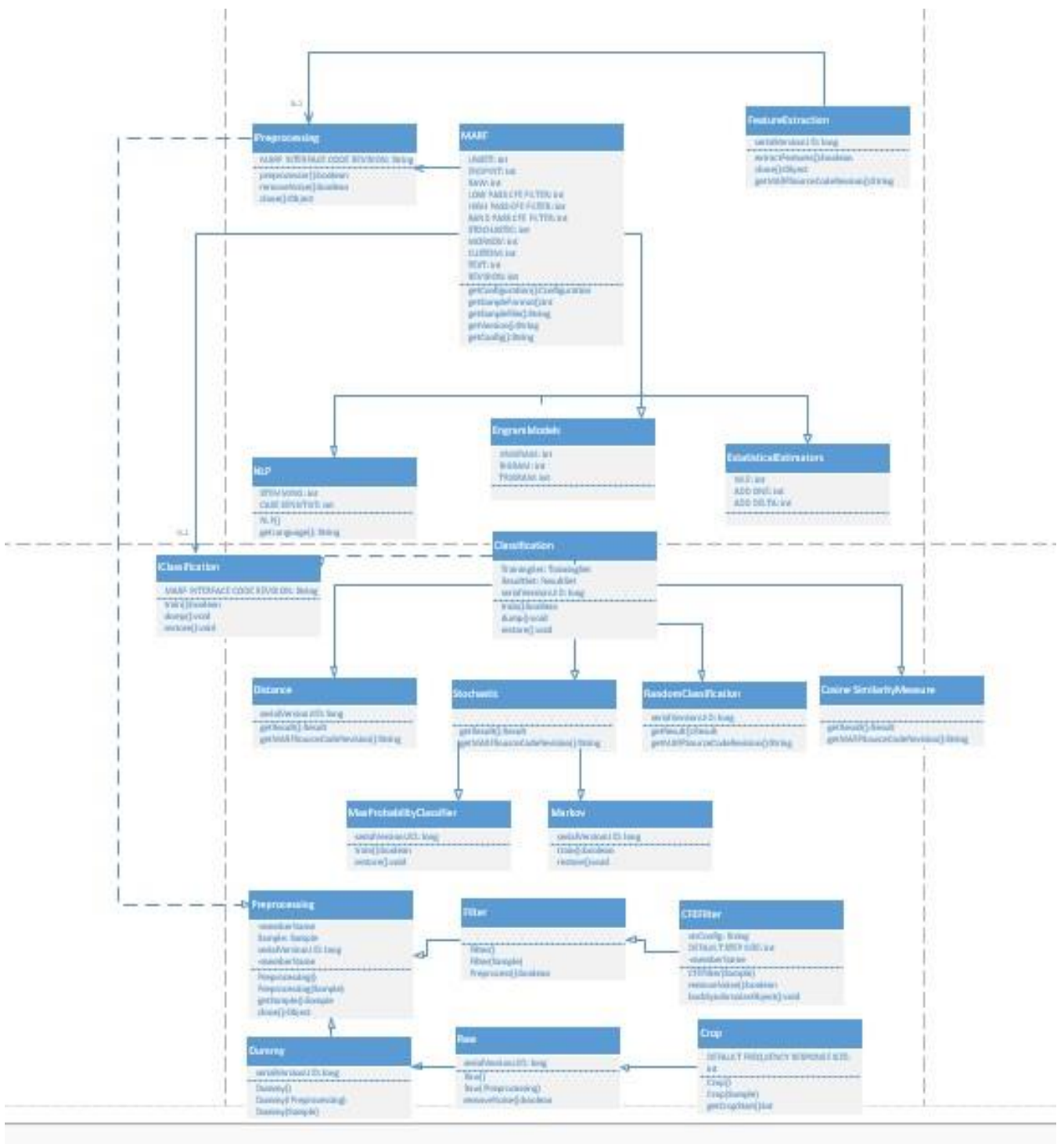

*Figure 19 Actual Classes in DMARF*

**Mapping between actual classes and conceptual classes for DMARF:-**

| CONCEPTUAL CLASSES | ACTUAL CLASSES | Differences |
|---|---|---|
| Sample preprocess | I Preprocessing | Sample preprocessing and Ipreprocessing removes unwanted components like noise and extracts a classes from feature extract class |
| Sample Feature Extract | Feature extraction | In conceptual classes sample feature extract is based on algorithms whereas feature extraction will import classes to IPreprocessing. |
| Sample Classification | Classification | Sample classification provides comparison and classification between two tests. Classification uses cosine similarity measure class to get the result from testing samples. |

**Relations between the Actual classes in DMARF:-**

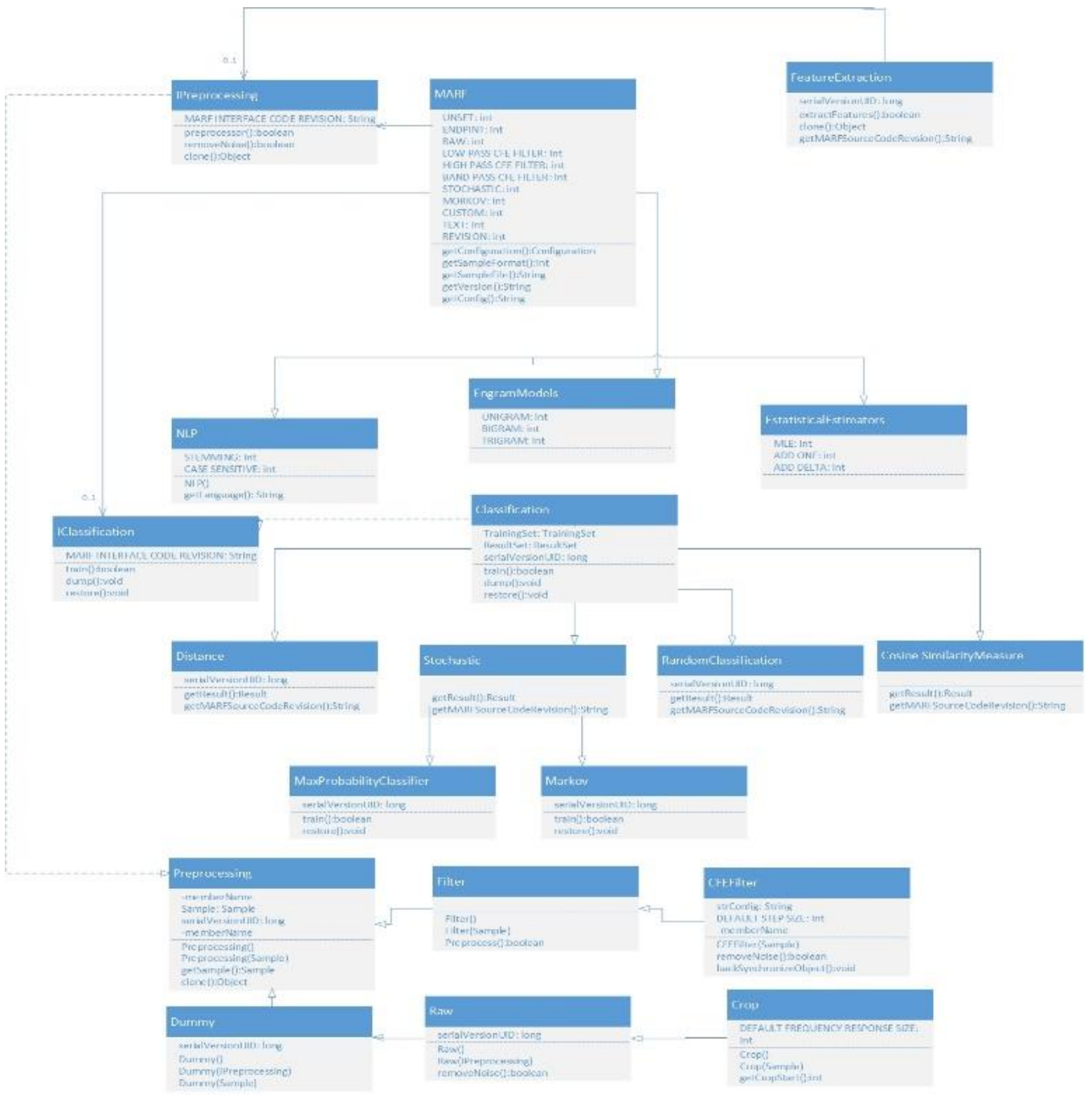

*Figure 20 Relations between the Actual classes in DMARF*

In the above diagram MARF is the main class. IPreprocessing extracts the classes from feature extraction and Iprepocessing sends the classes to main class MARF.MARF class inherits the NLP, Engram models and Estatistical Estimators.

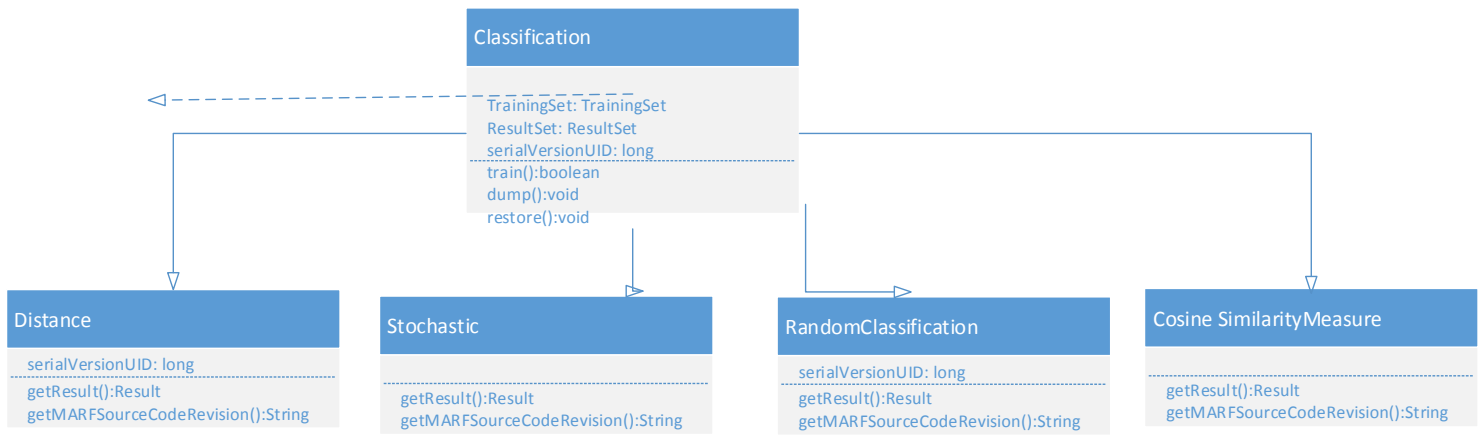

*Figure 21 simplified version of class interaction DMARF*

In above diagram the classification class is divided into distance, Stochastic, Random Classification and Cosine similarity to get the result from testing samples.

**Dmarf code relations between the classes:**

**FeatureExtraction:**

```
public abstract class FeatureExtraction
extends StorageManager
private static final long serialVersionUID = 7608607169147548576L;
public boolean extractFeatures()
        throws FeatureExtractionException
        {
                return
this.extractFeatures(this.oPreprocessing.getSample().getSampleArray());
        }
public Object clone()
        {
                FeatureExtraction oClone =
(FeatureExtraction)super.clone();
                oClone.adFeatures = (double[])this.adFeatures.clone();
                oClone.oPreprocessing = this.oPreprocessing;
                return oClone;
        }
public static String getMARFSourceCodeRevision()
        {
                return "$Revision: 1.36.4.2 $";
        }
```

**Ipreprocessing**

```
public interface IPreprocessing
extends Cloneable
{
                String MARF_INTERFACE_CODE_REVISION =
"$Revision: 1.7 $";
        boolean preprocess()
        throws PreprocessingException;

        boolean removeNoise()
        throws PreprocessingException;

                Object clone()
   throws CloneNotSupportedException;
}
```

2) *GIPSY*

**Actual classes in GIPSY:-**

RIPE (Run time programing environment): It interacts with controller class, textual editor class,DFG editor class to run the program.

GIPC (General Intentional Programing Complier): GIPC uses preprocessor (input stream) class and DFG analyzer which inherits the IComplier to execute the program.

GEE ( General Eduction Engine) :-This GEE class which is generated by GIPC class gives result to  the demand worker and demand dispatcher to execute the file. Exception handler class is used if any rectifications to be done.

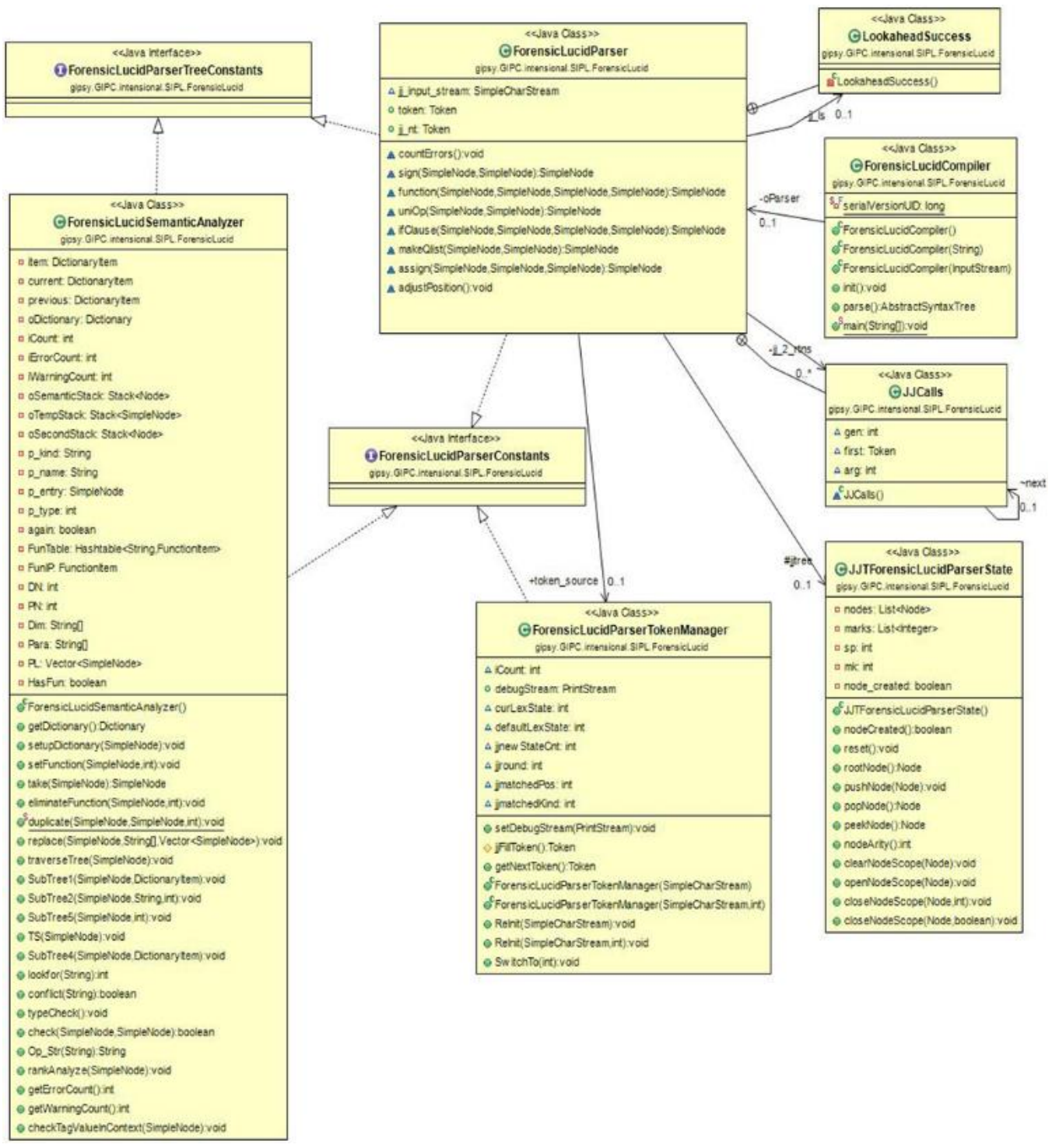

*Figure 22 GIPC class diagram*

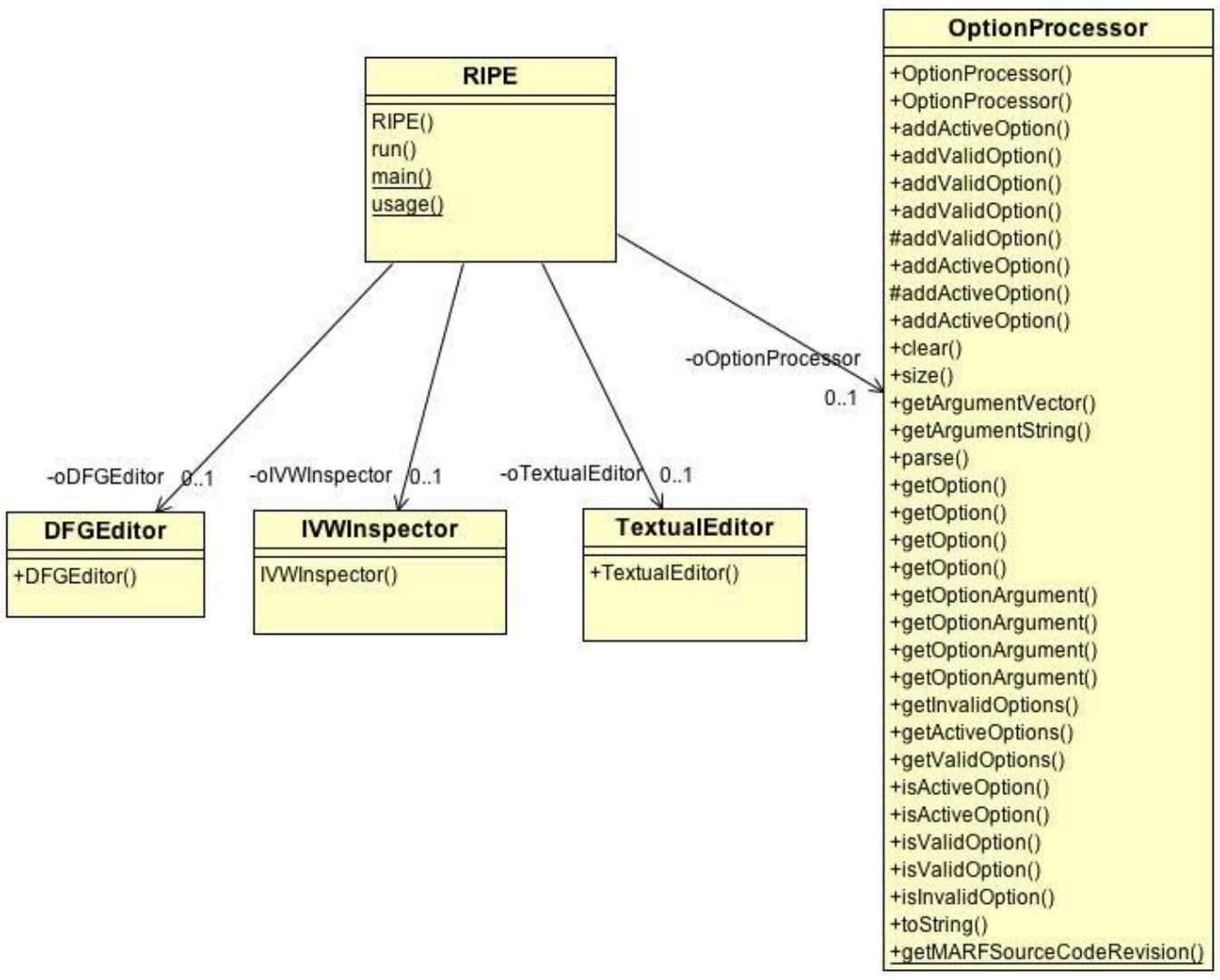

*Figure 23 RIPE class diagram*

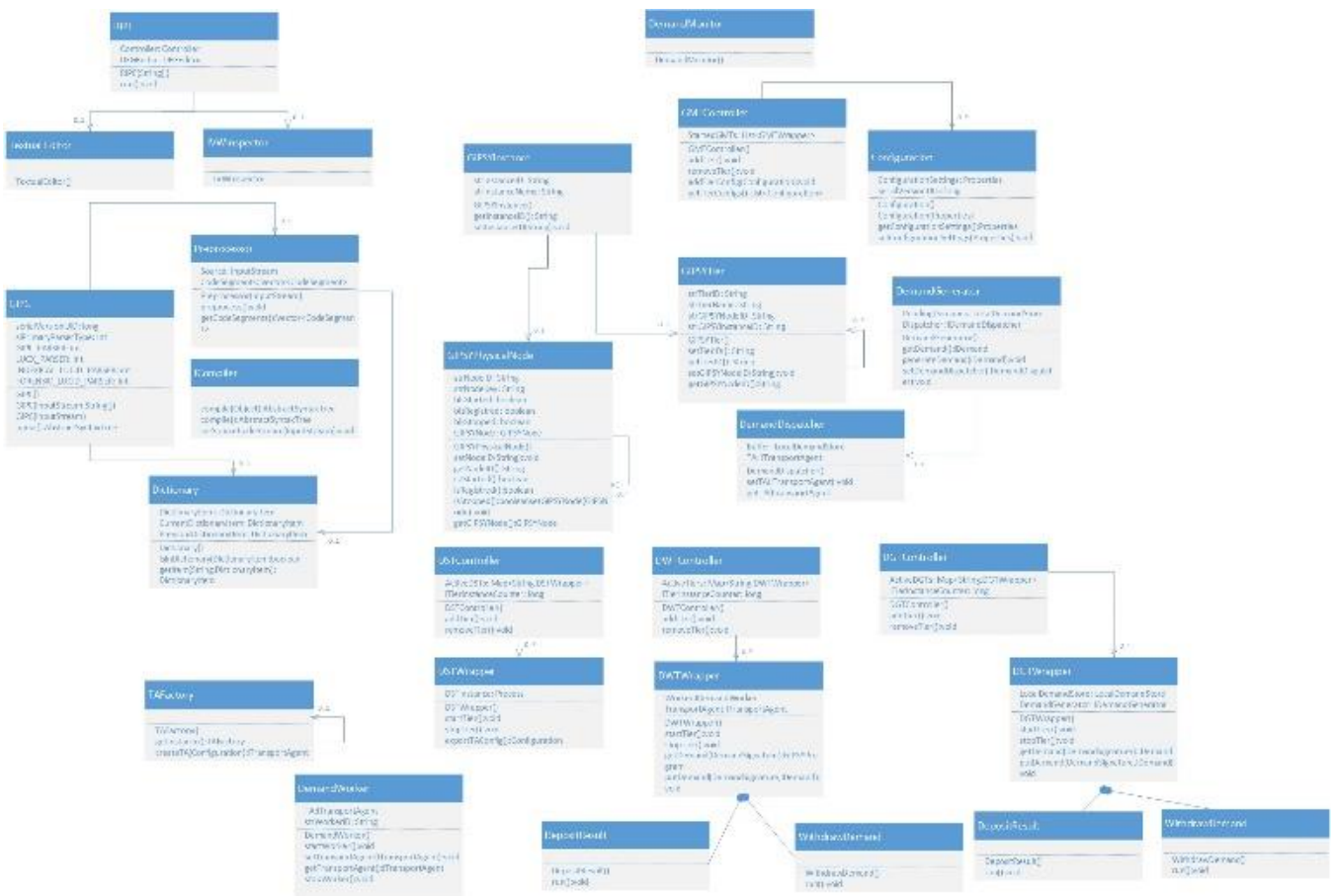

*Figure 24 Actual classes in GIPSY*

## Relationships between the actual system classes in GIPSY

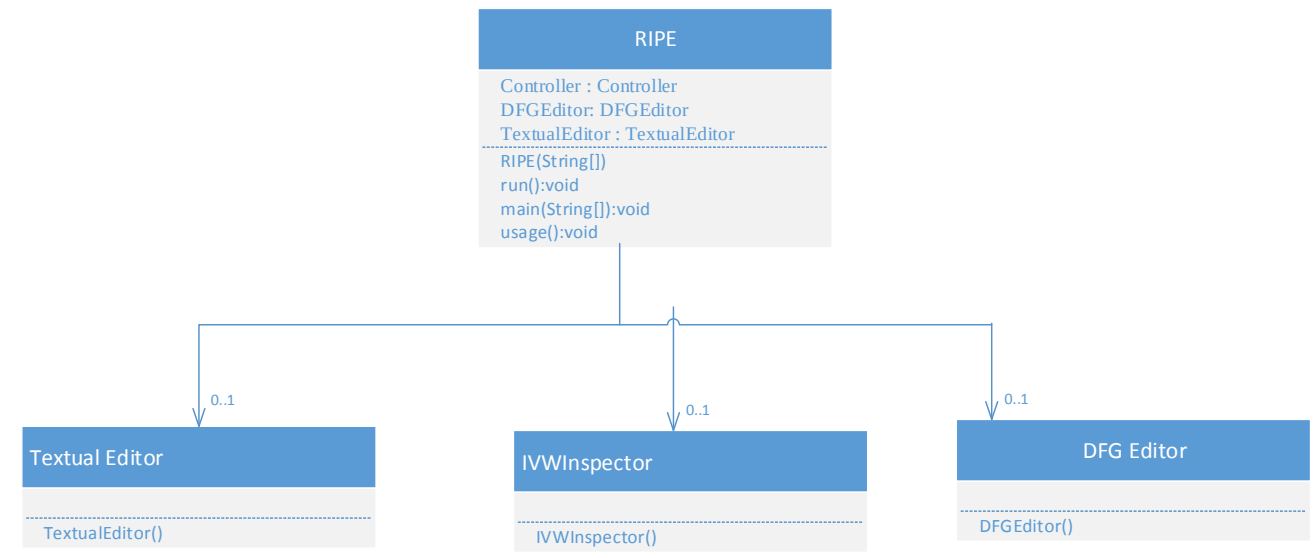

*Figure 25 simplified version of class interaction GIPSY*

To run the program in actual system RIPE user interacts with textual editor(),IVWinspector(),DFG editor() classes.

In below diagram GIPS uses Preprocessor, DFG analyzer ,Abstract syntax tree ,Icomplier .I complier is used to run the program in GIPC.GIPC inherits GEE and gives Demand dispatcher an demand worker classes.If any problem arises there is and exception handler class will handle and resolve the problems.

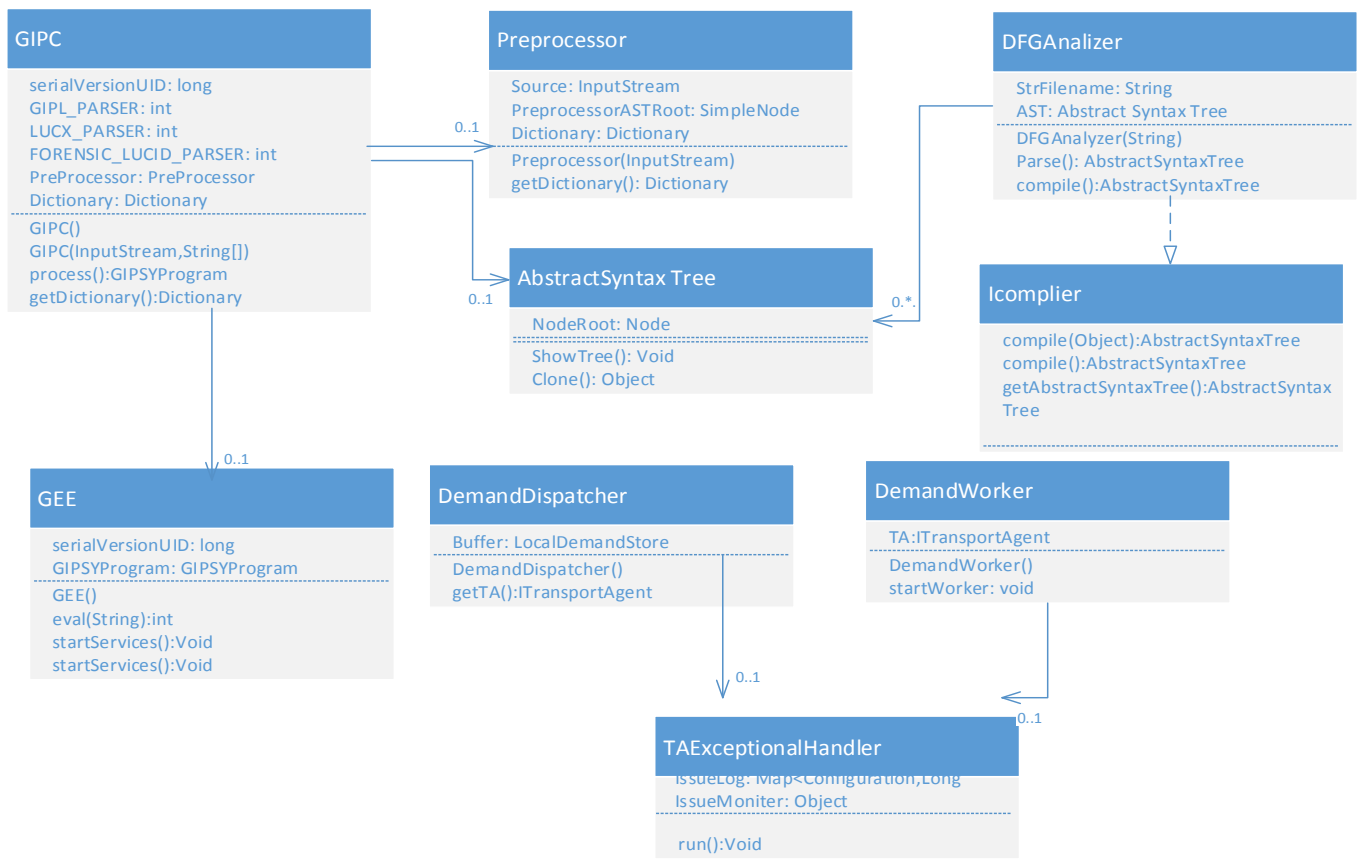

*Figure 26 actual system classes in GIPSY*

**Mapping between actual classes and conceptual classes for GIPSY:-**

| CONCEPTUAL CLASSES | ACTUAL CLASSES | Differences |
|---|---|---|
| C Complier | I Complier | Both C compiler and I complier are used to call the program in GIPSY |
| RIPE (Run time programming environment ) | RIPE | In conceptual classes may users interacts with RIPE to run the program where as in actual classes RIPE use to run text editor (), IVWinspector(), DFG editor (). |
| GEE( General Edcution Engine) | GEE | In conceptual class and Actual classes the classes are generated by GIPC |
| GIPC (General Intentional programming complier) | GIPC | In conceptual classes C Complier is used to compile the program file, where as in actual classes I complier is used to compile in GIPC. |

**Relations between two classes GIPC and GEE :**

**CODE FOR GIPC**

```
public class GIPC
extends IntensionalCompiler
private static final long serialVersionUID = 5412492426375587515L;
public static final int GIPL_PARSER = 0;
public static final int LUCX_PARSER = 2;
public static final int FORENSIC_LUCID_PARSER = 3;
private Preprocessor oPreprocessor = null;
private Dictionary oDictionary = null;

public GIPC()
        throws GIPCException
        {
                super();
                setupDefaultConfig();
                this.oObjectToSerialize = this.oGIPSYProgram;
        }
public GIPC(InputStream poSourceCodeStream, String[] argv)
        throws GIPCException
        {

        this.oOptionProcessor.addActiveOption(OPT_NO_FILENAME, "--
nofilename");
                setupConfig(argv);
                this.oSourceCodeStream = poSourceCodeStream;
        }
public Dictionary getDictionary()
        {
                return this.oDictionary;
        }
public GIPSYProgram getGIPSYProgram()
        {
                return this.oGIPSYProgram;
        }
public GIPSYProgram getGIPSYProgram()
        {
                return this.oGIPSYProgram;
        }
        public GIPSYProgram getGEER()
        {
                return getGIPSYProgram();
        }
```

**CODE FOR GEE:**

```
public class GEE
extends StorageManager
private static final long serialVersionUID = 2178955671888147327L;
private GIPSYProgram oGIPSYProgram = null;
        public GEE()
        {
                //super("GIPSY GEE $Revision: 1.39 $, TID = " +
BaseThread.getNextTID());
                this.bDumpOnNotFound = false;

                Debug.enableDebug();

                Debug.debug("Constructing " + this);

                this.oOptionProcessor.addValidOption(OPT_STDIN, "--
stdin");
                this.oOptionProcessor.addValidOption(OPT_ALL, "--
all");

        this.oOptionProcessor.addValidOption(OPT_THREADED, "--
threaded");
                this.oOptionProcessor.addValidOption(OPT_RMI, "--
rmi");
                this.oOptionProcessor.addValidOption(OPT_JINI, "--
jini");
                this.oOptionProcessor.addValidOption(OPT_DCOM, "--
dcom");
                this.oOptionProcessor.addValidOption(OPT_CORBA, "-
-corba");

                this.oOptionProcessor.addValidOption(OPT_DEBUG, "-
-debug");
                this.oOptionProcessor.addValidOption(OPT_HELP, "--
help");
                this.oOptionProcessor.addValidOption(OPT_GEE, "--
gee");

                this.oOptionProcessor.addValidOption(OPT_DGT, "--
dgt");
                this.oOptionProcessor.addValidOption(OPT_DST, "--
dst");
                this.oOptionProcessor.addValidOption(OPT_DWT, "--
dwt");
                this.oOptionProcessor.addValidOption(OPT_GMT, "--
gmt");
                this.oOptionProcessor.addValidOption(OPT_NODE, "--
node");

                Debug.debug("Constructed " + this);
        }
@Deprecated
        public int eval(String pstrContext)
        {
                if(this.oExecutor == null)
                {
                        this.oExecutor = new Executor();
                }

                System.out.println("GEER instance: [[[" +
this.oGIPSYProgram +"]]]");
                return
((Executor)this.oExecutor).execute(this.oGIPSYProgram.getDictionary(),
pstrContext);
        }
public void stopServices()
        throws GEEException
```

```
{
        /*
         * Stop all the potentially started services.
         */
        for(IMultiTierWrapper oTier: this.oTierWrappers)
        {
                oTier.stopTier();
        }
}
```

**Differnce as compared to PM2**

```
GIPSY
Conceptual Architectures              Actual system arechitectures

In conceptual class digram             In acutal systems Icomplier is used
C complier is used to call the         to call the program in GIPC
program in GIPC

In RIPE many users interacts with      In actual systems text editor(),DGF editor()
on RIPE to run the program.            and IVWinspector() are used to run the program.

GEE classes are generated by            GEE classes are generated by GIPC.Demanddispatcher
GIPC classes.Worker class is generated  and demandworker classes are generated if any problem
to get the result.                      expection handling is to be done for rectification.


DMARF

Conceptual Archeitectures                  Actual system archeitectures

Sample preprocessing is used to            Ipreprocessing extracts the classes from
detect unwanted noise .                    feature extract classes.

Sample feature extract is based on         In actual system feature extract the classes
algorithms .                               wil import from the Ipreprocessing.

Sample classification is used to test      Classification uses cosine similarity to get the
and clarify the tests .                    test result .
```

## IV. METHODOLOGY

### A. Refactoring

#### 1) Code Smells and System Level Refactoring's

After doing a bit of research based on identifying the bad code smell for GIPSY and DMARF projects, the whole project was inspected by JDeodorant application for identifying bad code and which can be refactor for better use.

There are 3 smells that are recognized they are
- Feature Envy
- God Class
- Type Checking

Feature Envy:

Feature envy is a technique that a particular package of data that might be a method or any kind of data is used by other class more frequently. For solving this problem we have to use either Move Method or Extract Method techniques. Here we can resolve by moving the method from a class to another class which is using this method more frequently in order to increase the cohesion. Here we can follow 2 techniques either moving complete method to the desired class or moving a particular part of a method such that it can be used by both the classes.

God Class:

God class are a type of large classes, it consists of many functionalities as a reason lot of code will be generated. God classes are often related to bad code which may affect the quality of a software. In object oriented systems, evolution of god class is known as an unhealthy system. God class is a class which knows too much and which does too much with many number of functionalities. We can restrict some functionalities and can achieve avoiding god class smell. We can limit the responsibilities of a certain class which is responsible for God Class.

Type Checking:

Type checking is done only through eclipse plugin. Type checking bad smells can be eliminated by using Replace conditional with Polymorphism or Replace code with State/Strategy refactoring's. Usually Type Checking smell arises due to complexity of conditional statements such as if, if else if, switch statements. Whenever we come across this type of smells we can resolve by introducing inheritance and polymorphism techniques. The methods which are leading to type checking smell are moved from that class and placed in subclasses in order to avoid Type Checking smell. We can achieve low coupling by replacing the code and by using polymorphism and by creating sub classes.

*a) DMARF*

(TYPE CHECKING) BAD SMELL 4:

In DMARF also the same, the method named addSymbol is used by other class marf.npl.Parsing.Token nearly 5 times by two functions namely getLexeme() and getPosition(). The class which contains the method is using it less than other class so it's not good for a class as it violates the rules.

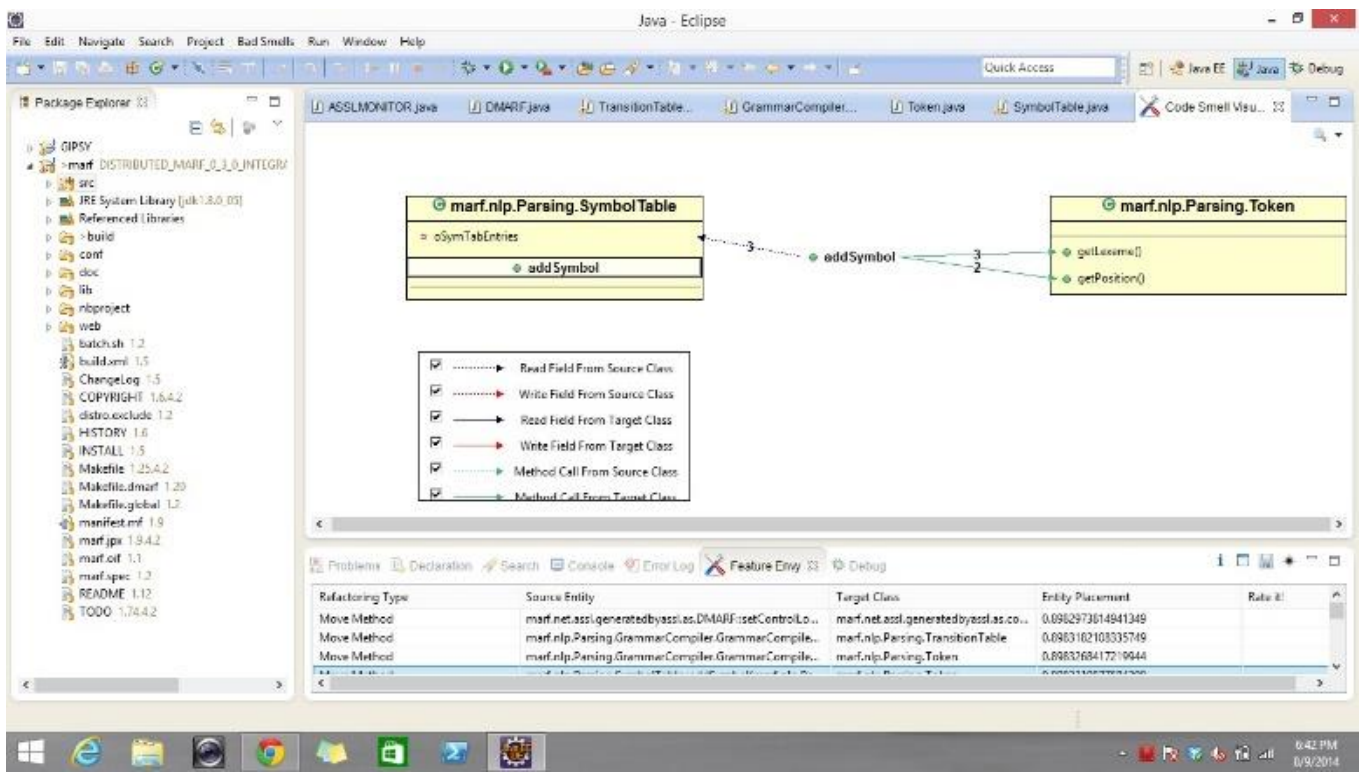

*Figure 27 Bad smell DMARF (TYPE CHECKING)*

GOD CLASS

BAD SMELL 7:

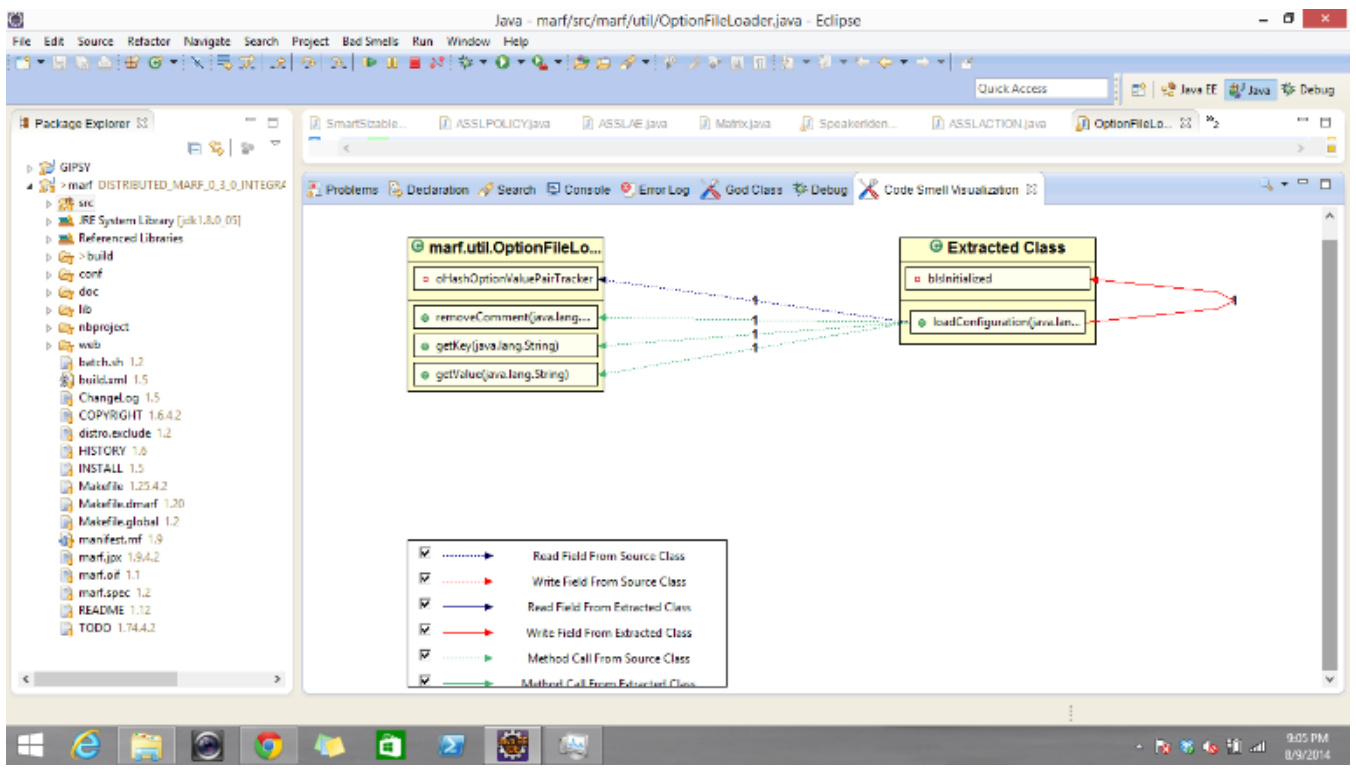

*Figure 28 bad smell DMARF GOD CLASS*

Here the method in Extracted classes has many functions as it also calls the methods which are in source class and also reads the fields from source class.

BAD SMELL 8:

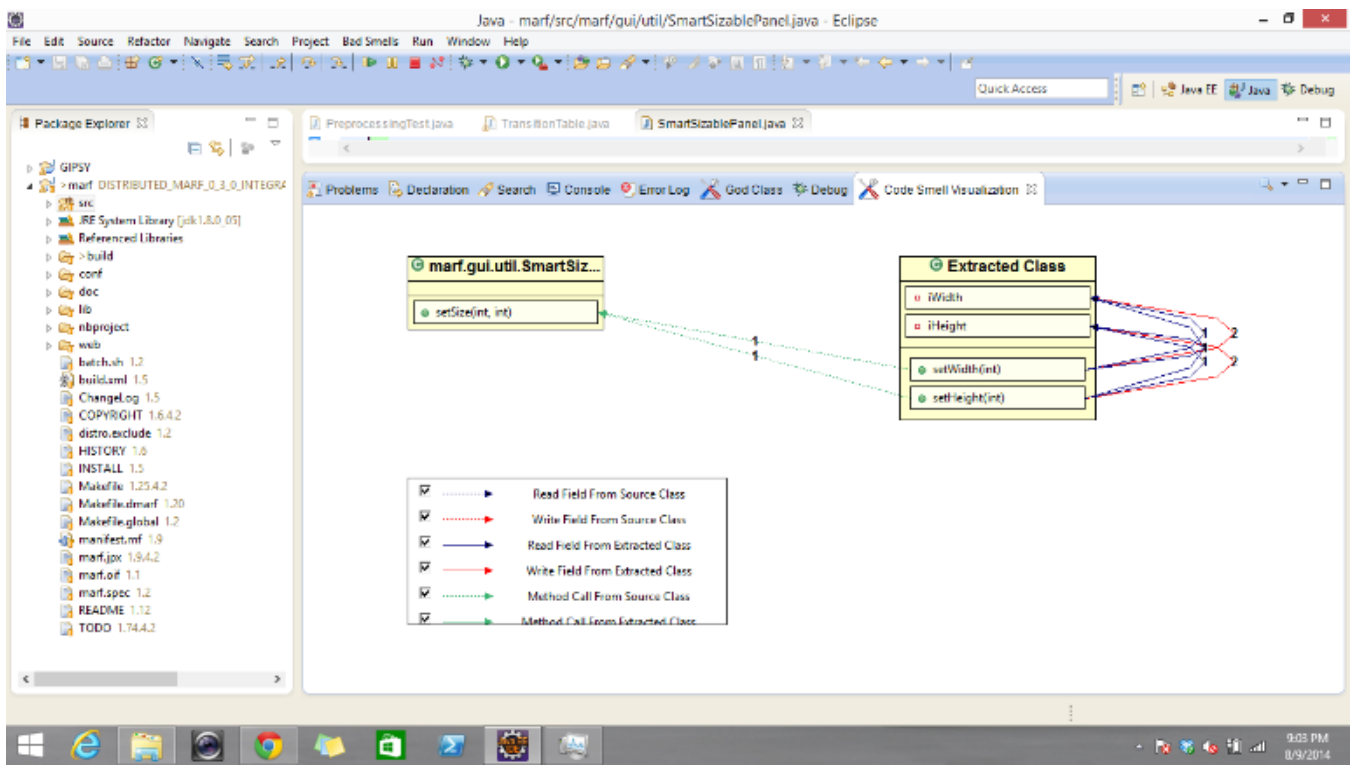

*Figure 29 bad smell DMARF GOD CLASS*

The methods which are in Extracted Class are used too much by its own class and the source class also. This may lead to system failure and cause unhealthy to the object oriented system and there is a high chances of risk factor. As the 2 methods in Extracted Class are used mostly 9 times by source class and extracted class. It interacts with source class for method calling and it is used by its own class for reading the field and writing the field. This tendency is not good in any system. The quality team has to check for the most risk smells which may lead to system failure and has to resolve them for a better purpose.

TYPE CHECKING

BAD SMELL 11:

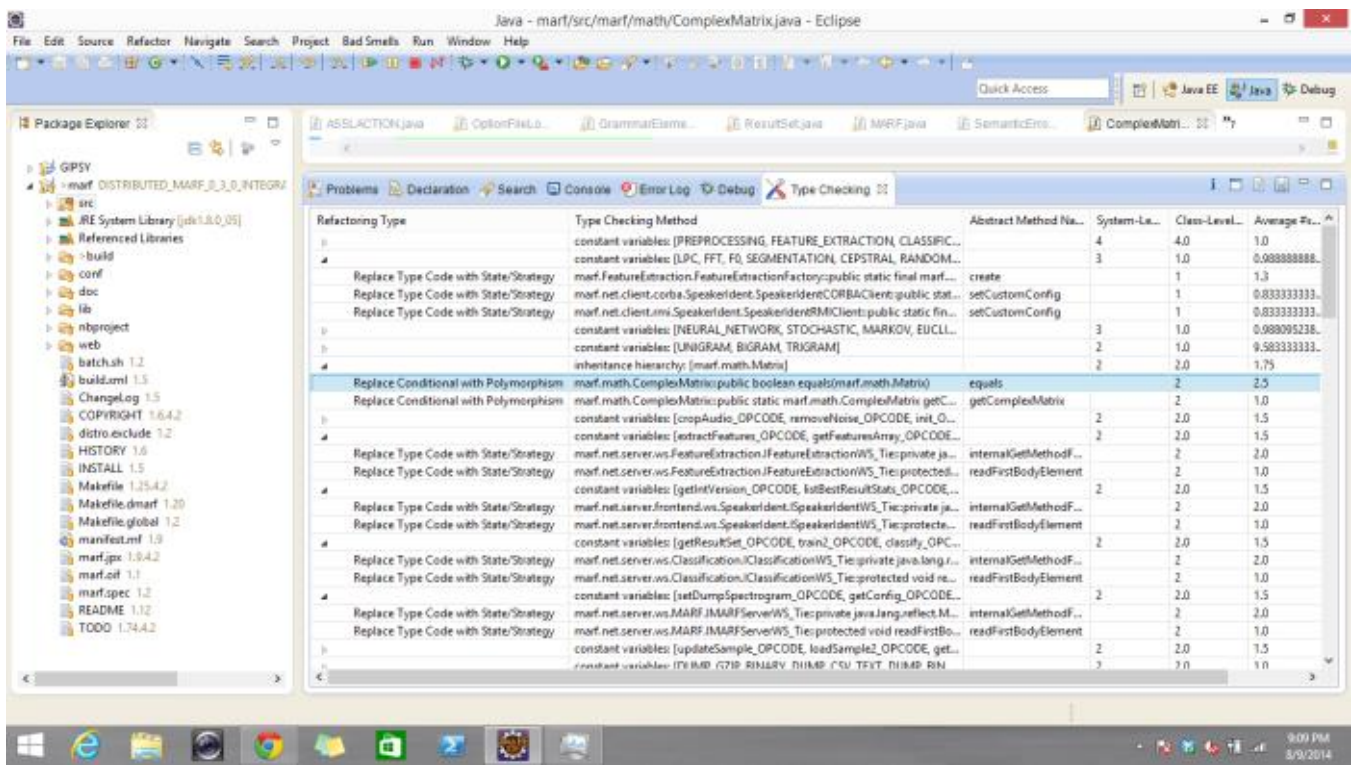

*Figure 30 bad smell DMARF TYPE CHECKING*

Here an equals abstract method is responsible for this type checking error in marf.math.ComplexMatrix class. This smell can be removed by replacing the conditional with polymorphism refactoring technique.

*b) GIPSY*

BAD SMELL 1:On basis of feature envy identification, the method named (commonTokenAction) in IndexicalLucidParserTokenManager.java file can be shifted to GIPC.util.Token class, as the method is not being called in its parent class. It seems that commonTokenAction method is more interested in other class rather than its parent class.

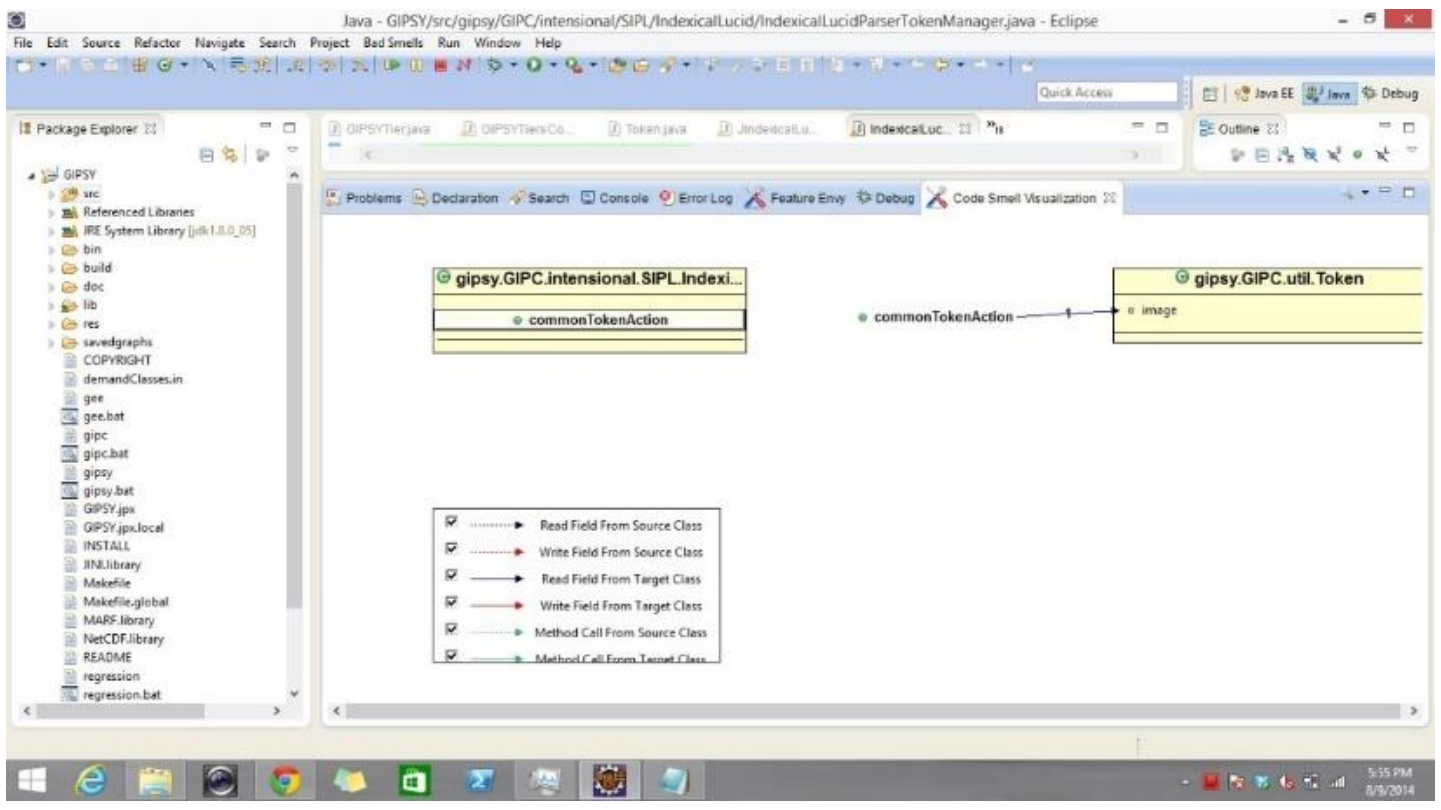

*Figure 31 bad smell GIPSY Feature Envy*

BAD SMELL 2:

The method named SubTree5 in gipsy.GIPC.SemanticAnalyzer class is mostly used by the other class gipsy.GIPC.intensional.SimpleNode rather than the parent class. The gipsy.GIPC.intensional.SimpleNode uses the SubTree5 method 7 times, whereas the parent class uses just once.

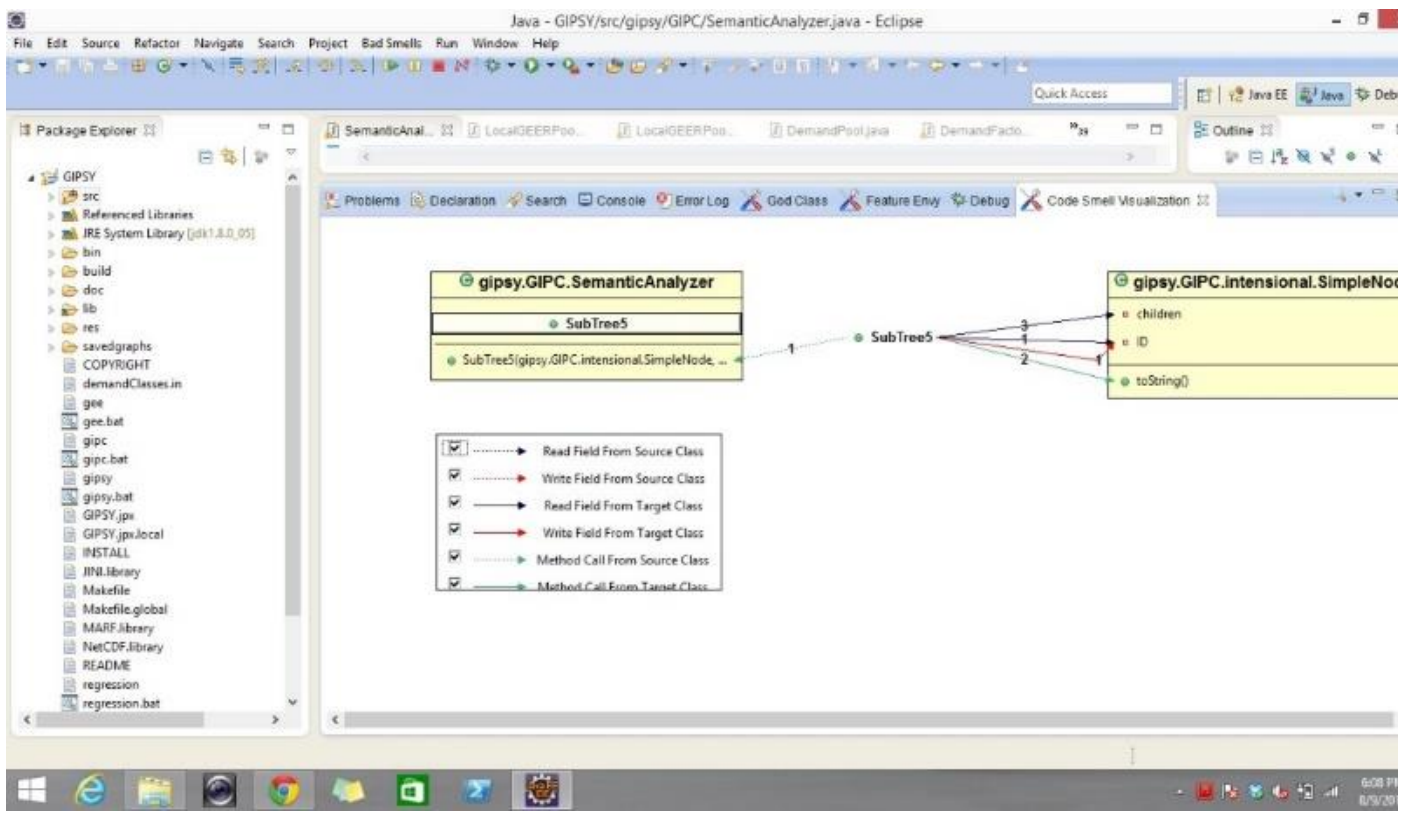

*Figure 32 Bad smell GIPSY Feature Envy*

BAD SMELL 3:

SubTree4 which is a method in gipsy.GIPC.SemanticAnalyzer is used only once, whereas gipsy.GIPC.intensional.SimpleNode uses it 13 times. The functions toString() and getImage() itself uses it 8 times.

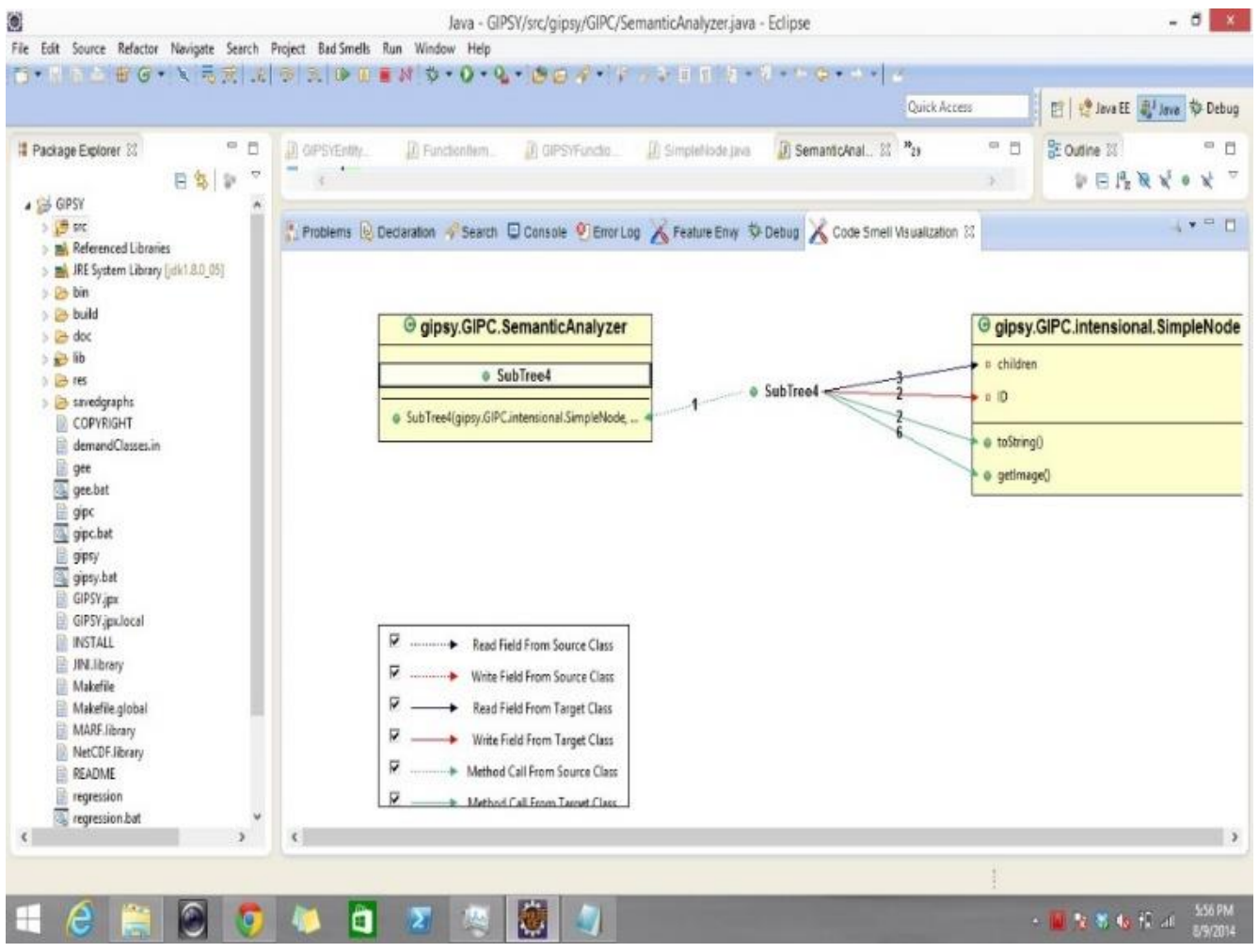

*Figure 33 bad smell GIPSY Feature Envy*

**UML diagram representing Feature Envy:**

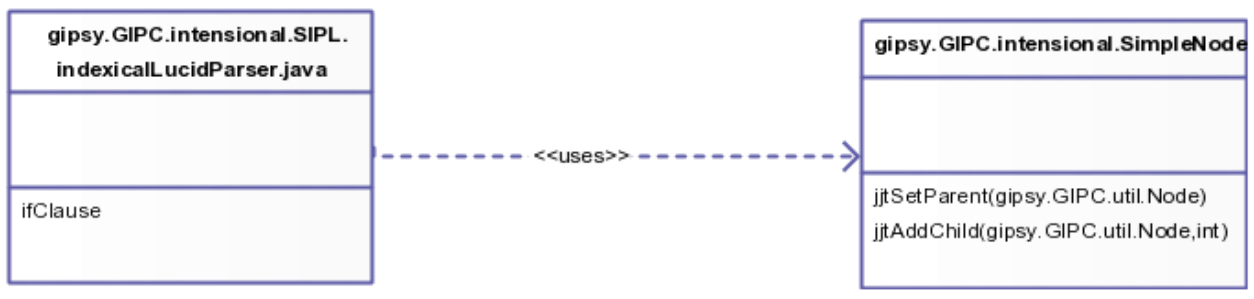

GOD CLASS

BAD SMELL 5:

Here you can see in Extracted class consists of many functionalities which not only uses the resources in that class but also in gipsy.GEE.IDP.Deman class also. This class may lead to system failure in future.

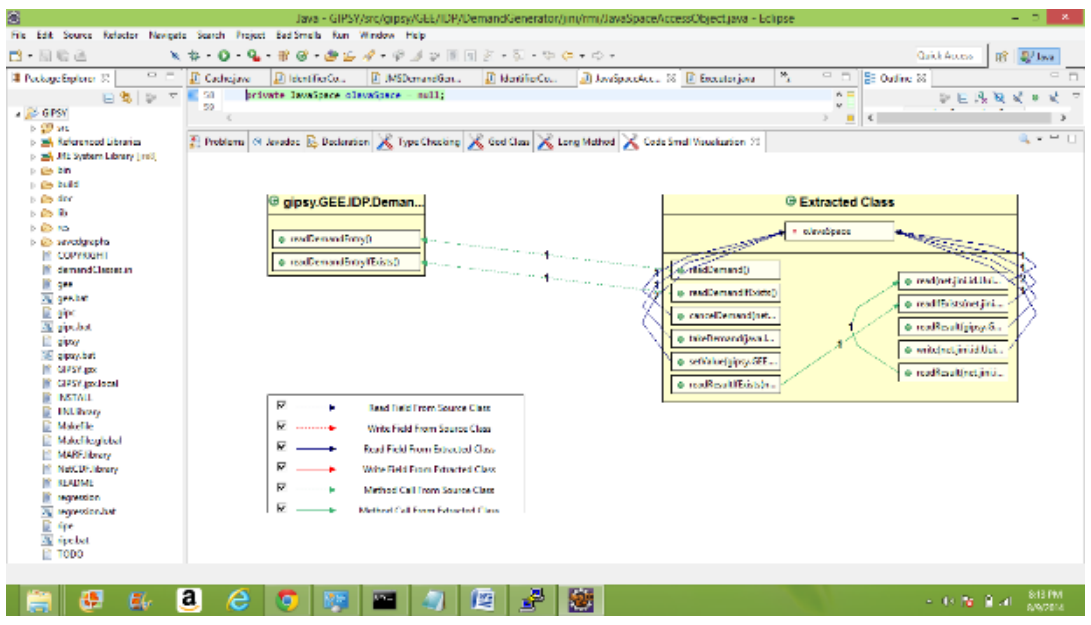

*Figure 34 bad smell GIPSY GOD CLASS*

BAD SMELL 6:

In a similar way here also in Extracted Class there are many functionalities which are used more than once as a result god class smell is generated and may damage the system.

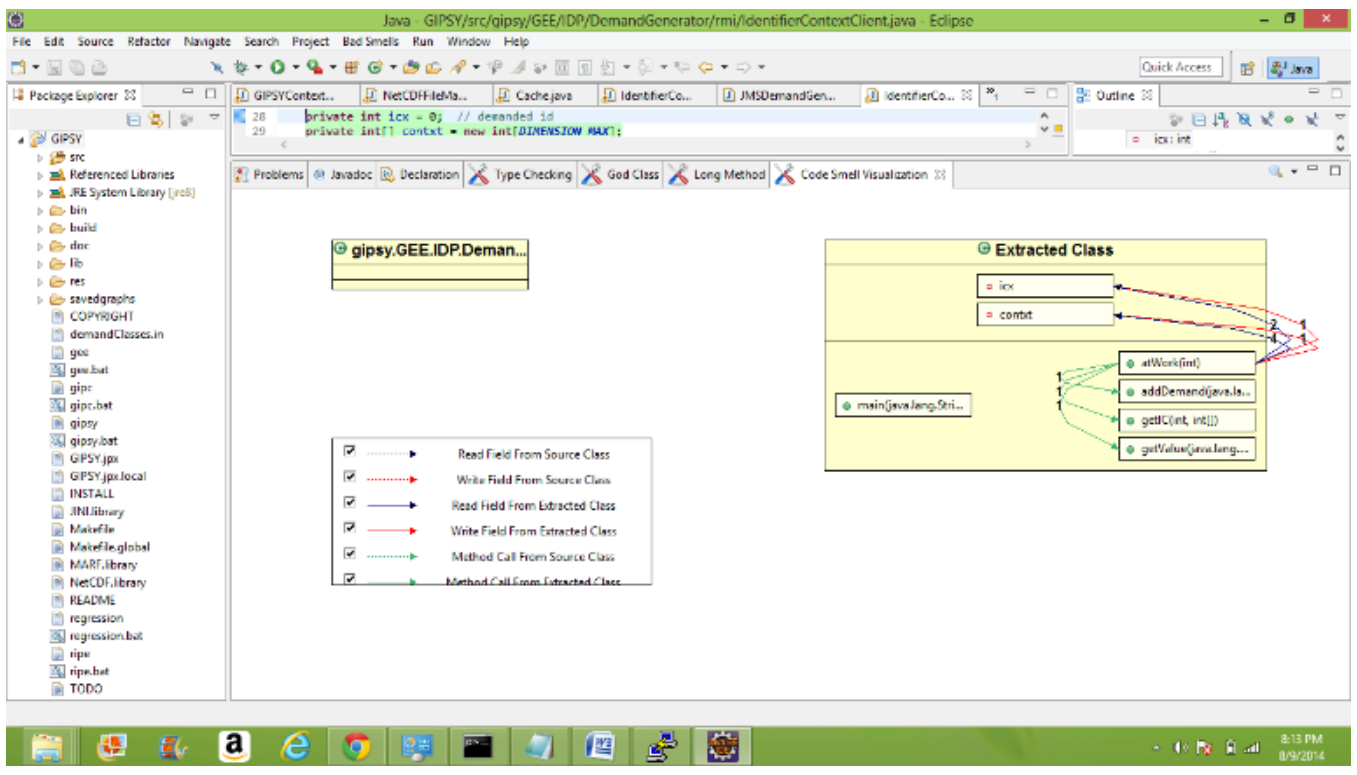

*Figure 35bad smell GIPSY GOD CLASS*

TYPE CHECKING

BAD SMELL 9:

In the figure we can see clearly that the Type checking smell is raised in gipsy.GEE.multitire.GIPSYNode method which is called by public void run (). Here we can avoid this smell by using Replace Conditional with Polymorphism refactoring technique by adding the type checking smell code to the inheritance hierarchy.

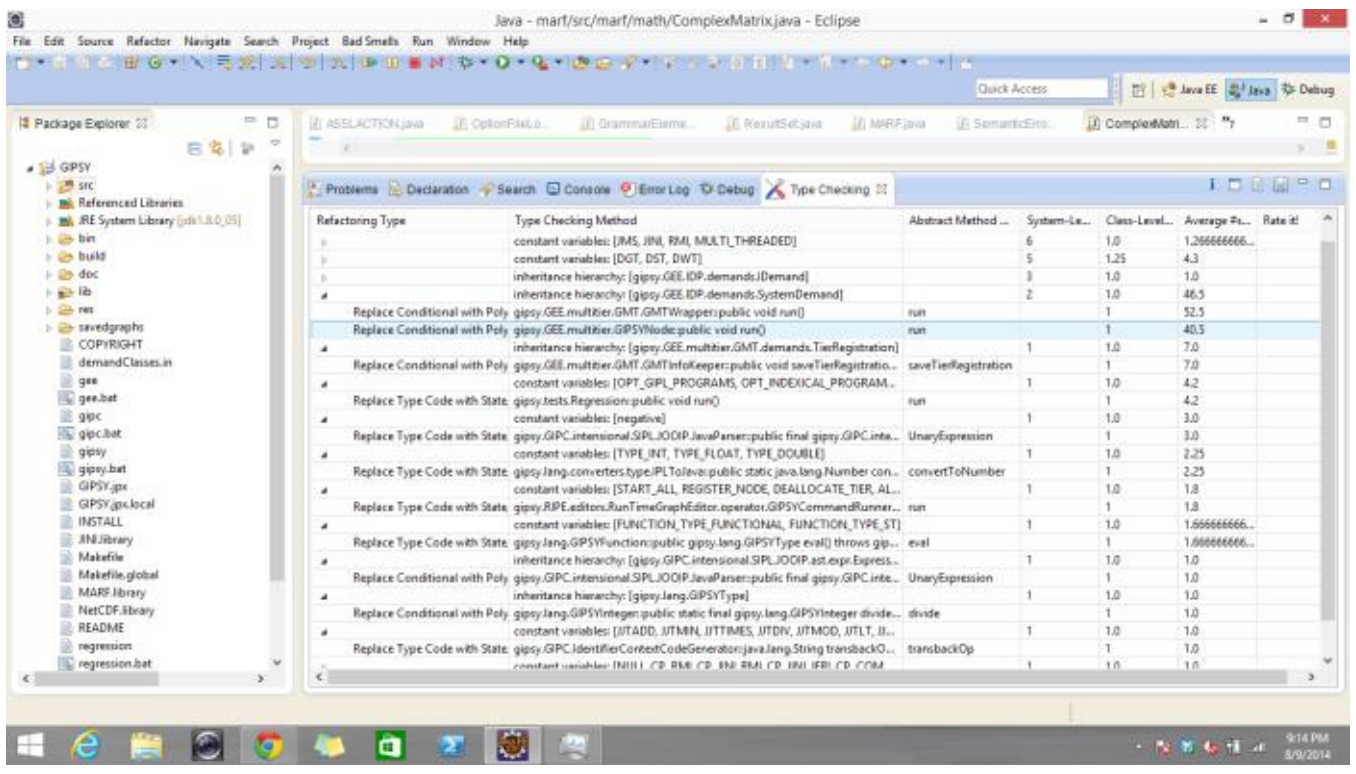

*Figure 36 bad smell GIPSY TYPE CHECKING*

**BAD SMELL 10:**

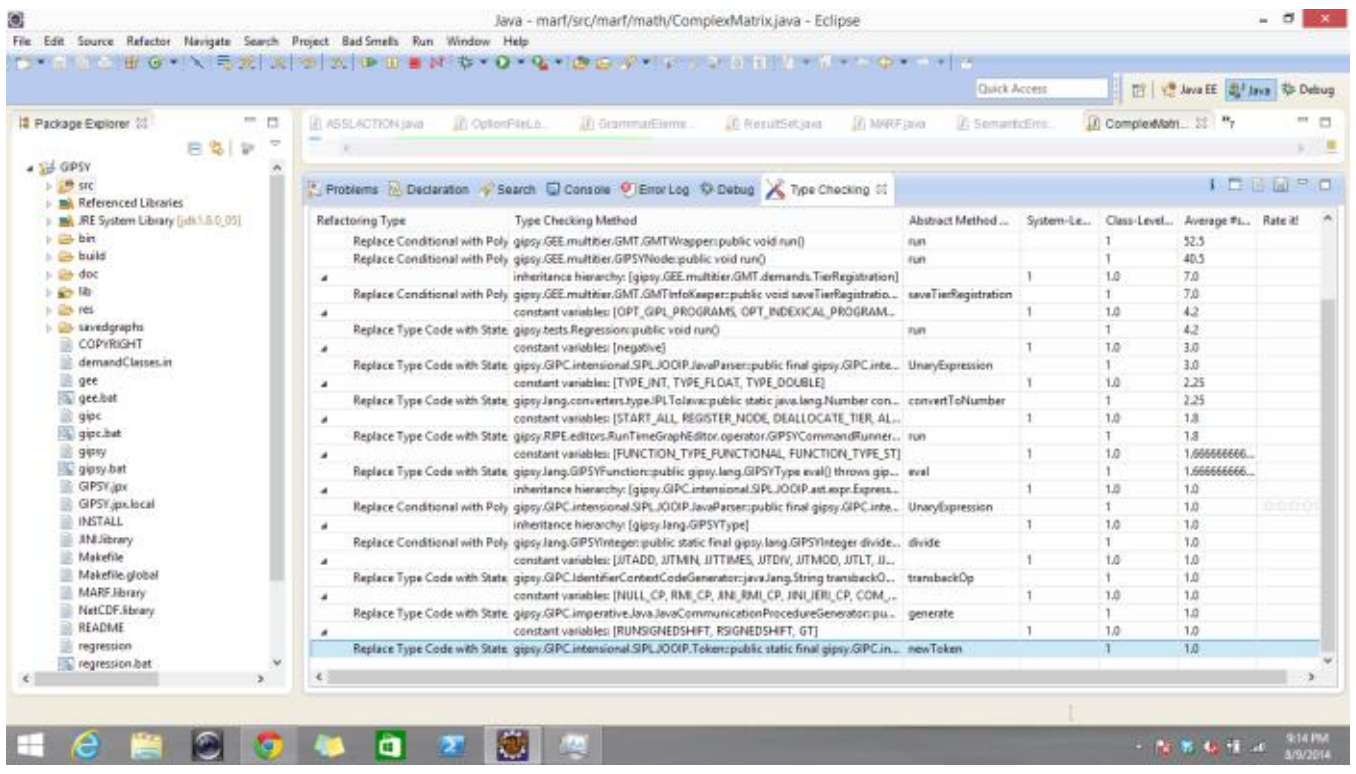

*Figure 37 bad smell GIPSY TYPE CHECKING*

Here we can see that there is a type checking smell in class named gipsy.GIPC.intensional.SIPL.JOOIP.Token class which is due to an abstract method named newToken here we can avoid type checking by using Replace Type Code with state/strategy refactoring technique. The conditional classes which are responsible for this smell are moved from that class and placed in another sub class.

*2) Specific Refactoring's*

Actually we found 6 refactoring techniques for bad smells we found. The refactoring techniques are Move Method, Extract Method, Extract Class, Extract Subclass and Replace conditional with Polymorphism, Replace code with State/Strategy

Move Method is all about creating a new method with similar body in a class which uses the method more frequently. In Extract Method the large method is further divided into sub methods where other classes can be able to use a particular method rather than using the whole large method. In Extract Class moving the relevant fields of a class to a new class is done, but in Extract Subclass a sub class is created and relevant sub set of functionalities are moved to that class. Replace conditional with polymorphism and Replace code with state/strategy is about using polymorphism and inheritance technique.

We will try to implement all these refactoring methods as of now. If there are any changes we will mention in PM4.

**Interrelation among the Refactoring's**:

All these three smells are interrelated. If we are performing a refactoring techniques on a particular method 1st by moving a method to a particular class which is using more frequently rather than the source class here we are achieving cohesion and by limiting the responsibilities of the class we are avoiding the god class smell as the class has many functionalities and many things which are done by the class are restricted. And by introducing polymorphism and creating sub classes the coupling factor can be achieved.

*a) DMARF*
*b) GIPSY*

As of in PM3 the feature envy code smells are restructured in order to avoid the chance of risk in future.

Feature Envy:

Feature envy is a technique if a particular method is used by other class more frequently rather than its own class then we come across this feature envy smell. For achieving high cohesion we have to move that particular method to a class which is using this method more frequently. For solving this problem we have to use either Move Method or Extract Method techniques.

Bad smell 1 and bad smell 2 are considered for resolving and for restructuring the system. In bad smell 1 it is clear that the method of class IndexicalLucidParserTokenManager.java is not used at all instead it is used by another class called GIPC.util.Token. Here we have to move that method from source class to this GIPC.util.Token class such that there will be high coupling and risk factor can be reduced.

In bad smell 2 it is clear that method named SubTree5 in gipsy.GIPC.SemanticAnalyzer class is used 7 times by gipsy.GIPC.intensional.SimpleNode, whereas source class uses only once. Now the method SubTree5 is moved from source class such that the risk can be reduced and the mostly using class can easily access that method

While refactoring, the method void commonTokenAction() of gipsy.GIPC.intensional.SIPL.IndexicalLucid. IndexicalLucidParserTokenManager.java is being only used by gipsy.GIPC.util. Token.java, so the method will be moved to target java file with some of the changes in arguments to display the desired result.

The main class is

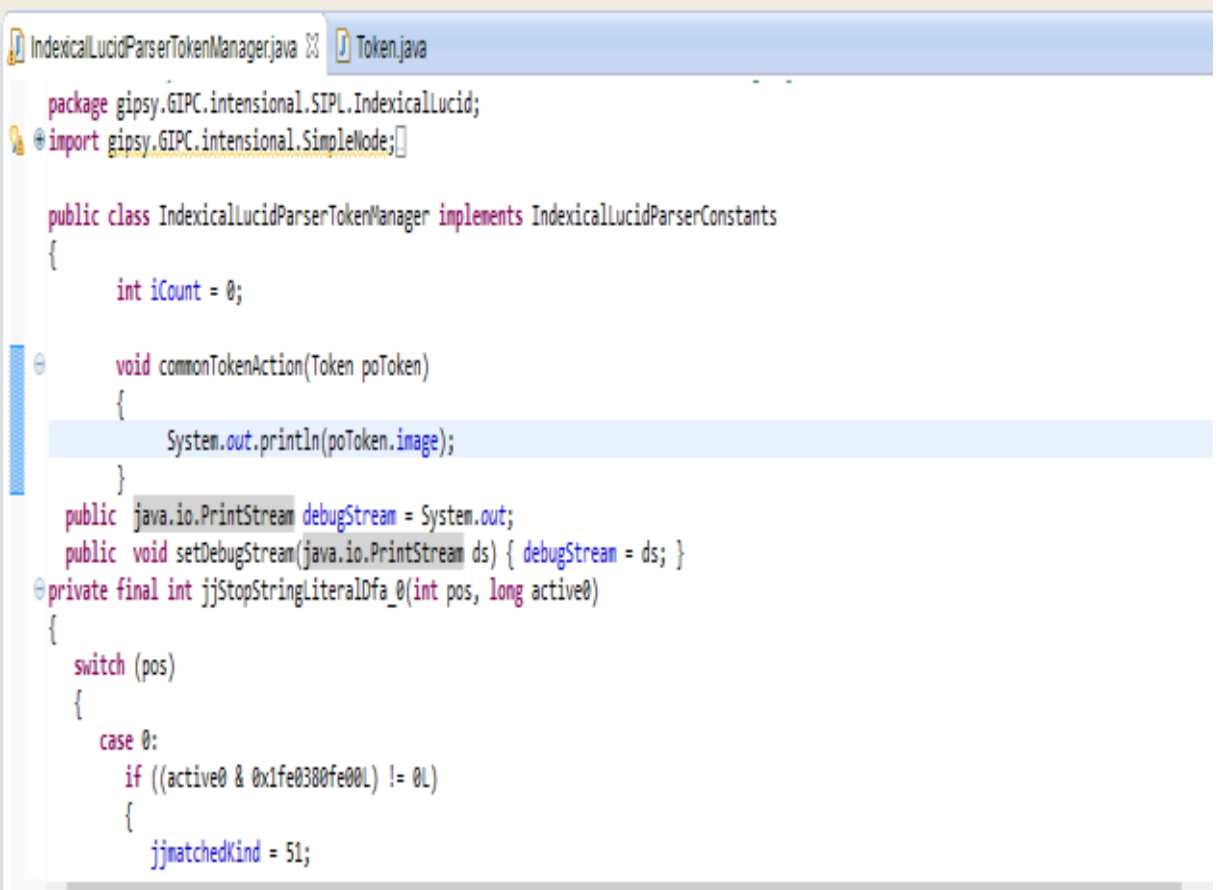

*Figure 38 Orignal Class*

After refactoring, the final result is as below

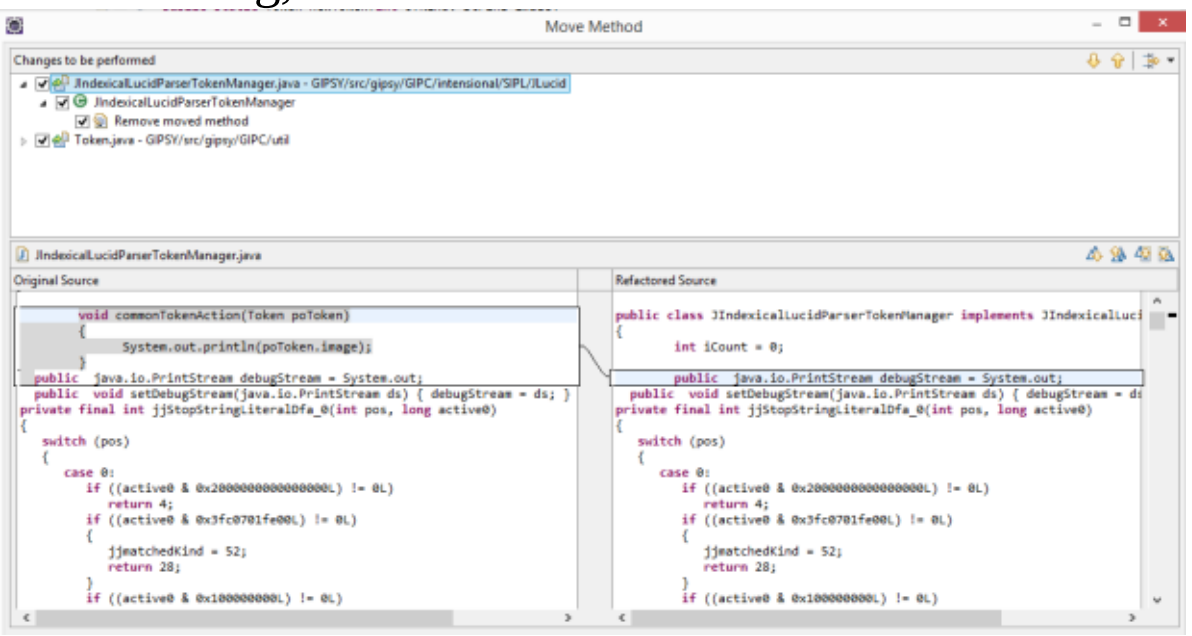

*Figure 39 Refactoring Result*

Second refactoring, main class is gipsy.GIPC. SemanticAnalyzer.java and method name is subTree5() that has to be moved to gipsy.GIPC.intention.SimpleNode.java.
The code to be moved is

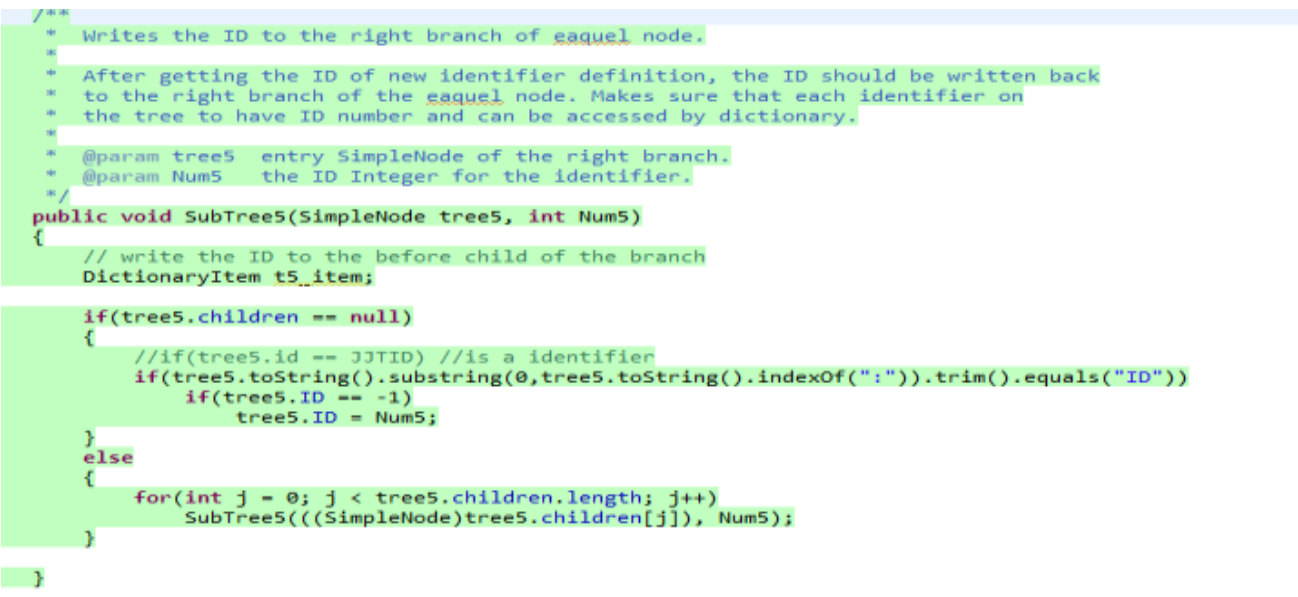

*Figure 40 Refactoring Result*

Refactoring results in

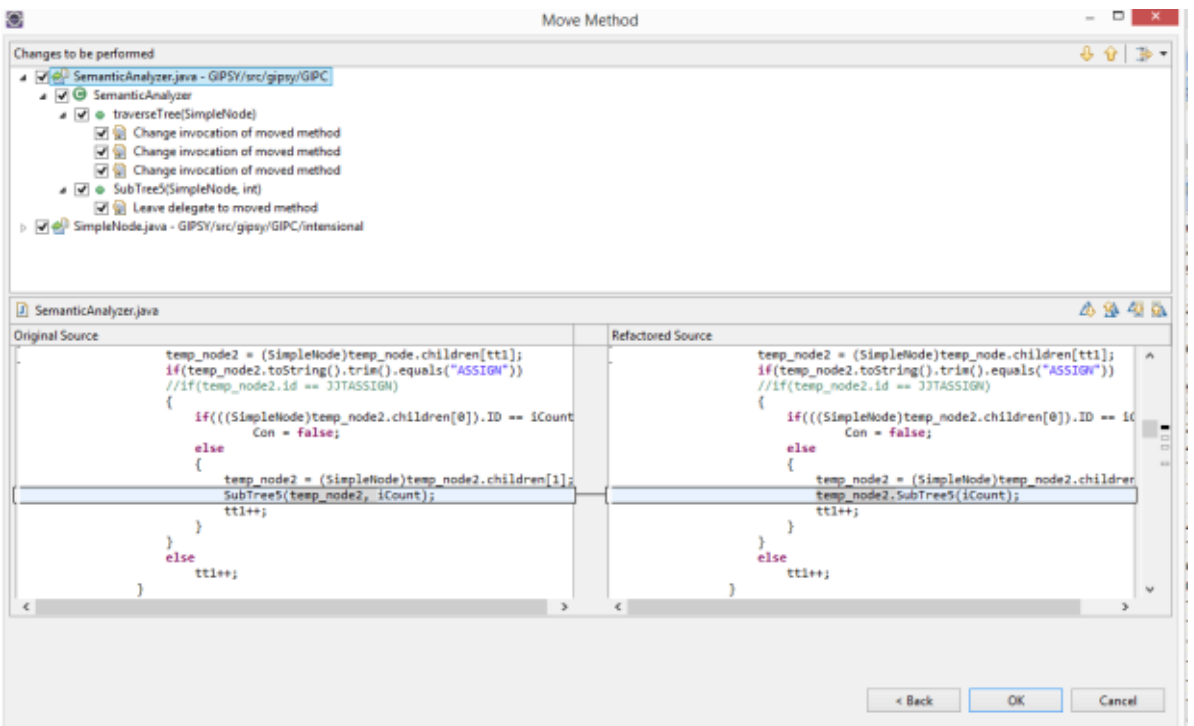

*Figure 41 Refactoring Result*

**Test cases in GIPSY and DMARF:**

There are few test cases related to GIPSY in gipsy.test folder. There are already few components tested they are expected, GEE, GIPC, gipsy, jooip, junit, lucid, lucx, translation_rules. We also have a regression.java file of test cases. All these components have tests done already and every module related to them are tested.

In DMARF there are also few test cases in test.java which is used for testing with respect to junit application. The ,main folder for testing is junit under source code which comprise of preprocessing, math, stats, storage and versiontest,java, these different folders namely preprocessing or math or stats contains the testing file for unit. Every folder has a test file for that particular unit.

**Refectoring suggestions**

For refactoring, let us provide the example for type checking bad smell. In this the class that is for our concern is gipsy.GEE.multitier.GIPSYNode.java and the method that we have to look into is void run(). It contains conditional statements which can be replaced with converting based polymorphism concept. The source code of the class is below with the problematic method (void run()) highlighted by making it bold. Due to space constraint, comments are trimmed.

```
@Override
		public void run()
		{
			// Instantiate the TAException Handler and started.
			this.oTAExceptionHandler =
				new TAExceptionHandler(this.strNodeID, this.strGMTTierID, this.oRegDSTTA);
			((NodeController) this.oDWTController).setTAExceptionHandler(this.oTAExceptionHandler);
			((NodeController) this.oDGTController).setTAExceptionHandler(this.oTAExceptionHandler);

			this.oTAExceptionHandler.start();

			System.out.println(MSG_PREFIX + "Node started!");
```

```
while(true)
{
	SystemDemand oDemand = null;;

	// Wait for demands sent to this Node
	while(oDemand == null)
	{
		try
		{
			System.out.println(MSG_PREFIX + "Waiting for a new system demand ...");
			oDemand = (SystemDemand) this.oSysDSTTA.getDemand(this.strNodeID);
		}
		catch (DMSException oException)
		{
			try
			{
				this.oSysDSTTA = this.oTAExceptionHandler.fixTA(this.oSysDSTTA, oException);
			}
			catch (InterruptedException oInterrptedException)
			{
				oInterrptedException.printStackTrace(System.err);
			}
		}
	}

	if(oDemand instanceof TierAllocationRequest)
	{
		TierAllocationRequest oRequest = (TierAllocationRequest)oDemand;
		TierIdentity oTierIdentity = oRequest.getTierIdentity();
		TierAllocationResult oResult = null;
		Configuration oTierConfig = oRequest.getTierConfig();
		int iNumOfInstances = oRequest.getNumberOfInstances();

		boolean bIsReallocation = false;
		String strTierID = oTierConfig.getProperty(IMultiTierWrapper.WRAPPER_TIER_ID);

		if(strTierID != null)
		{
		bIsReallocation = true;
		}

		try
		{
			if(iNumOfInstances <= 0)
			{
				System.out.println(MSG_PREFIX + "Received an illegal TierAllocationRequest!");
				throw new MultiTierException("Wrong number of instances specified " + iNumOfInstances + "!");
			}
			TierRegistration[] aoTierRegs = new TierRegistration[iNumOfInstances];
			IMultiTierWrapper oTier;

			for(int i = 0; i<aoTierRegs.length; i++)
			{
				oTier = null;

				/*
				 * A shallow clone of the underlying hash table is used. This
				 * is effective for immutable objects such as the String properties
				 * and values, as well as sharable object values such as the TA
				 * configuration for DGT and DWT.
				 */

				oTierConfig = (Configuration) oRequest.getTierConfig().clone();

				switch(oTierIdentity)
				{
				case DST:

					if(i == 0)

					{

						System.out.println(MSG_PREFIX + "Received a DSTAllocationRequest");

					}

					oTier = this.oDSTController.addTier(oTierConfig);

					oTier.startTier();

					DSTRegistration oDSTReg = new DSTRegistration(this.strNodeID, oTier.getTierID(),
							oTier.getConfiguration(), ((DSTWrapper)oTier).exportTAConfig(),
							this.strGMTTierID);

					if(!bIsReallocation)

					{

						int iMaxActiveConnection = Integer.parseInt(oTierConfig.getProperty(DSTWrapper.MAX_ACTIVE_CONNECTION));

						oDSTReg.setMaxActiveConnection(iMaxActiveConnection);

					}

					aoTierRegs[i] = oDSTReg;
```

```
break;
case DGT:
if(i == 0)
{
    System.out.println(MSG_PREFIX + "Received a DGTAllocationRequest");
}

oTier = this.oDGTController.addTier(oTierConfig);
oTier.startTier();

aoTierRegs[i] = new DGTRegistration(this.strNodeID,
        oTier.getTierID(), this.strGMTTierID);
break;

case DWT:
if(i == 0)
{
    System.out.println(MSG_PREFIX + "Received a DWTAllocationRequest");
}

oTier = this.oDWTController.addTier(oTierConfig);
oTier.startTier();

aoTierRegs[i] = new DGTRegistration(this.strNodeID,
        oTier.getTierID(), this.strGMTTierID);
break;
}
System.out.println(MSG_PREFIX + oTierIdentity + oTier.getTierID()
    + " was allocated and started!");
}

oResult = new TierAllocationResult(aoTierRegs, null);
oResult.setSignature(oRequest.getSignature());
while(true)
{
try
{
System.out.println(MSG_PREFIX +
        "Sending TierAllocationResult enclosing tier registrations ...");
this.oSysDSTTA.setResult(oResult);
break;
}
catch (DMSException oException)
{
try
{
    this.oSysDSTTA = this.oTAExceptionHandler.fixTA(this.oSysDSTTA, oException);
    System.out.println(MSG_PREFIX + "TierAllocationResult sent!");
}
catch (InterruptedException oInterrptedException)
{
    oInterrptedException.printStackTrace(System.err);
}
}
}
}
catch(MultiTierException oException)
{
oResult = new TierAllocationResult(null, oException);
oResult.setSignature(oRequest.getSignature());
while(true)
{
try
{
System.out.println(MSG_PREFIX + "Sending allocation result enclosing exception ...");
this.oSysDSTTA.setResult(oResult);
System.out.println(MSG_PREFIX + "TierAllocationResult sent!");
break;
}
catch (DMSException oDMSException)
{
try
{
    this.oSysDSTTA = this.oTAExceptionHandler.fixTA(this.oSysDSTTA, oDMSException);
```

```
			}

			catch (InterruptedException oInterrptedException)

			{

				oInterrptedException.printStackTrace(System.err);

			}
							}
						}
					}
				}
				else if(oDemand instanceof
TierDeallocationRequest)
				{
					TierDeallocationRequest
oRequest = (TierDeallocationRequest)oDemand;

					String[] astrTierIDs =
oRequest.getTierIDs();
					TierIdentity oTierIdentity=
oRequest.getTierIdentity();


			System.out.println(MSG_PREFIX + "Received a
DSTDeallocationRequest");

					for(int i = 0;
i<astrTierIDs.length; i++)
					{

			System.out.println(MSG_PREFIX  + oTierIdentity +
astrTierIDs[i] + " deallocated!");
					}

					TierDeallocationResult oResult
= new TierDeallocationResult(null);

			oResult.setSignature(oRequest.getSignature());
					while(true)
					{
						try
						{

			System.out.println(MSG_PREFIX + "Sending deallocation
result ...");

			this.oSysDSTTA.setResult(oResult);

			System.out.println(MSG_PREFIX + "TierDeallocationResult
sent!");
							break;
						}
						catch
(DMSException oDMSException)
						{
							try
							{

			this.oSysDSTTA =
this.oTAExceptionHandler.fixTA(this.oSysDSTTA, oDMSException);
							}
							catch
(InterruptedException oInterrptedException)
							{

			oInterrptedException.printStackTrace(System.err);
							}
						}
					}
```

```
				}
			}
		}

		public static Configuration loadFromFile(String
pstrFileName)
		throws IOException
		{
			FileInputStream oFileInStream = new
FileInputStream(pstrFileName);
			Configuration oConfig = new Configuration();
			oConfig.load(oFileInStream);
			oFileInStream.close();
			return oConfig;
		}

		public INodeController getDSTController()
		{
			return oDSTController;
		}

		public String getNodeID()
		{
			return strNodeID;
		}

		public void setNodeID(String pstrNodeID)
		{
			this.strNodeID = pstrNodeID;
		}

		public ITransportAgent getRegistrationDSTTA()
		{
			return oRegDSTTA;
		}

		public void setRegistrationDSTTA(ITransportAgent
poRegistrationDSTTA)
		{
			this.oRegDSTTA = poRegistrationDSTTA;
			this.oRegDSTTAConfig =
this.oRegDSTTA.getConfiguration();
		}

		public ITransportAgent getSystemDSTTA()
		{
			return oSysDSTTA;
		}

		public void setSystemDSTTA(ITransportAgent
poSystemDSTTA)
		{
			this.oSysDSTTA = poSystemDSTTA;
			this.oSysDSTTAConfig =
this.oSysDSTTA.getConfiguration();
		}

		public INodeController getGMTController()
		{
			return this.oGMTController;
		}

		public String getHostName()
		{
			return this.strHostName;
		}
}
```

## Reverse Engineering Tool

The Object aid is a light-footed and lightweight code visualization device for the Eclipse IDE. It demonstrates our Java source code and libraries in live UML class and succession graphs that naturally redesign as our code changes. All the diagrams which we generated for the gipsy and Dmarf are by using the reverse engineering tool i.e Objet aid. Below are some the working screen shots which we used to generate the class diagrams of the GIPSY and DMARF.

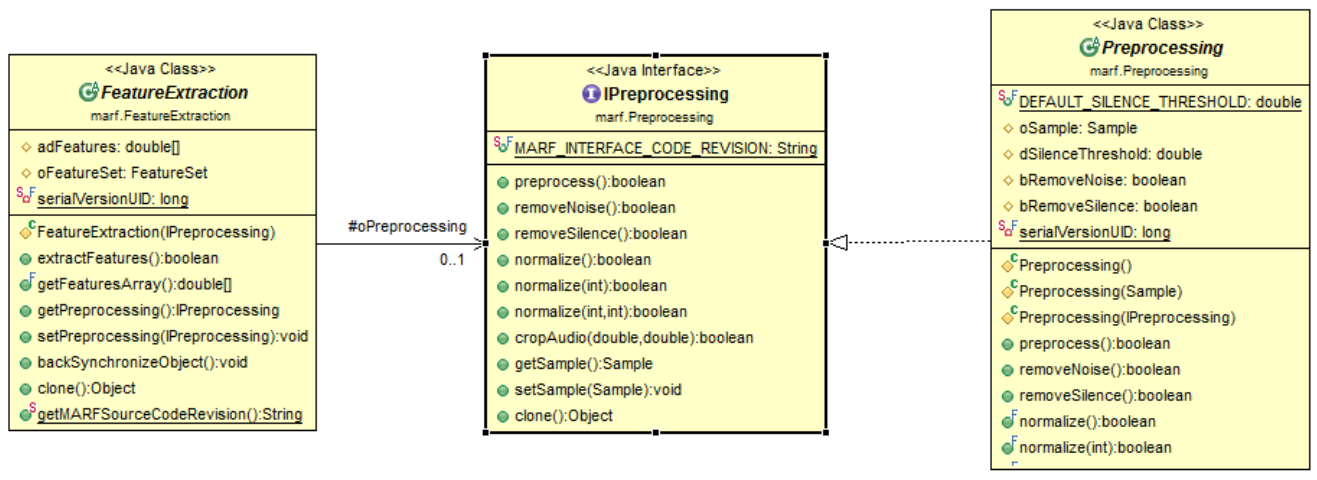

*Figure 42 UML diagram from Reverse Engineering Tool*

*B. Identification of Design Patterns*

*1) DMARF*

1) Factory Method
2) Template Method

*2) GIPSY*

3) Observer
4) Singleton
5) Decorator

Below is the brief description of all the pattern mention above: In a software system there might be many design patterns according to the problem raised. A design pattern is a solution for a set of problems that may occur during the development of the software. If a problem is meant to be repeated throughout the project then the design patterns will be solving the issue. These patterns are language independent. Any developer while developing a software will be facing few problems in common, they have to know the solution right away so these patters will help them a lot. We have to be care full while developing a pattern. Some patterns may lead to another pattern. Knowing design patterns will change a lot for developing code.
For this particular process, we have used a jar file for recognizing the patterns in the source code.

Individually we are doing 5 patterns. 3 in GIPSY and 2 in DMARF, they are

**Observer:** In any object oriented programming there is a need of object in every class. This observer patterns deals mainly with the objects and their dependencies, it define the one to many dependency between objects. Advantage of this observer pattern is that whenever an object is changed to another state all the other objects which are dependent on this main object are informed and updates automatically.

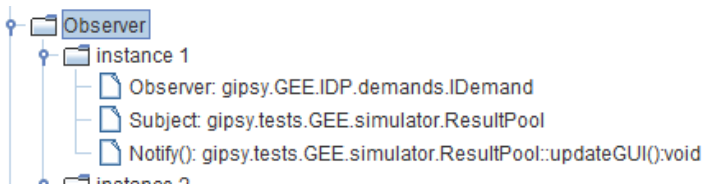

Code is

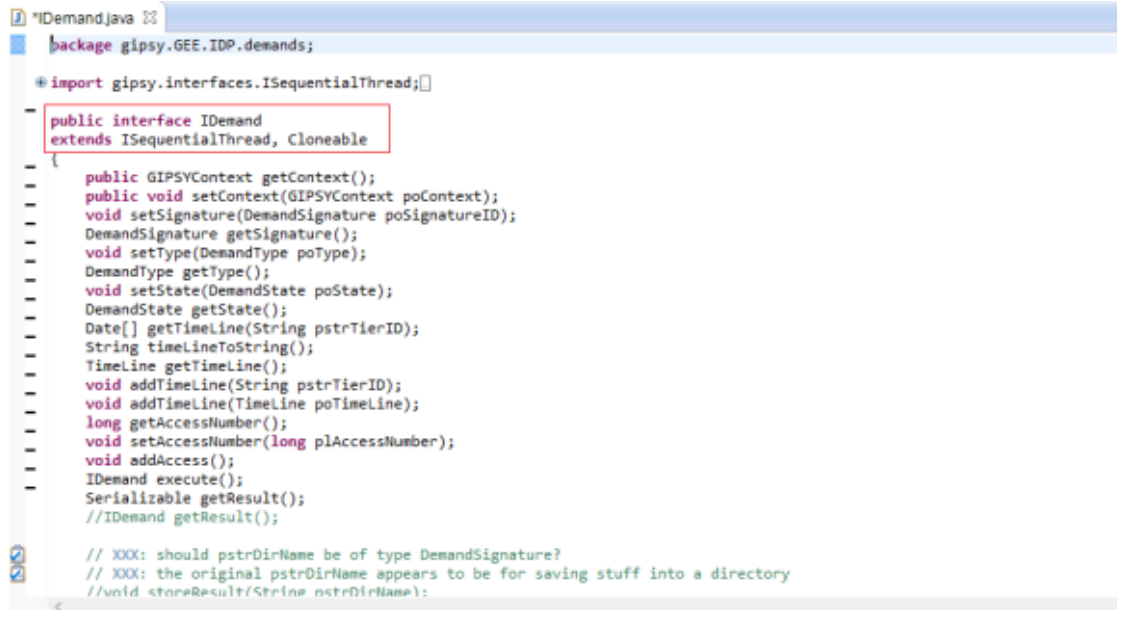

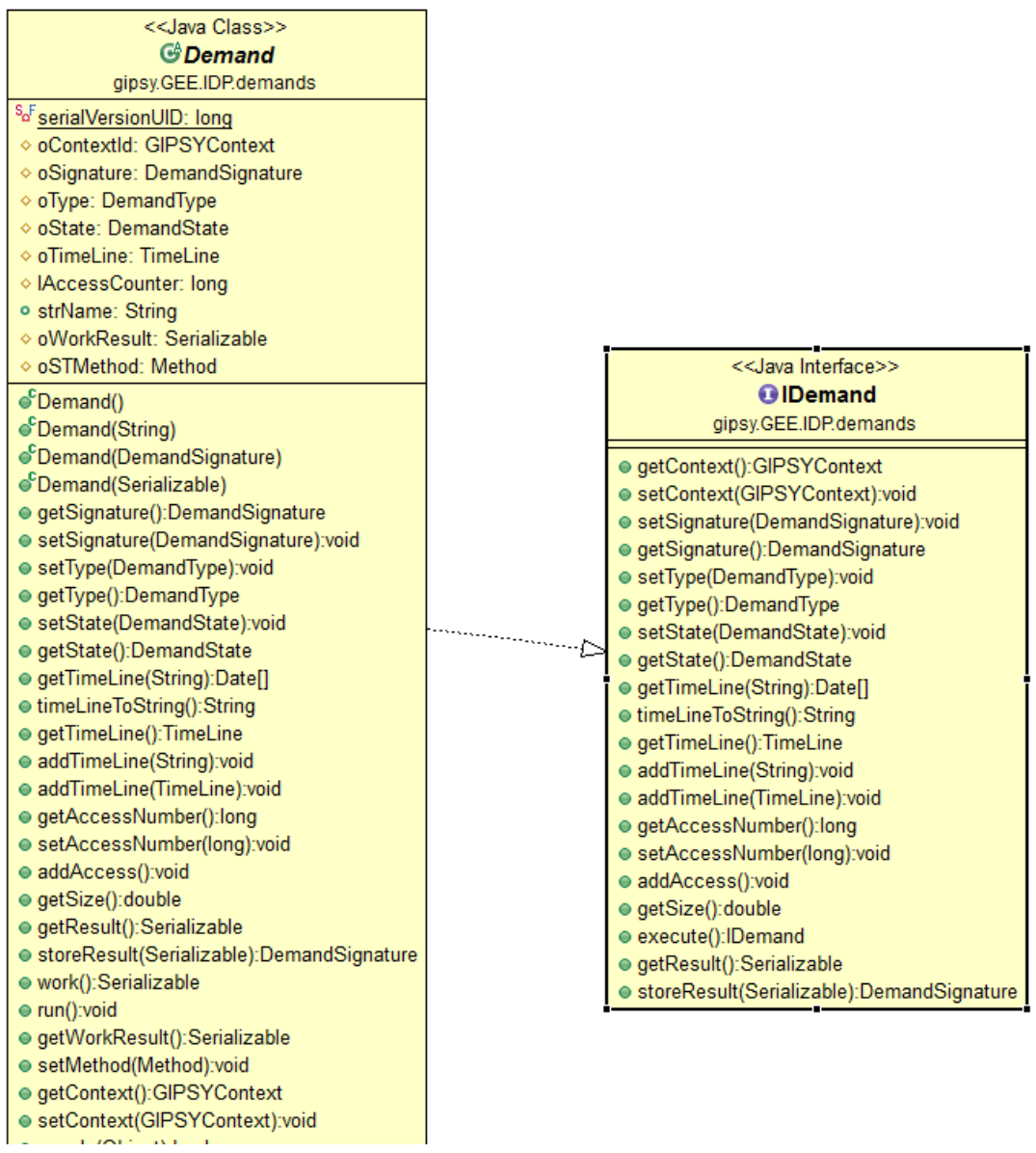

*Figure 43 Observer*

**Singleton:**

Singleton pattern is mainly concerned with making the instance of the class, it makes sure that it creates only one instance for one class. Though it creates only one instance it makes sure that it has access globally. By creating one instance at every point of time there is one instance variable.

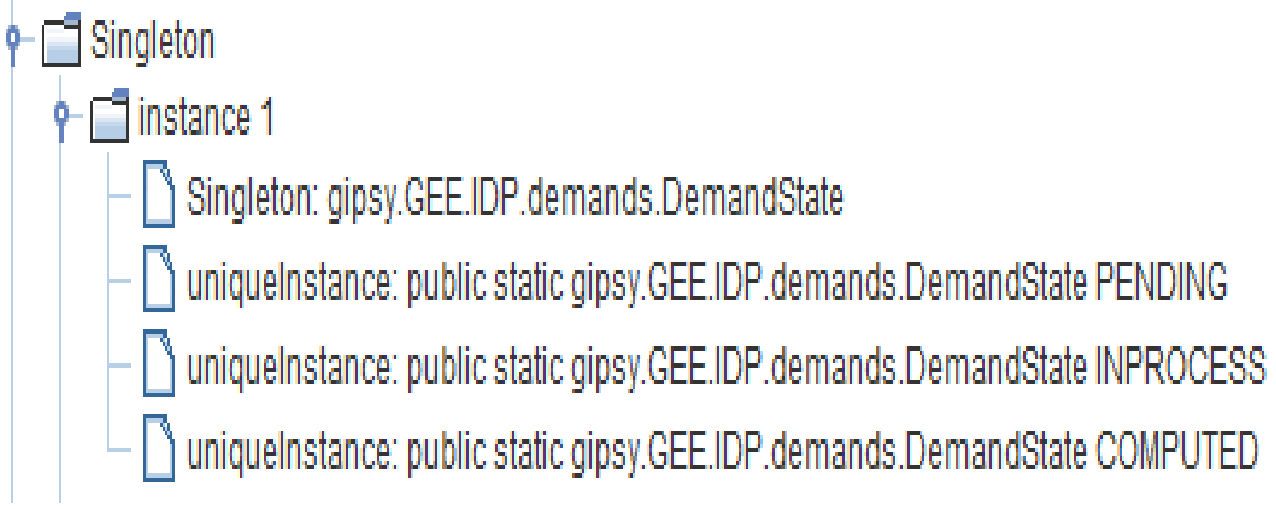

*Figure 44*

The code is

```
package gipsy.GEE.IDP.demands;
import java.io.Serializable;
public class DemandState
implements Serializable
{
	private static final long serialVersionUID = 1000L;
	private static final String STATE_PENDING = "pending";
	private static final String STATE_INPROCESS = "inprocess";
	private static final String STATE_COMPUTED = "computed";
	private String strState = "";
	public static final DemandState PENDING = new DemandState(STATE_PENDING);
	public static final DemandState INPROCESS = new DemandState(STATE_INPROCESS);
	public static final DemandState COMPUTED = new DemandState(STATE_COMPUTED);
	private DemandState(String pstrNewState)
	{
		this.strState = pstrNewState.toLowerCase();
	}
	public boolean isPending()
	{
		if(this.strState.equals(STATE_PENDING))
		{
			return true;
		}
		else
		{
			return false;
		}
	}
	public boolean isInProcess()
	{
		if(this.strState.equals(STATE_INPROCESS))
		{
			return true;
		}
		else
		{
			return false;
		}
	}
	public boolean isComputed()
	{
		if(this.strState.equals(STATE_COMPUTED))
		{
			return true;
		}
		else
		{
			return false;
		}
	}
	public String toString()
	{
		if(this.strState.equals(STATE_PENDING))
		{
			return "STATE_PENDING";
		}
		else if(this.strState.equals(STATE_INPROCESS))
		{
			return "STATE_INPROCESS";
		}
		else
			return "STATE_COMPUTED";
	}
}
```

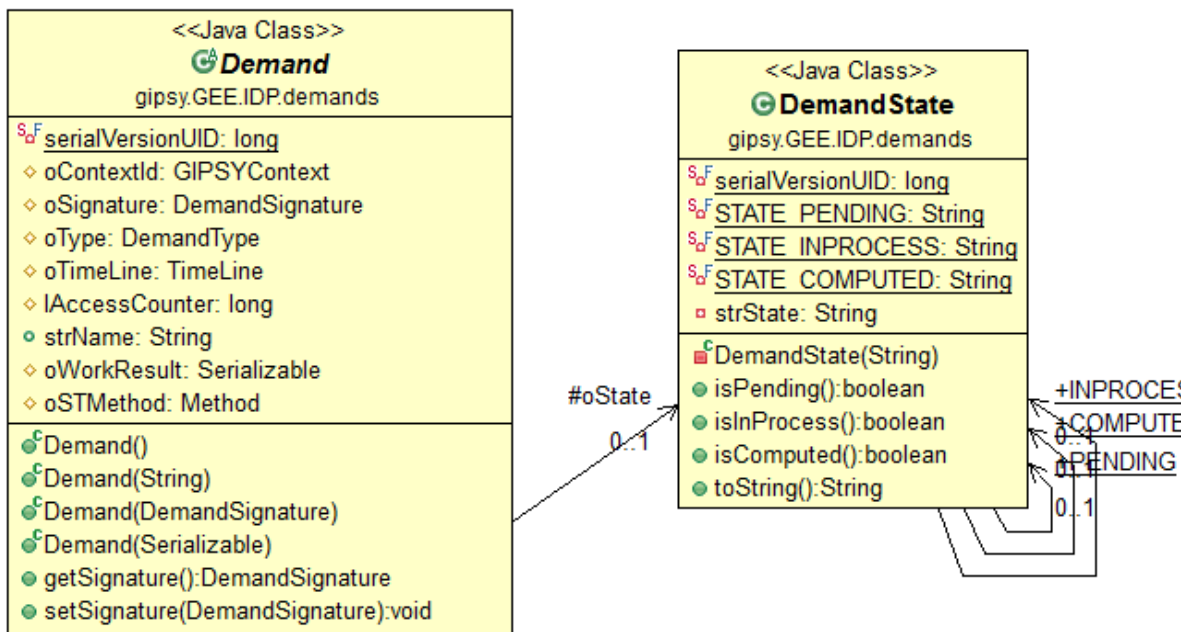

*Figure 45 Singleton*

**Decorator:**

Decorator can be able to add few more features to object dynamically. Actually through inheritance we can add few responsibilities statically but it's not possible dynamically. We can also remove some unnecessary responsibilities dynamically through decorator.

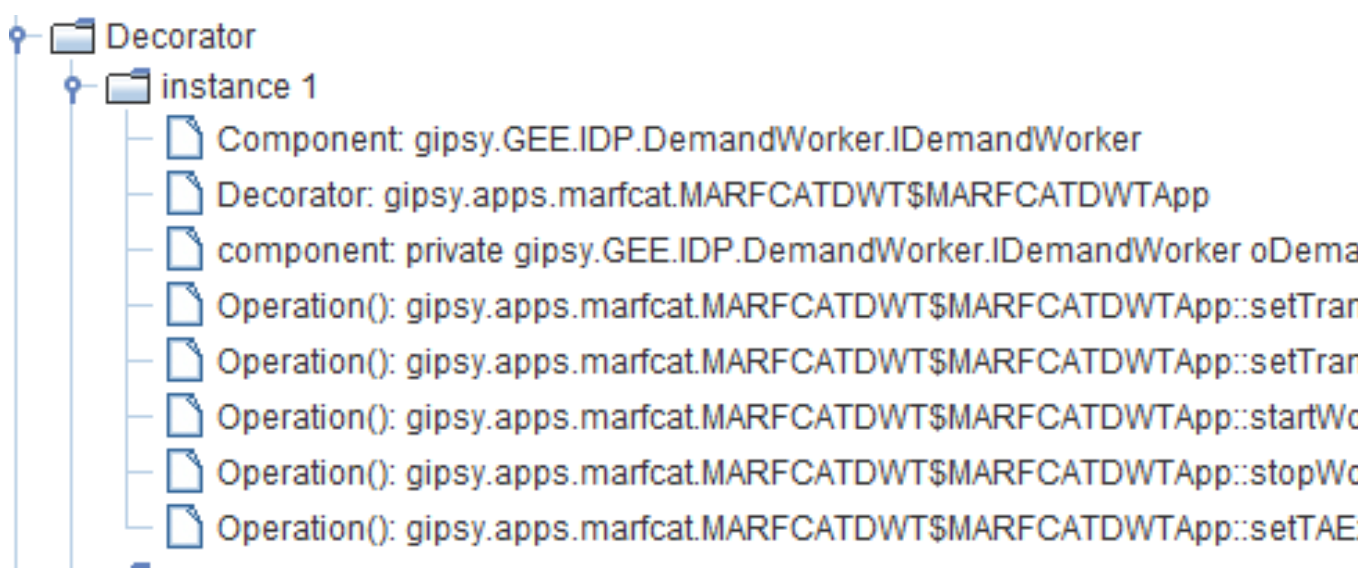

*Figure 46*

The code

```
package gipsy.GEE.IDP.DemandWorker;

import gipsy.GEE.IDP.ITransportAgent;

public interface IDemandWorker
extends Runnable
{
	/**
	 * Set TA which will be used by the worker.
	 *
	 * @param poDMFImp might be Jini, JMS...
	 */
	void setTransportAgent(EDMFImplementation poDMFImp);

	void setTransportAgent(ITransportAgent poTA);

	void setTAExceptionHandler(TAExceptionHandler poTAExceptionHandler);

	void startWorker();

	void stopWorker();
}

// EOF
```

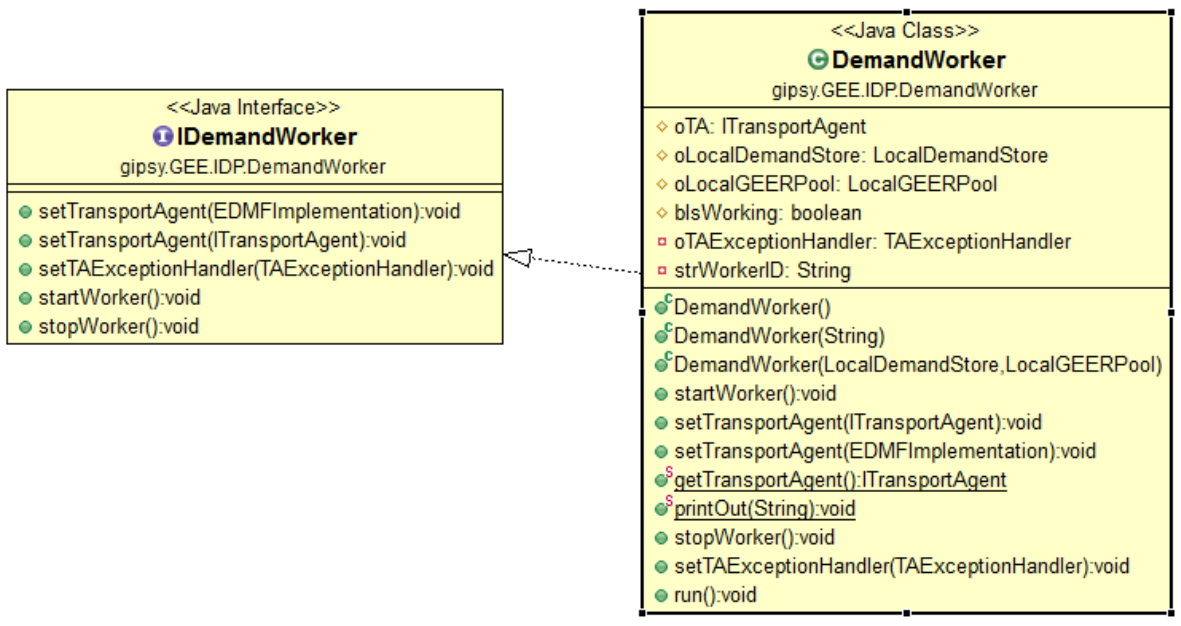

*Figure 47 Decorator*

## Factory Method

Main functionality of this factory method pattern is to define an interface for any object which is to be created. It lets subclass to decide for initiation. It also makes sure that for any newly created object it creates some reference and a common interface.

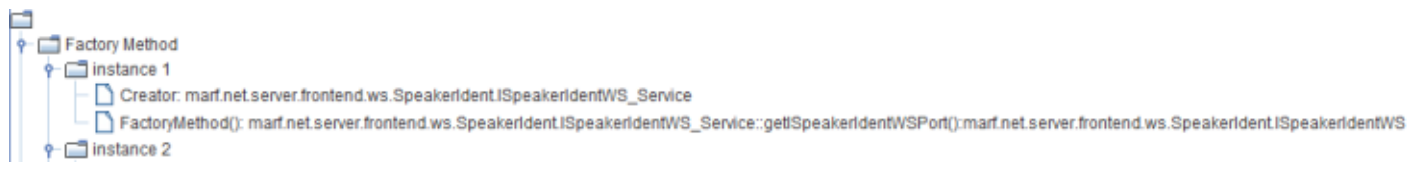

The source code for Factory method Pattern is displayed below.

```
ISpeakerIdentWS_Service.java

// This class was generated by the JAXRPC SI, do not edit.

package marf.net.server.frontend.ws.SpeakerIdent;

import javax.xml.rpc.*;

public interface ISpeakerIdentWS_Service extends javax.xml.rpc.Service {
    public marf.net.server.frontend.ws.SpeakerIdent.ISpeakerIdentWS getISpeakerIdentWSPort() throws ServiceException;
}
```

UML Diagram:

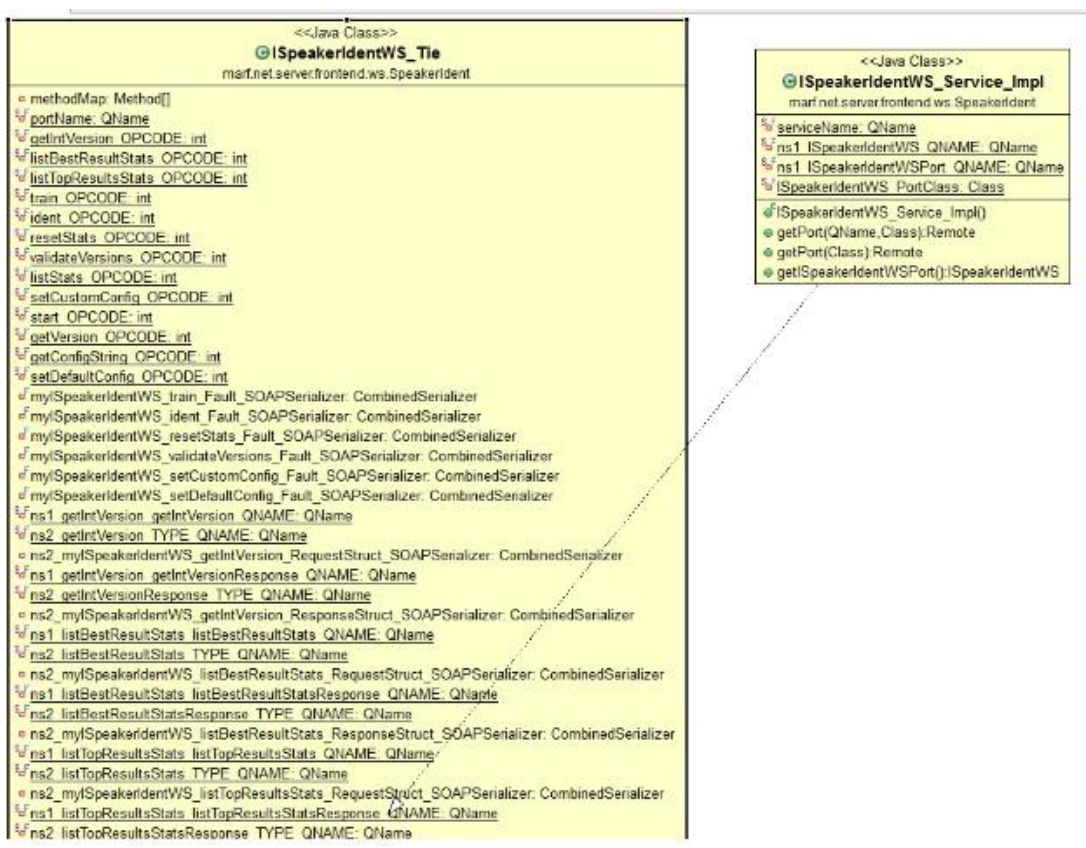

*Figure 48 Factory method*

## Template Method

It defines skeleton of the algorithm in a method. Which is named after template method. Without changing the structure of an algorithm it redefines certain steps.

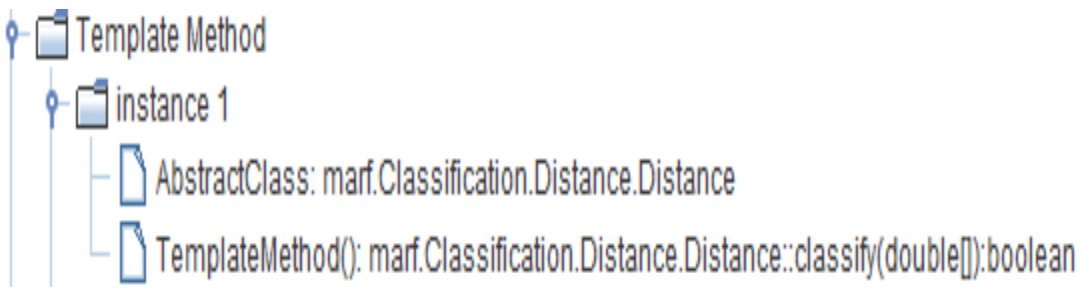

The code is

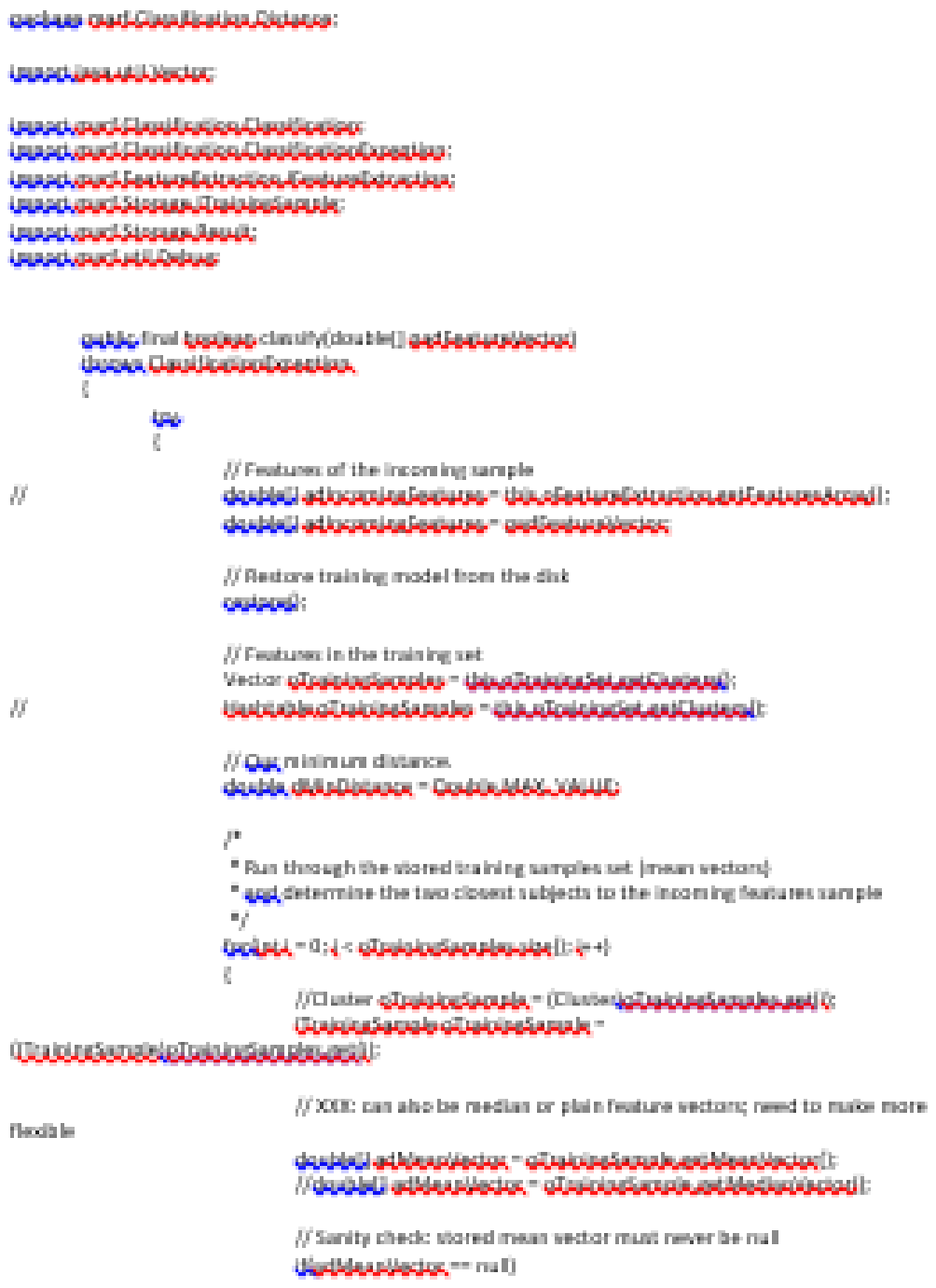

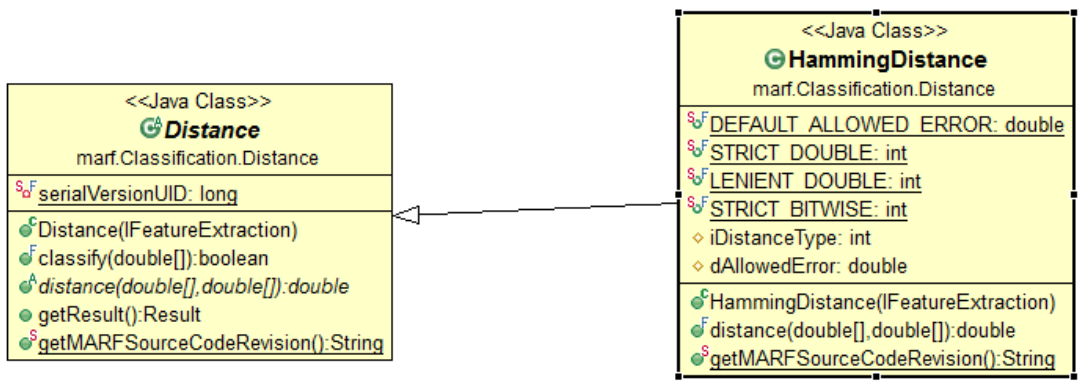

*Figure 49 Template method*

### *C. Implementation*

#### *1) Refactoring Changesets and Diffs*

The bad smell identified in earlier part is refactored in the following section.

##### *a) DMARF*

## Refactoring 1

## marf.nlp.Parsing. SymbolTable.java

```
package marf.nlp.Parsing;
import java.util.Enumeration;
import java.util.Hashtable;
import marf.util.NotImplementedException;

public class SymbolTable
{
	private Hashtable oSymTabEntries = null;
	protected Hashtable oSymTabIndex = new Hashtable();
	protected SymbolTable oParentSymTab = null;
	protected String strName = "";

	public int addSymbol(Token poToken)
	{
		if(this.oSymTabEntries.contains(poToken.getLexeme()))
		{
			SymTabEntry oSymTabEntry = (SymTabEntry)oSymTabEntries.get(poToken.getLexeme());

			oSymTabEntry.addLocation(poToken.getPosition());
		}
		else
		{
			SymTabEntry oSymTabEntry = new SymTabEntry(poToken);

			oSymTabEntry.addLocation(poToken.getPosition());

			this.oSymTabEntries.put(poToken.getLexeme(), oSymTabEntry);
		}

		return 0;
	}
}
```

## Test Case 1

The test class used to test the method addSymbol is below:

```
package marf.tests.nlp.Parsing;
import java.util.Hashtable;
import java.util.Iterator;
import java.util.Map;
import java.util.Set;
import marf.nlp.Parsing.SymbolTable;
import marf.nlp.Parsing.Token;
import junit.framework.TestCase;

public class SymbolTableTest
{
		public static void main(String[] args)
		{
				try
				{
						Hashtable oSymTabEntries = new
Hashtable();
						Token token = new
Token("HashTableEntryTest",2,1,2);
						SymbolTable test = new SymbolTable();
						System.out.println(token.getLexeme());
						System.out.println(token.getPosition());
						System.out.println(test.addSymbol(token));

		System.out.println(test.getSymTabEntries());// Getting entry of
Hashtable
				}
				catch(Exception e)
				{
						System.err.print(e.getMessage());
				}

		}

}
```

**The result of test case is:**

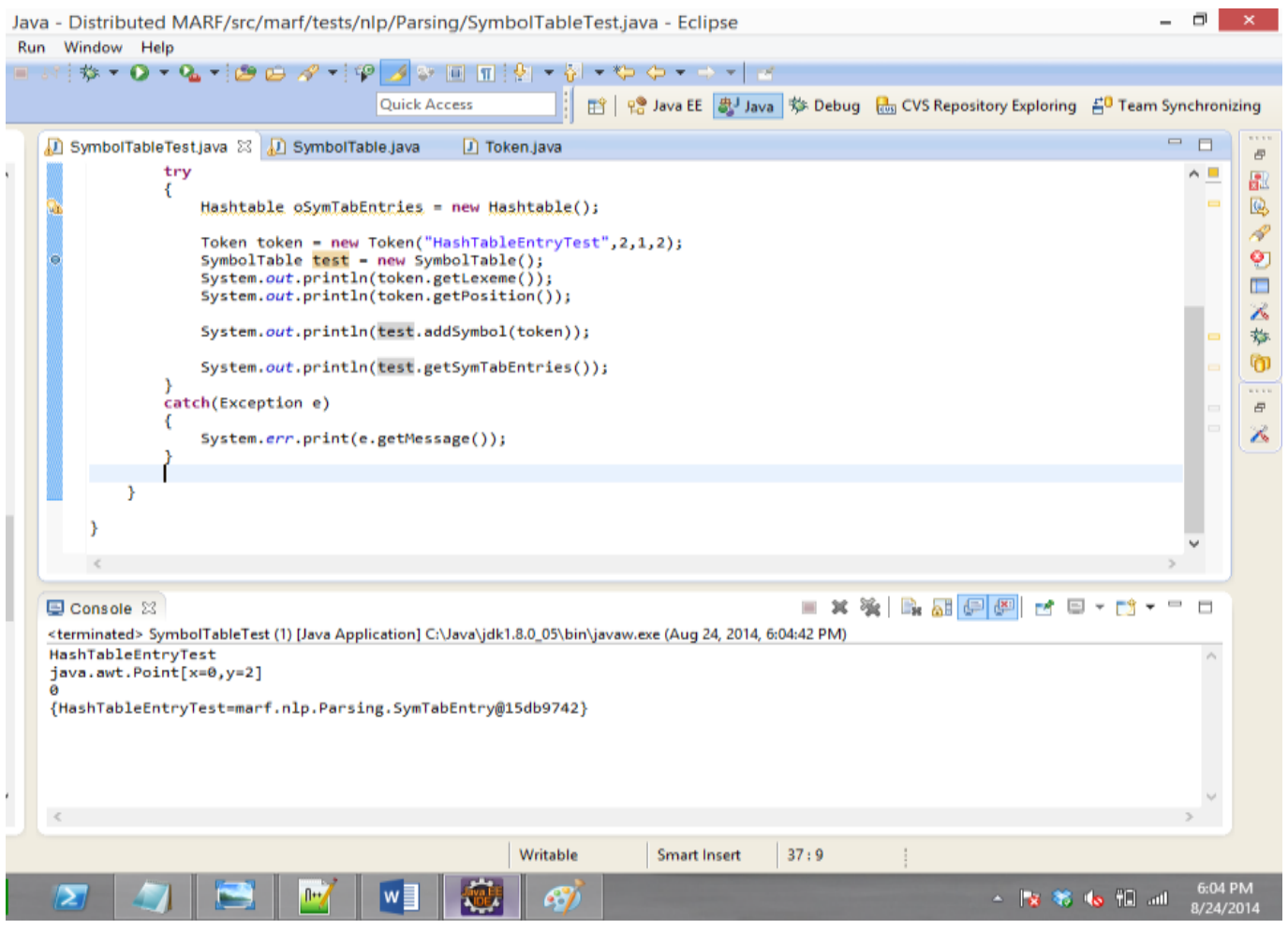

After refactoring and moving method addSymbol to marf.nlp.Parsing.Token.java, the refactored code is

```
public int addSymbol(SymbolTable poSymTable)
		{

		if(poSymTable.oSymTabEntries.contains(this.getLexeme()))
				{
						SymTabEntry oSymTabEntry =
(SymTabEntry)poSymTable.oSymTabEntries.get(this.getLexeme());

		oSymTabEntry.addLocation(this.getPosition());
				}
				else
				{
						SymTabEntry oSymTabEntry = new
SymTabEntry(this);

		oSymTabEntry.addLocation(this.getPosition());

		poSymTable.oSymTabEntries.put(this.getLexeme(),
oSymTabEntry);
				}

				return 0;
		}
```

The test case result is :

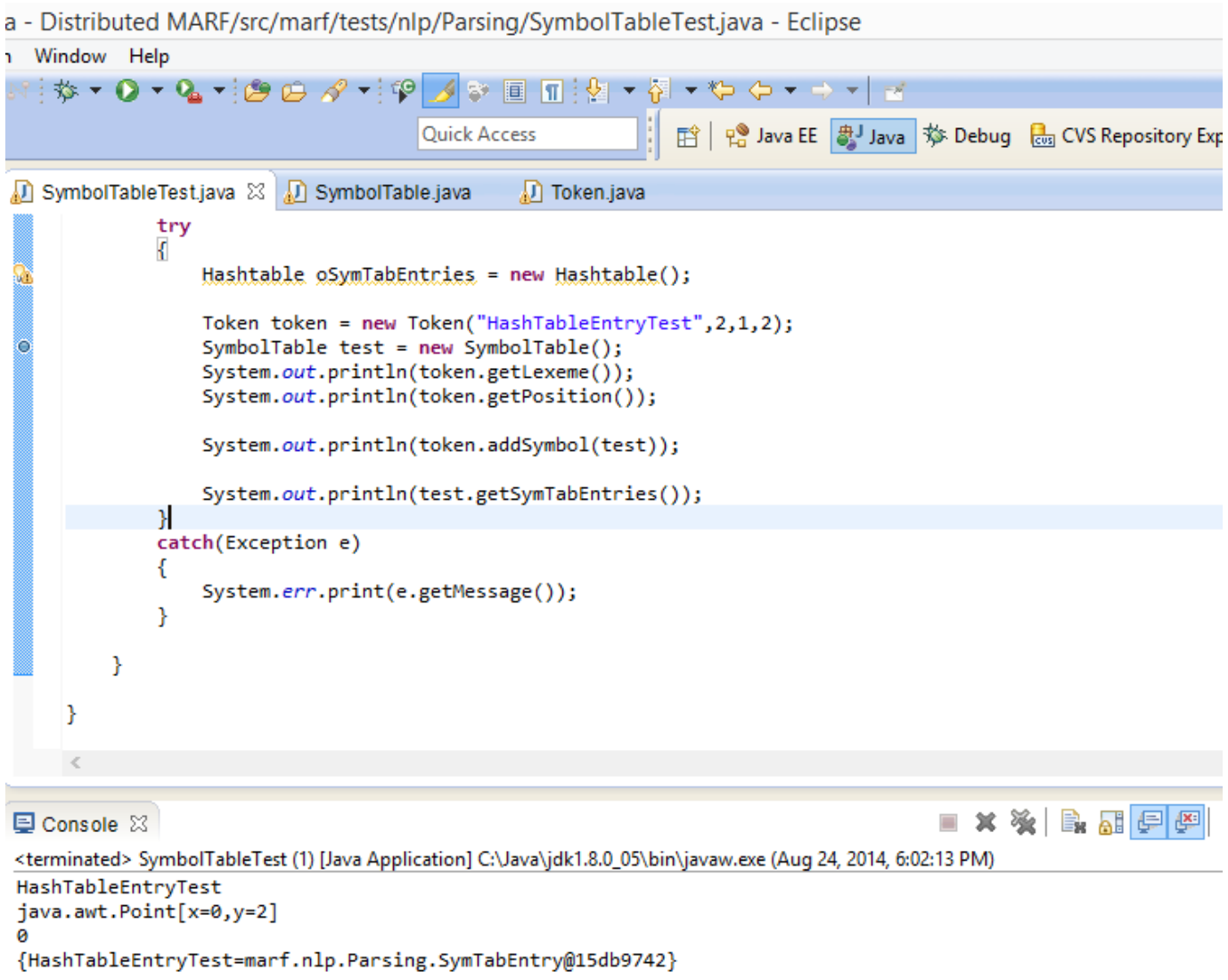

### *b) GIPSY*

## Refactoring 2

### gipsy.GIPC.intensional.SIPL.IndexicalLucid.IndexicalLucidParserTokenManager.java

```
/* Generated By:JJTree&JavaCC: Do not edit this line.
IndexicalLucidParserTokenManager.java */
package gipsy.GIPC.intensional.SIPL.IndexicalLucid;
import gipsy.GIPC.intensional.SimpleNode;
import gipsy.GIPC.util.Node;
import gipsy.GIPC.util.ParseException;
import gipsy.GIPC.util.SimpleCharStream;
import gipsy.GIPC.util.Token;
import gipsy.GIPC.util.TokenMgrError;
import gipsy.interfaces.AbstractSyntaxTree;
import marf.util.Debug;

public class IndexicalLucidParserTokenManager implements
IndexicalLucidParserConstants
{
	int iCount = 0;

	public void commonTokenAction(Token poToken)
	{
		System.out.println(poToken.image);
	}

  public  java.io.PrintStream debugStream = System.out;
  public  void setDebugStream(java.io.PrintStream ds) { debugStream = ds; }
private final int jjStopStringLiteralDfa_0(int pos, long active0)
```

```
{
  switch (pos)
  {
    case 0:
      if ((active0 & 0x1fe0380fe00L) != 0L)
      {
        jjmatchedKind = 51;
        return 28;
      }
      if ((active0 & 0x1000000000000000L) != 0L)
        return 4;
      if ((active0 & 0x80000000L) != 0L)
        return 50;
```

## Test Case Used

The test class used to test the method commonTokenAction is below and moreover the function is made public for testing purpose.

```
package gipsy.tests.GIPC.intensional.SIPL.IndexicalLucid;

import gipsy.GIPC.intensional.SimpleNode;
import
gipsy.GIPC.intensional.SIPL.IndexicalLucid.IndexicalLucidParserTokenMan
ager;
import gipsy.GIPC.util.Node;
import gipsy.GIPC.util.ParseException;
import gipsy.GIPC.util.SimpleCharStream;
import gipsy.GIPC.util.Token;
import gipsy.GIPC.util.TokenMgrError;
import gipsy.interfaces.AbstractSyntaxTree;

public class IndexicalLucidParserTokenManagerTest
{

        /**
         * @param args
         */
        public static void main(String[] args)
        {
                try
                {

                        int kind=1;
                        String testString="test String";
                        Token test=new Token(kind,testString);  //
Token object with constructor initialized,
                        IndexicalLucidParserTokenManager oParser
=new IndexicalLucidParserTokenManager(null);
                        oParser.commonTokenAction(test);
                        // test.commonTokenAction(); /* used for
testing method after move
                }
                catch(Exception e)
                {
                        System.err.print(e.getMessage());
                }

        }

}
```

## The result of the test is displayed below:

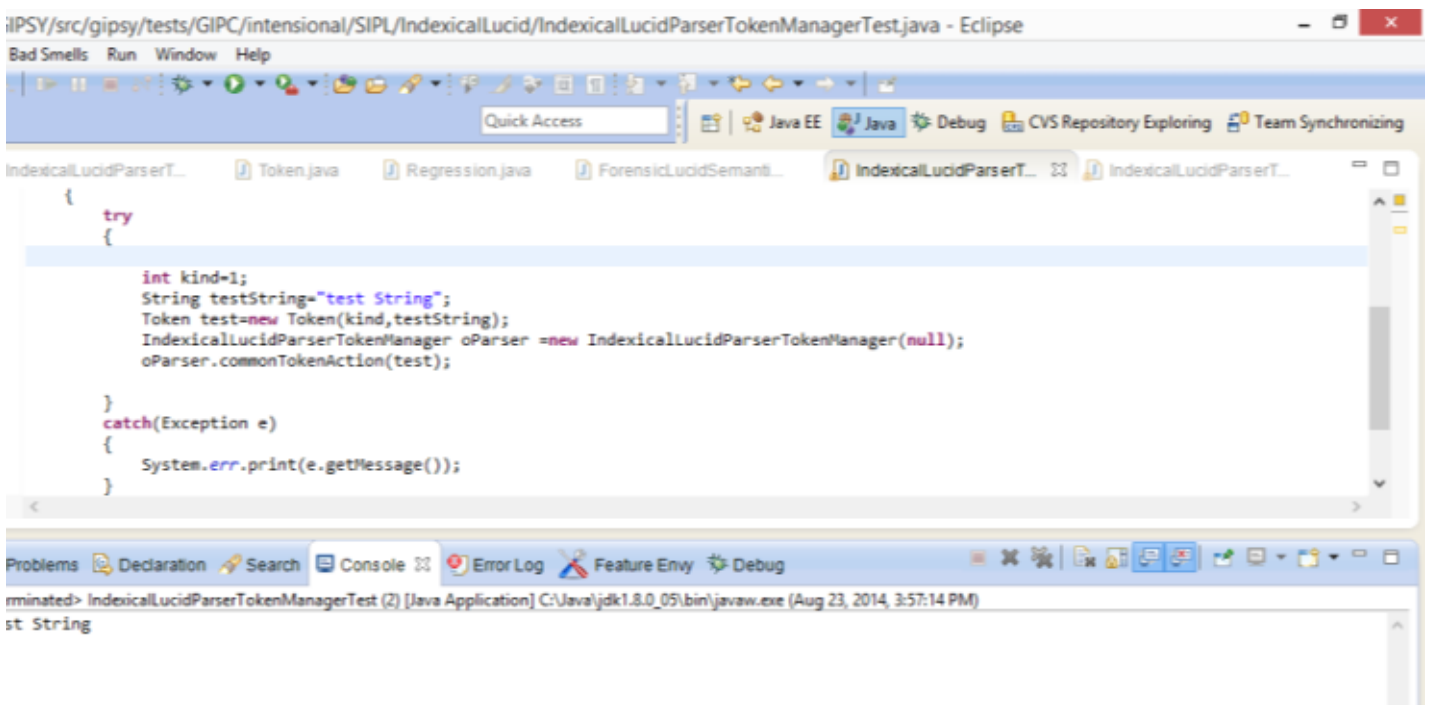

After moving the method to **gipsy.GIPC.util.Token.java and refactoring the method, the test case result is displayed below:**

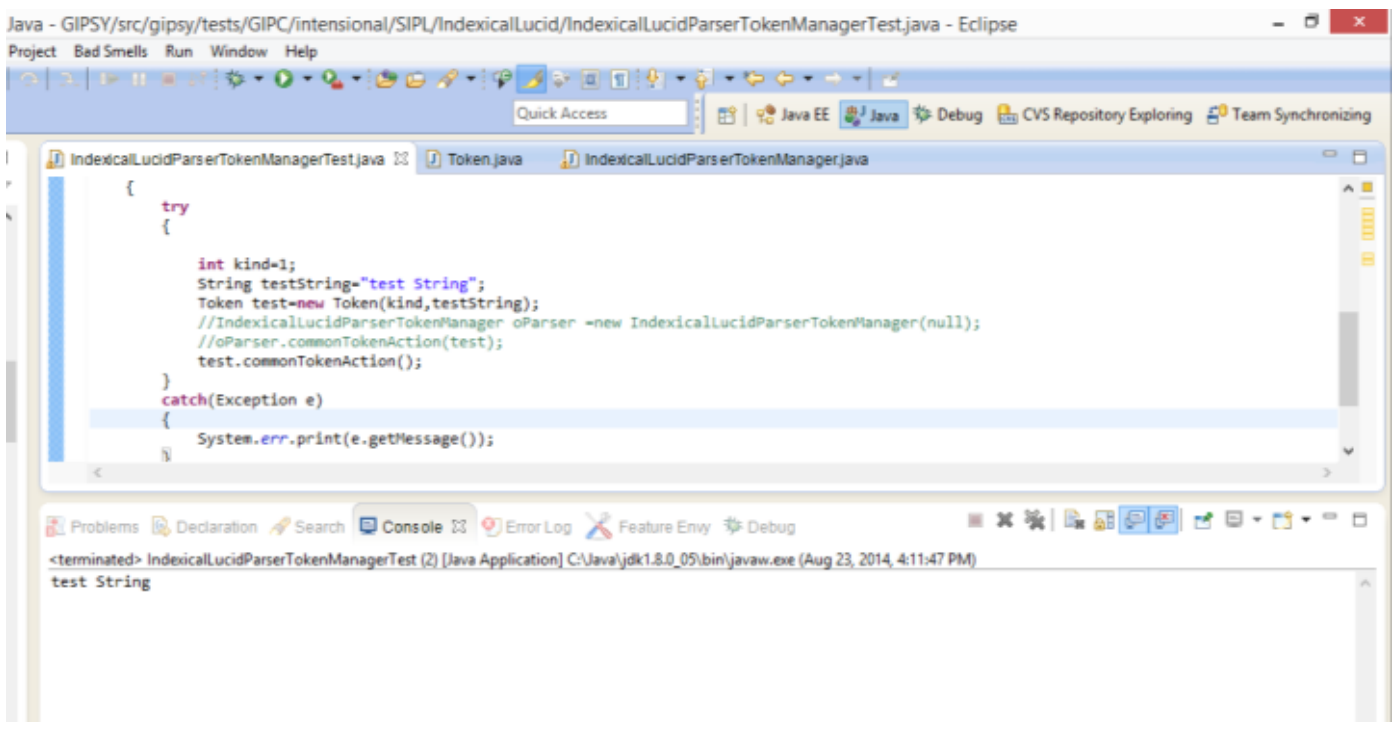

## Refactoring 3 GOD CLASS of GIPSY

```
package gipsy.GEE.IDP.DemandGenerator.rmi;

import gipsy.GEE.CONFIG;
public class IdentifierContext
extends IdentifierContextClient
implements CONFIG
{
        /**
         * each icx object has its specific context.
         */
        private int[] aContext = new int[DIMENSION_MAX];
        private Object oValue;

        public IdentifierContext()
        {
                super();
        }

        /**
         * TODO: Use arraycopy()
         */
        public void init(int[] paContext)
        {
                for(int i = 0 ; i < DIMENSION_MAX; i++)
                        this.aContext[i] = paContext[i];
        }

        public Object cal()
        {
                return null;
        }
}

// EOF
```

## Test Case 3

```
package gipsy.tests.GEE.IDP.DemandGenerator.rmi;

import gipsy.GEE.IDP.DemandGenerator.rmi.IdentifierContext;
import gipsy.GIPC.util.Token;

public class IdentifierContextTest
{
		public static void main(String[] args)
		{
				try
				{
						int[] max= {1,2,3,1,2,1,2};
						IdentifierContext test =  new IdentifierContext();
						test.init(max); // displays number of items in an array


				}
				catch(Exception e)
				{
						System.err.print(e.getMessage());
				}

		}

}
```

## Test case result:

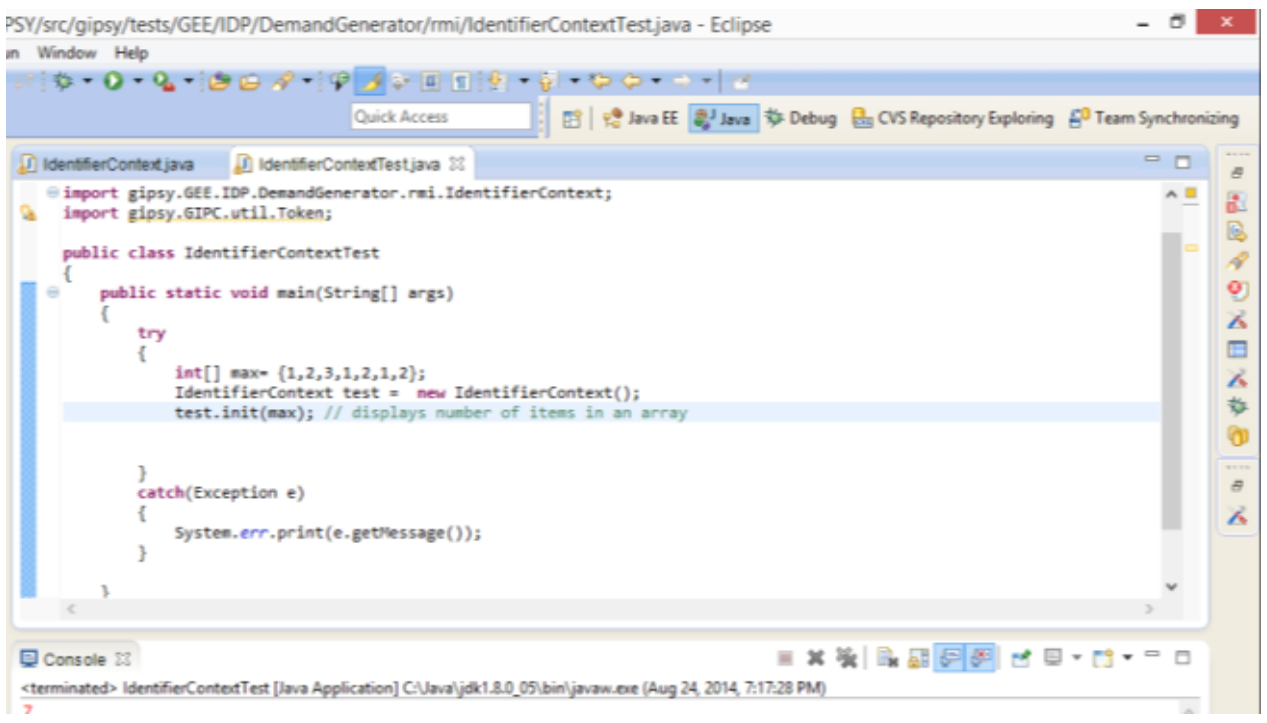

After refactoring make a new class and call the method. New class name IdentifierNumberArray.java and new refactored class is

```
package gipsy.GEE.IDP.DemandGenerator.rmi;

import gipsy.GEE.CONFIG;
public class IdentifierContext
extends IdentifierContextClient
implements CONFIG
{

		private IdentifierNumberArray identifier = new IdentifierNumberArray();
		private Object oValue;

		public IdentifierContext()
		{
				super();
		}
		public void init(int[] paContext)
		{
				identifier.init(paContext);
		}
		public Object cal()
		{
				return null;
		}
}

// EOF
```

## The test case result is:

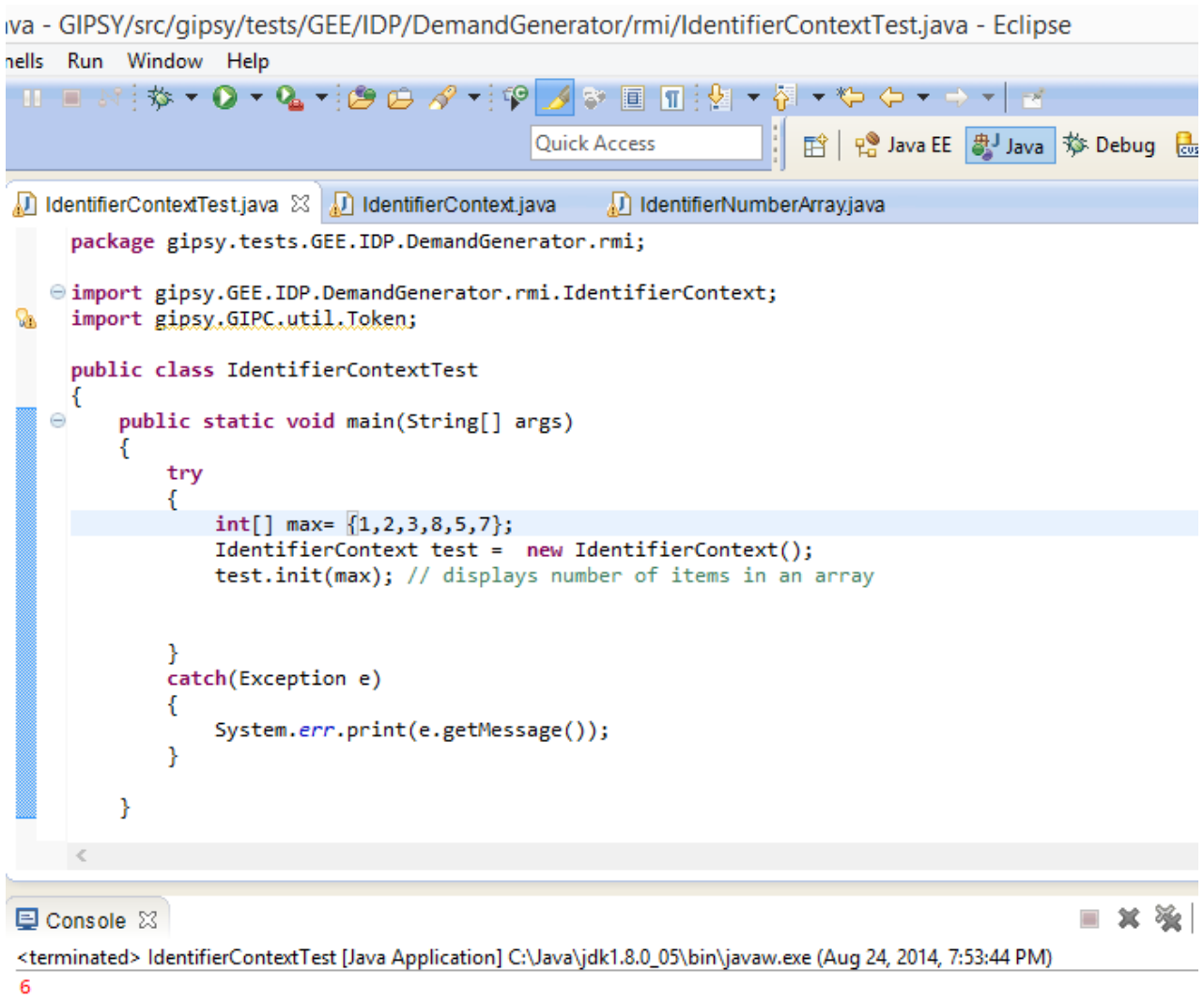

### 2) *Changelog (DMARF and GIPSY)*

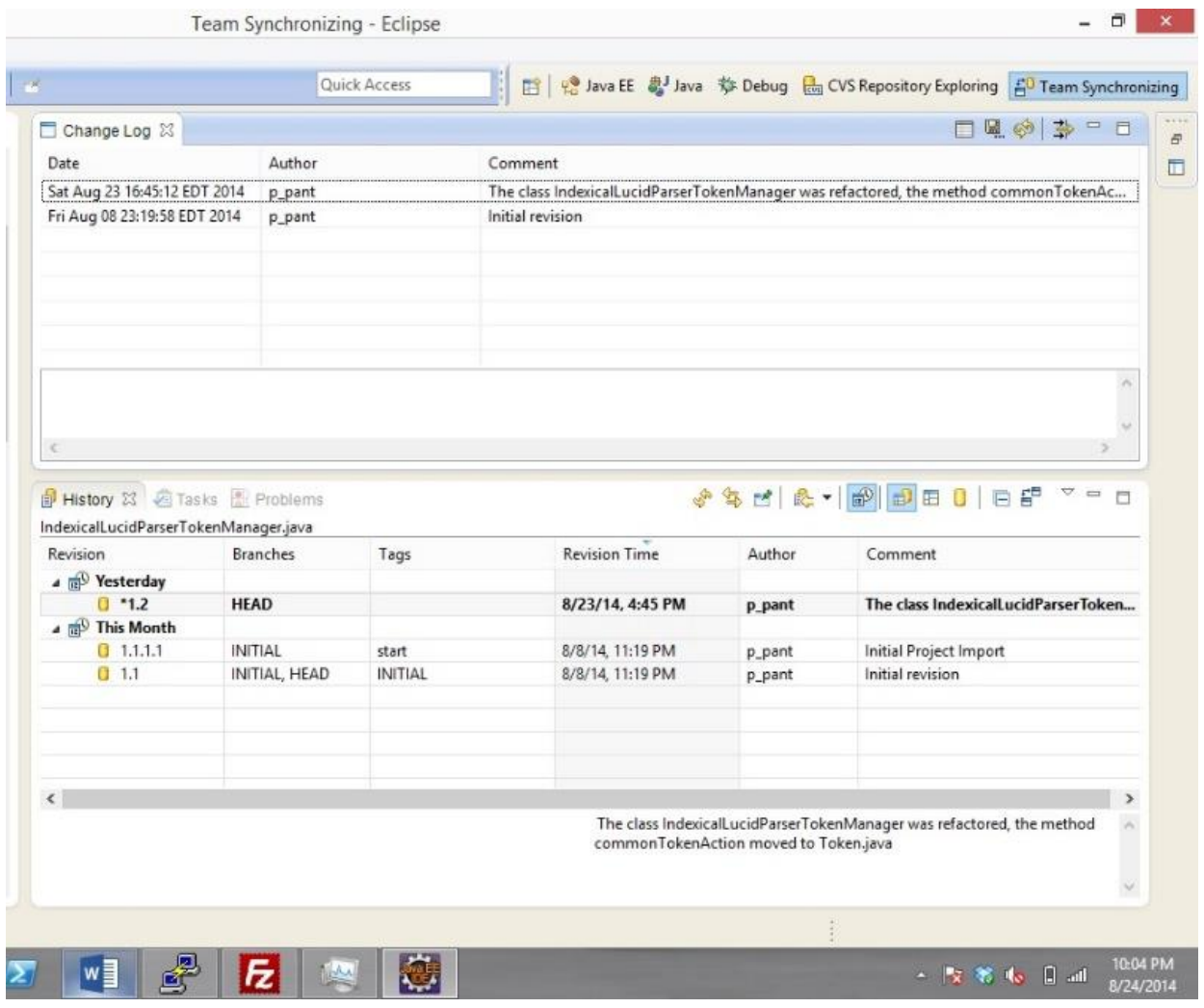

*Figure 51 Gipsy Changelog 1*

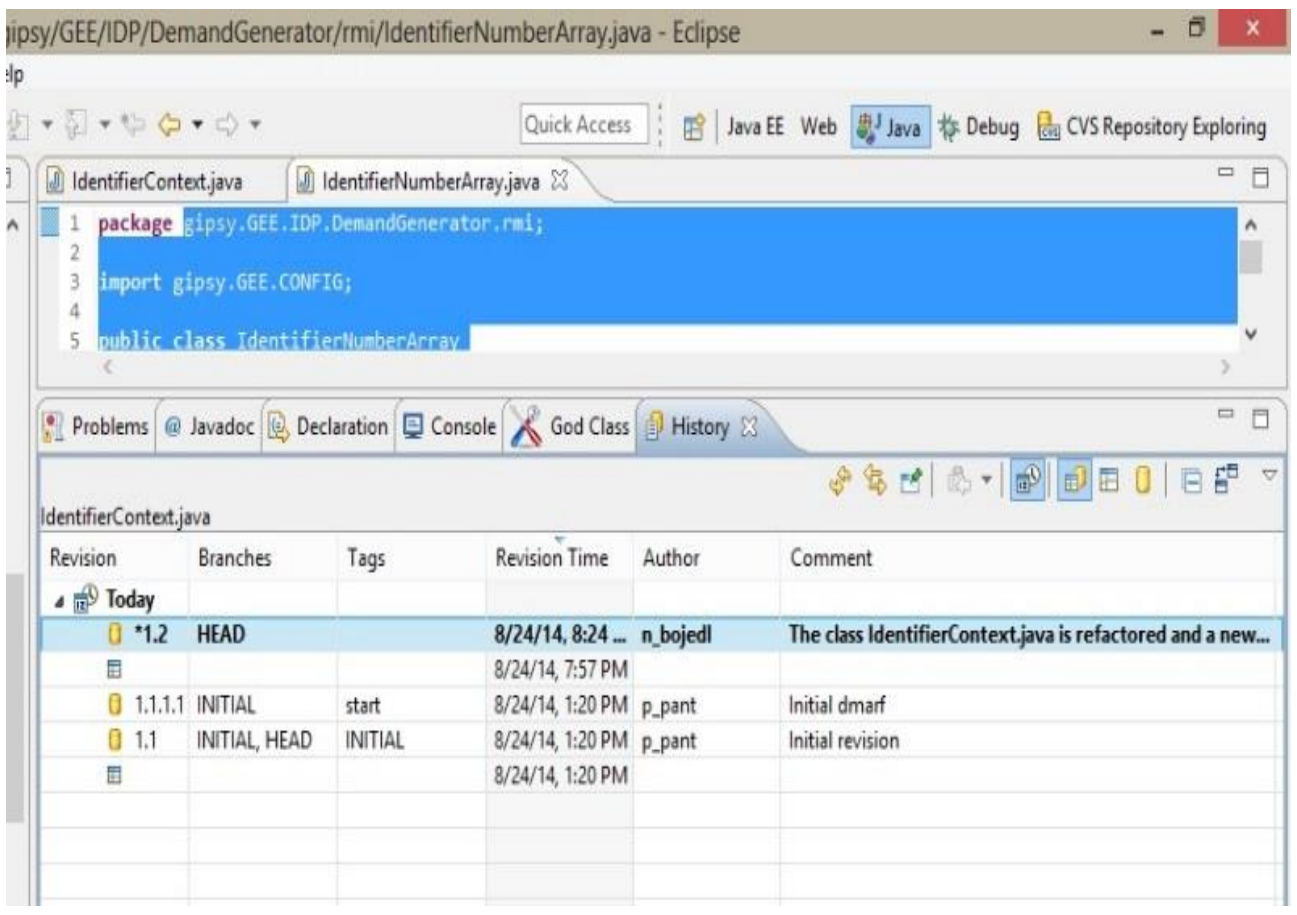

*Figure 50 Gipsy Changelog 2*

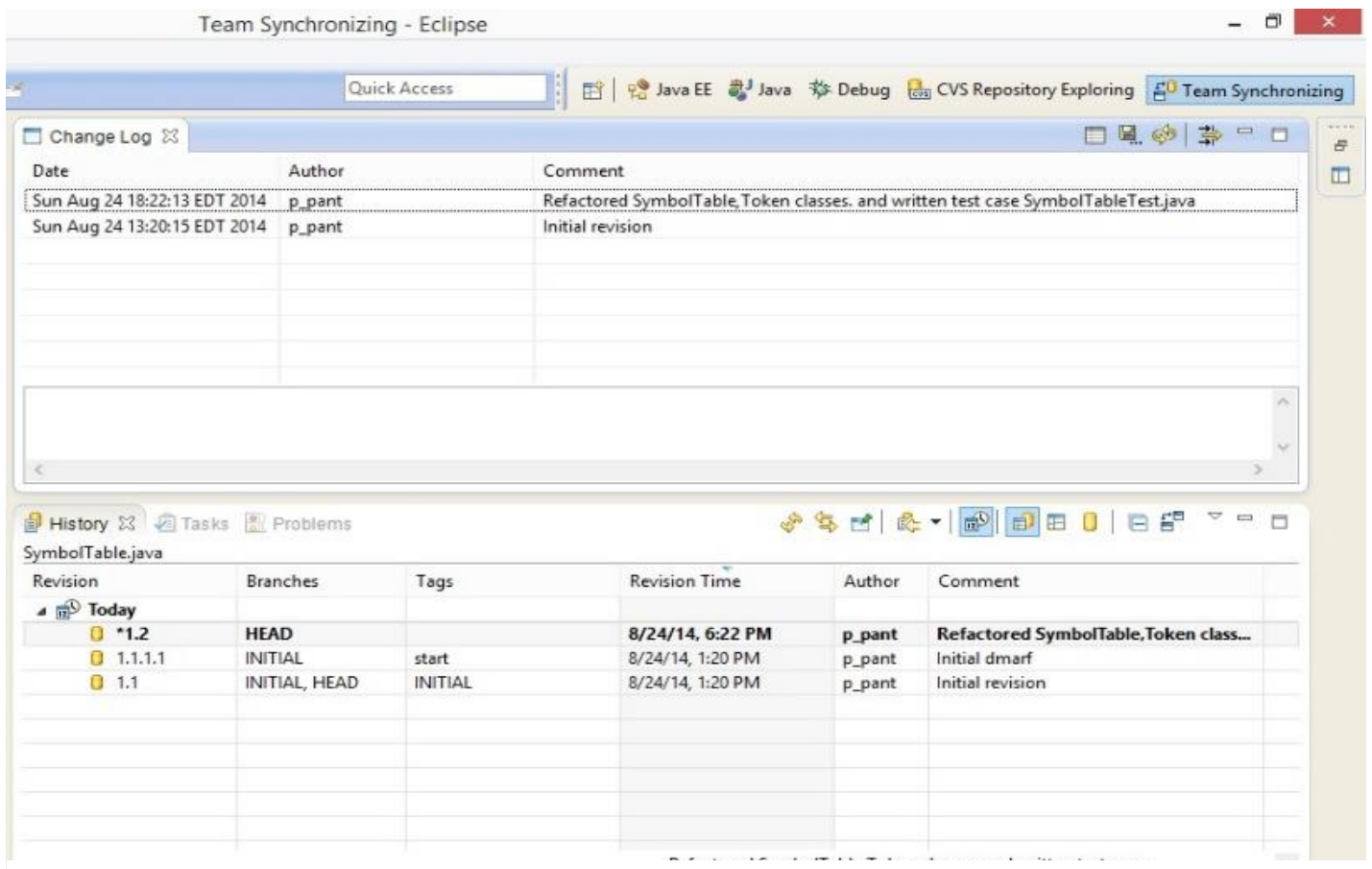

*Figure 52 DMARF Changelog 1*

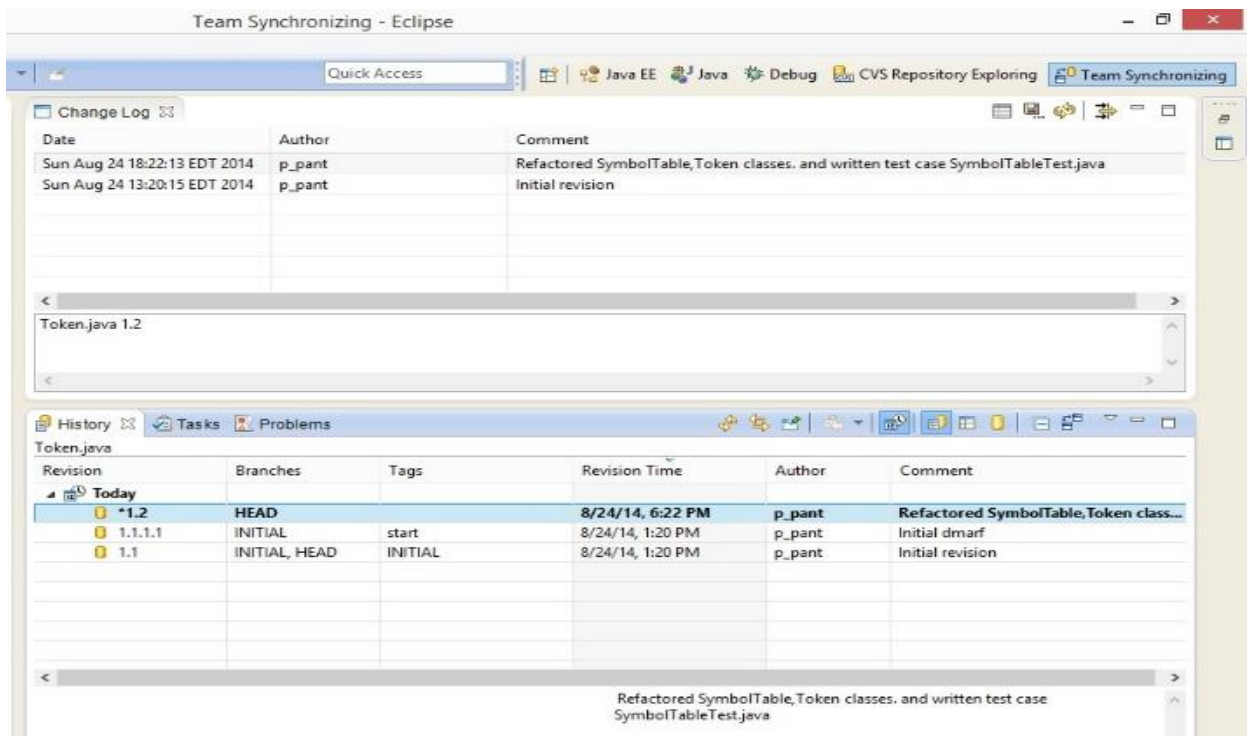

*Figure 53 DMARF Changelog 2*

## 3) Diff (DMARF and GIPSY)

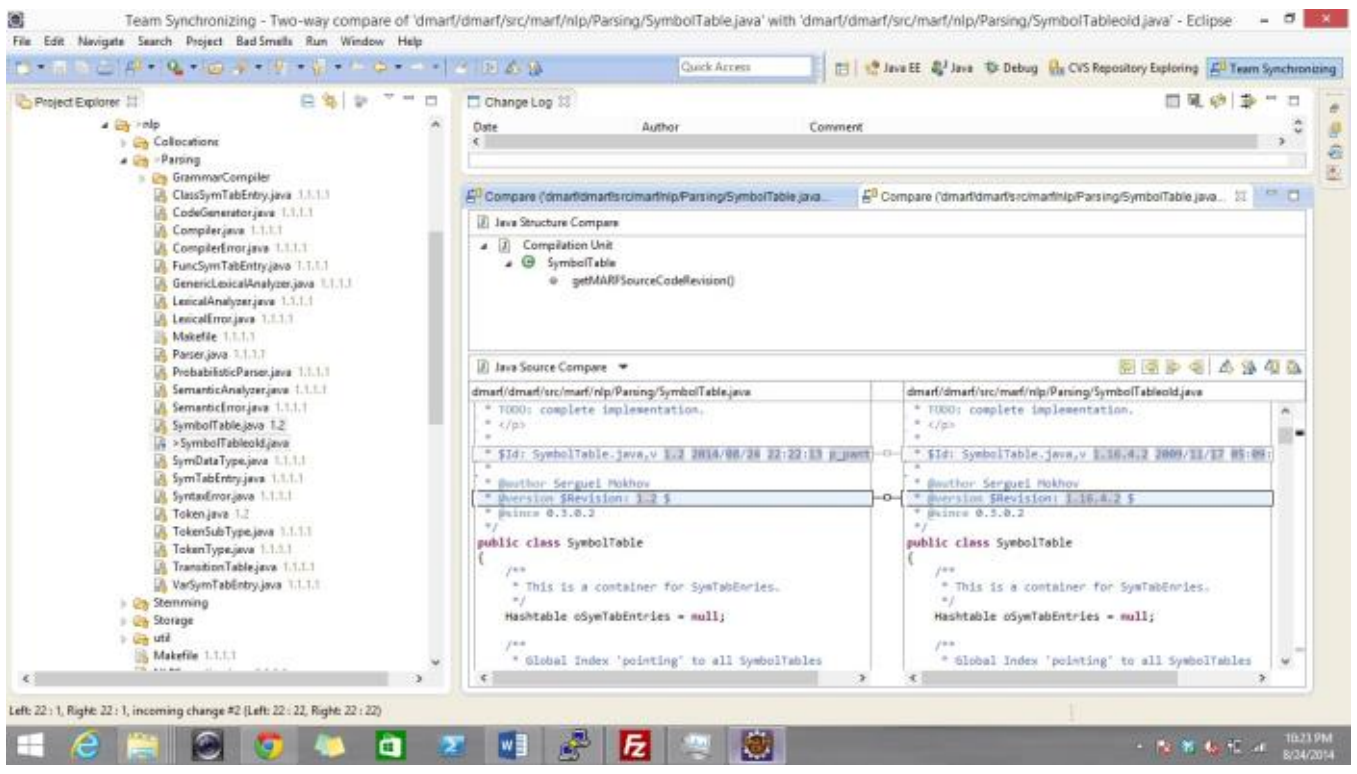

*Figure 54 DMARF Diff*

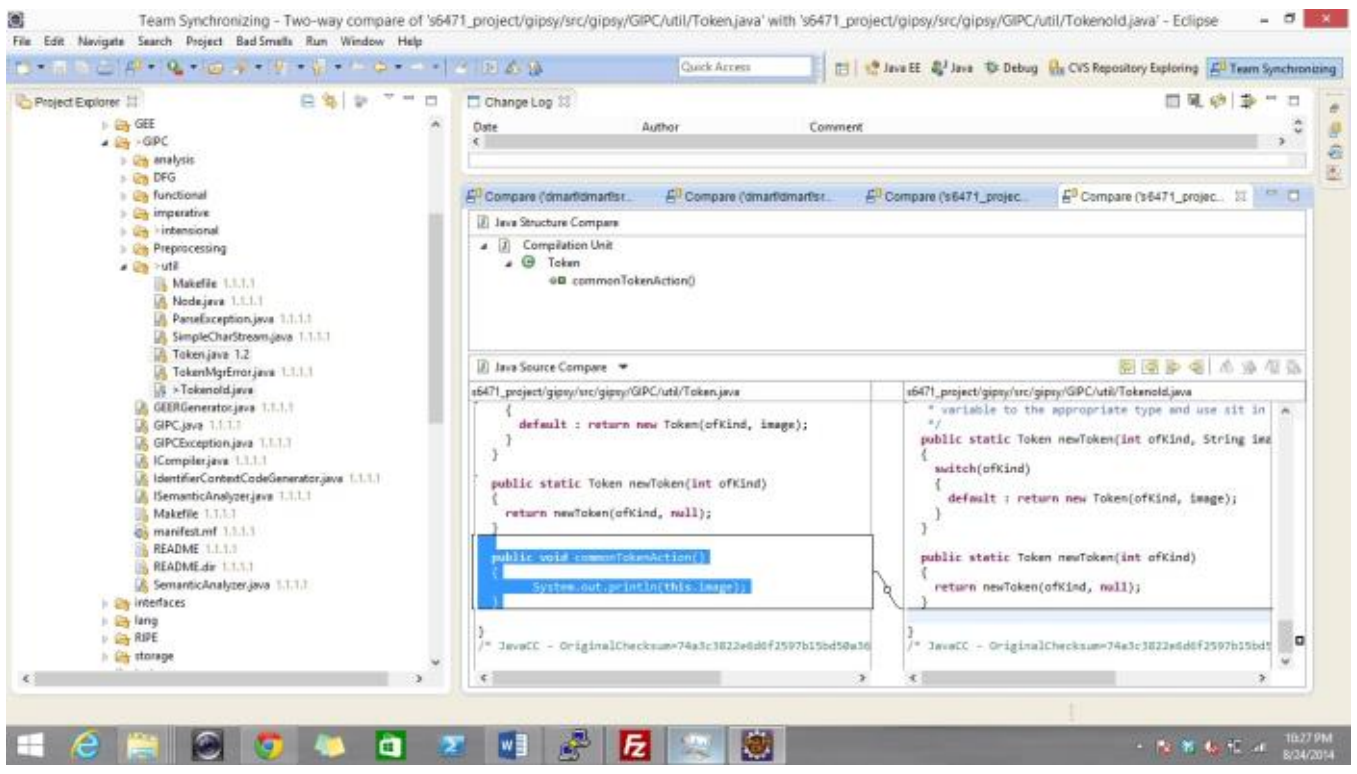

*Figure 55 GIPSY Diff*

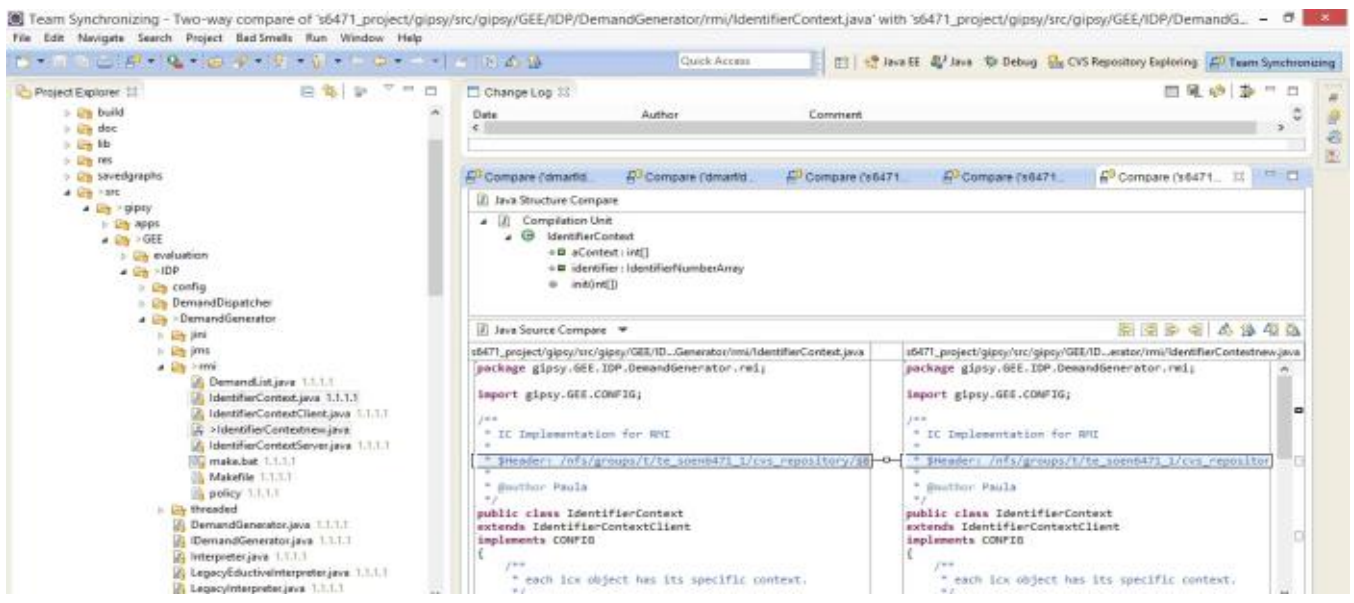

*Figure 56 Gipsy Diff 2*

## V. Tools used

For effective CVS workflow we use CVS flow in Eclipse, hence all the commits, diffs patch set and change log are calculated within eclipse. Also we used Jdeodorant to find bad smell, below is the small description on Jdeodorant

Jdeodorant is an Eclipse plug-in which is used to identify the design problems in the software , known as bad smells namely Feature-Envy, God class, State Checking and Long method.

Other software measurement tools were also used such as software metric plugin for eclipse to measure line of code and other aspects of DMARF and GIPSY framework.

## VI. Conclusion

Based on this document we came to conclude the frame works GIPSY and DMARF**.**The implementation of GIPSY will enable us to realize the solutions for unified frame work. Intensional languages like GIPSY has supported for compilation of Jlucid objective lucid. In GIPSY we used 3 compliers (General Intensional program complier) GIPC, General Eduction Engine (GEE), Generic Eduction Engine Resources (GEER) compliers to execute a GIPSY. GIPSY can be used in for application of cyber forensics.MARF is mainly used in recognition of the voice which consist of database of speakers. We came to know DMARF is used in Speech recognition for software developers, security based like lock with speech voice.

This paper also reviews about code detectors for analyzing the code smells. It also describes about requirements and design specifications like personas, actors and stake holders. We found nearly six refactoring methods for the code smells. Test cases are done for GISPY and DMARF for every module in the GIPSY, but in DMARF there are only few test cases written in test in which Junit application is used for test. We identified design patterns in both GIPSY & DMARF and Reverse Engineering tools as well.

The learning was immense from this project, starting from how to use CVS in collaborative way to more advance technical stuff including software patterns analysis, software refactoring techniques and writing test cases to see before and after constancy.

## Glossary

*Table 5 Terminology*

| | |
|---|---|
| DMARF | Distributed Modular Audio Recognition Framework |
| NLP | Natural Language Processing |
| GIPSY | The general intensional programming system |
| MARF | Modular Audio Recognition Framework |
| ASSL | Autonomic System Specification Language |
| Eclipse | Open Source IDE |
| CodePro | Eclipse Plugin |
| LOC | Line of code |
| SonarQube | An open source software quality measurement platform |
| OSS | Open Source Software System |
| SFAP | self-forensic autonomous property |
| JOOIP | Java based Object- Oriented Intensional Programming |
| AS | autonomic multi-tier system architecture0 |
| SLOs | specifying service-level objectives |
| | self-configuration, self-healing, self- |
| GIPC | General Intensional Programming Compiler |
| GEE | General Eduction Engine |
| AE | autonomic element |
| ME | ASSL management element |
| GEER | GEE engine resources |
| AC | Autonomic computing |
| ABLE | Agent Building and Learning Environment |
| ASTRM | Autonomic System Timed Reactive Model |
| ANTS | Autonomous Nano Technology Swarm |
| DST | Demand Store Tier |
| DMS | Demand Migration System |
| DG | Demand generator |
| DWT | Demand Worker Tier |
| DGT | Demand Generator Tier |
| GMT | GIPSY Manager Tier |
| GMs | GIPSY managers |
| GNs | GIPSY Nodes |
| ASIP | AS-Level interaction Protocol |
| NM | Node Manager |
| Domain Model | Simplifying problem domain |
| UML | Unified Markup Language |
| Use Case | Representing bi directional associations. |
| GMs | GIPSY managers |
| AGIPSY | Autonomous GIPSY |
| DWT | discreet wavelet transform |
| MARFL | Modular Audio Recognition Framework |
| DST | demand store tier |
| PS-DGT | problem-specific generator tiers |
| PS-DWT | problem-specific worker tiers |

## Appendix

Appendix A (Mapping case study to individual list)

*Table 6 Case Study to Individual Mapping (DMARF)*

| Name | Case Study Title |
|---|---|
| Dipesh Walia | Self-Optimization Property In Autonomic Specification Of Distributed MARF With ASSL [1] |
| Pankaj Kumar Pant | Towards a Self-Forensics Property in the ASSL Toolset [2] |
| Venkata Neela | Towards Security Hardening of Scientific Demand-Driven and Pipelined Distributed Computing Systems [3] |
| Naveen Kumar | Managing Distributed MARF with SNMP [4] |
| Ram Kunchala | Autonomic Specification of Self-Protection for Distributed MARF with ASSL [5] |

*Table 7 Case Study to Individual Mapping (GIPSY)*

| Name | Case Study Title |
|---|---|
| Dipesh Walia | An Interactive Graph-Based Automation Assistant: A Case Study to Manage the GIPSY's Distributed Multi-tier Run-Time System [6] |
| Pankaj Kumar Pant | Towards Autonomic GIPSY paper [7] |
| Venkata Neela | Self-Forensics Through Case Studies of Small-to-Medium Software Systems [8] |
| Naveen Kumar | Advances in the Design and Implementation of a Multi-Tier Architecture in the GIPSY Environment with Java [9] |
| Ram Kunchala | The GIPSY Architecture [10] |

*Table 8 Pattern Identification (Individual)*

| Name | GIPSY | DMARF |
|---|---|---|
| Dipesh Walia | Observer | - |
| Pankaj Kumar Pant | | Factory Method |
| Venkata Neela | Singleton | - |
| Naveen Kumar | Decorator | - |
| Ram Kunchala | - | Template Method |

## Appendix B (Snapshot of calculation results)

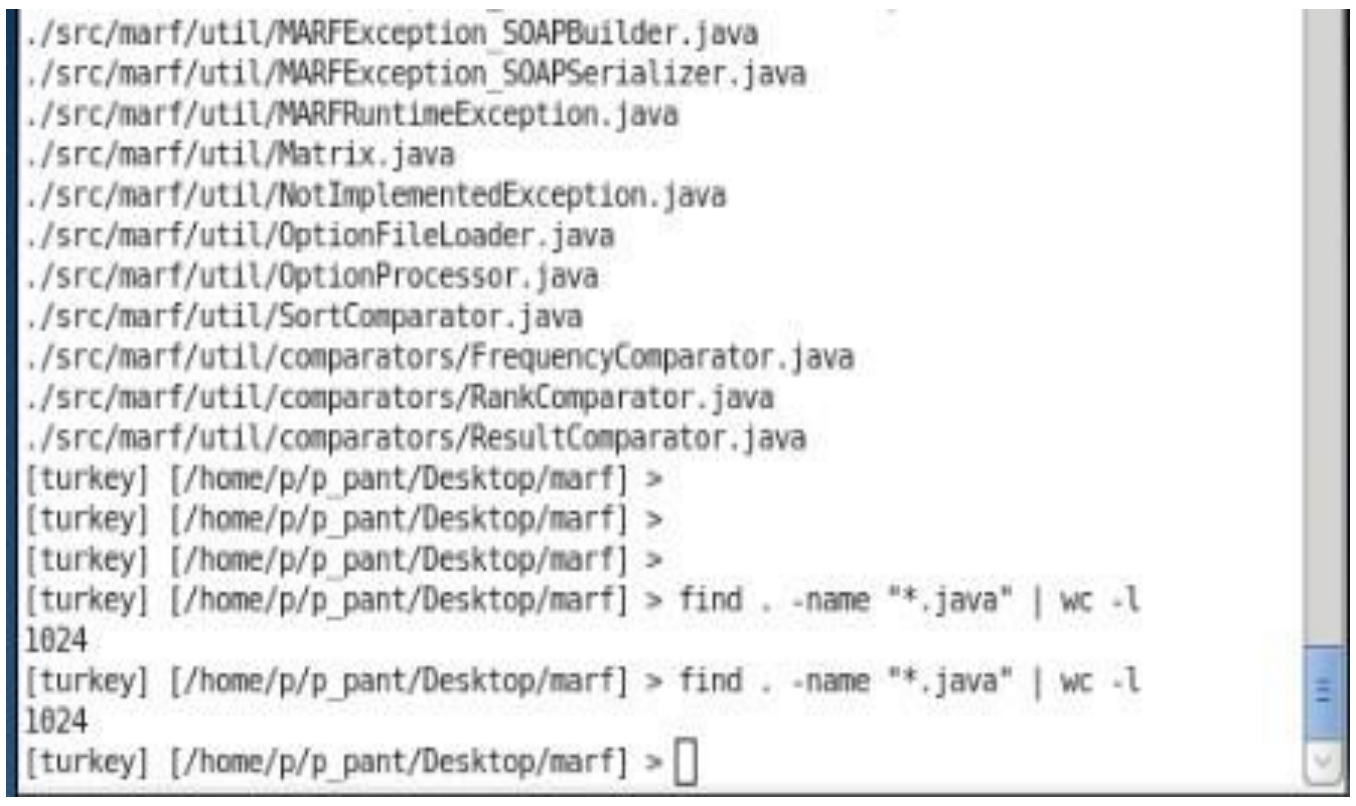

*Figure 57 Total Number of Java Files DMARF (Linux)*

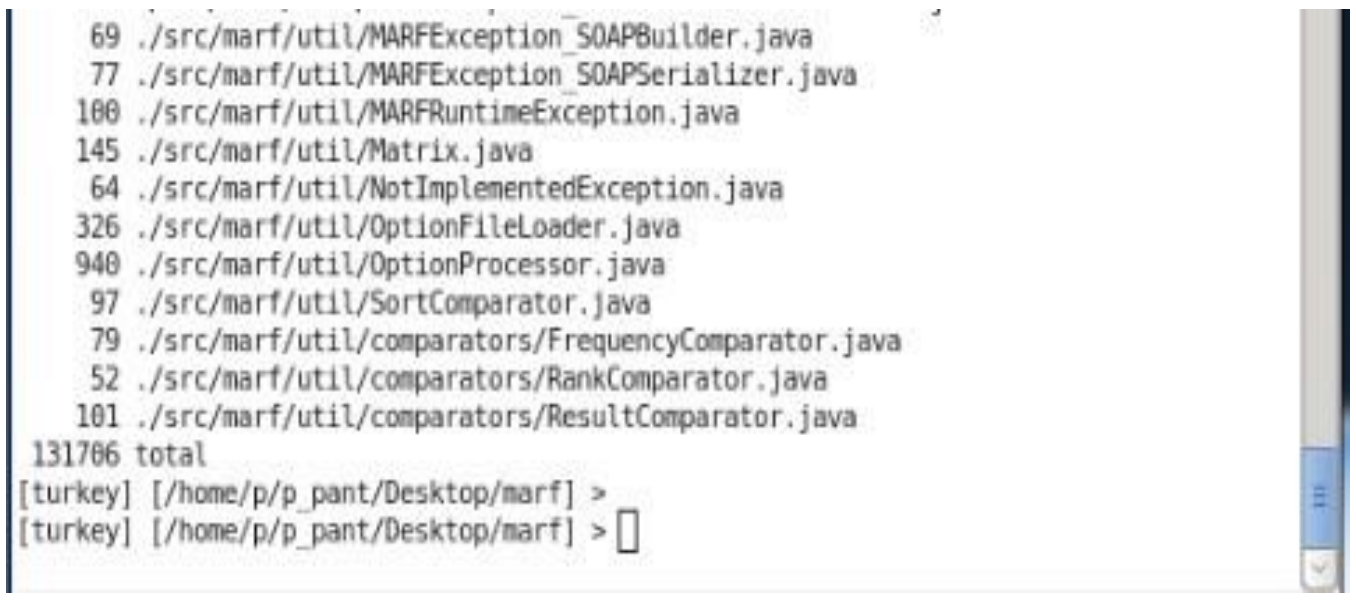

*Figure 58 Total Line of Code DMARF (Linux)*

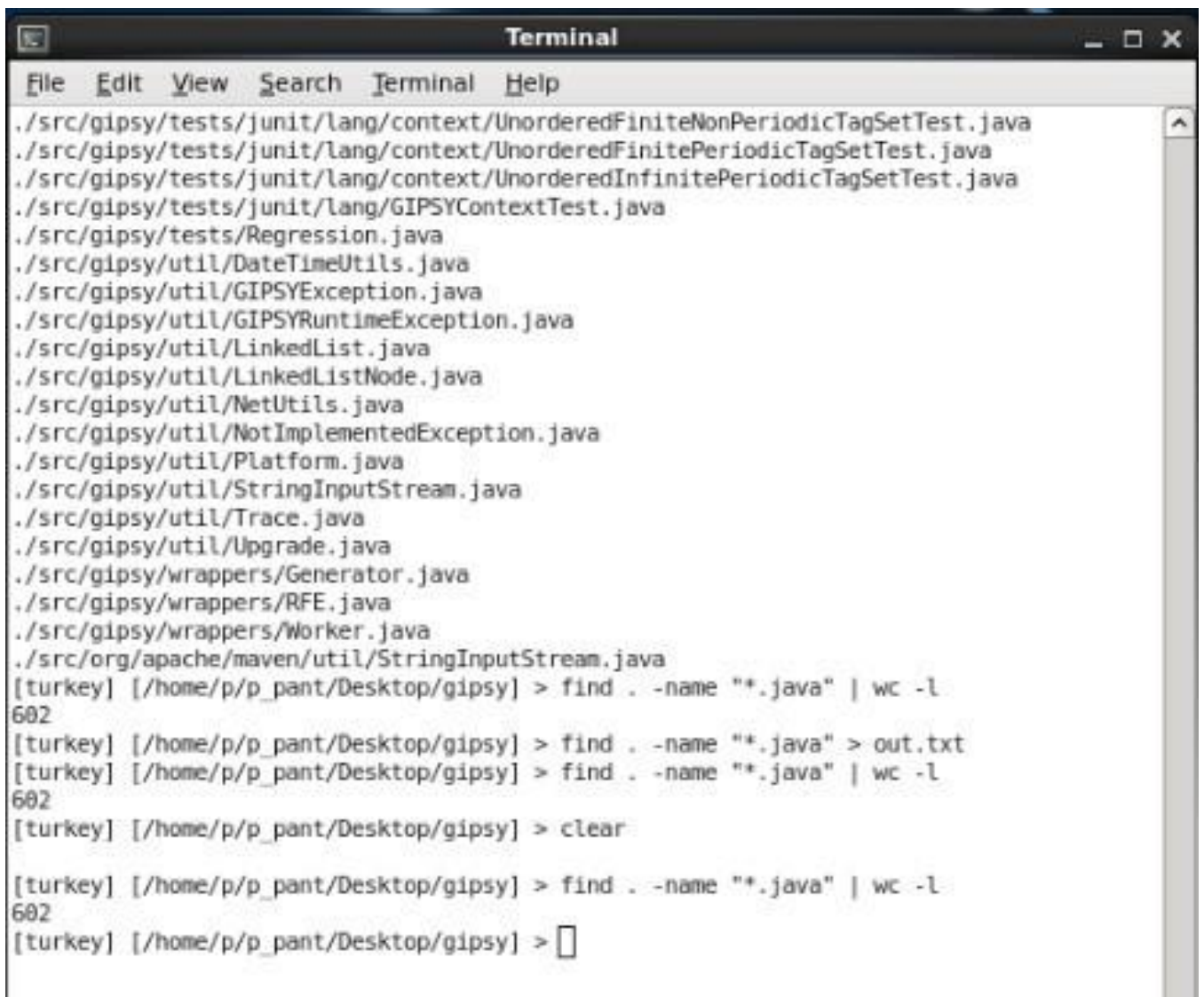

*Figure 59 Total Number of Java files GIPSY*

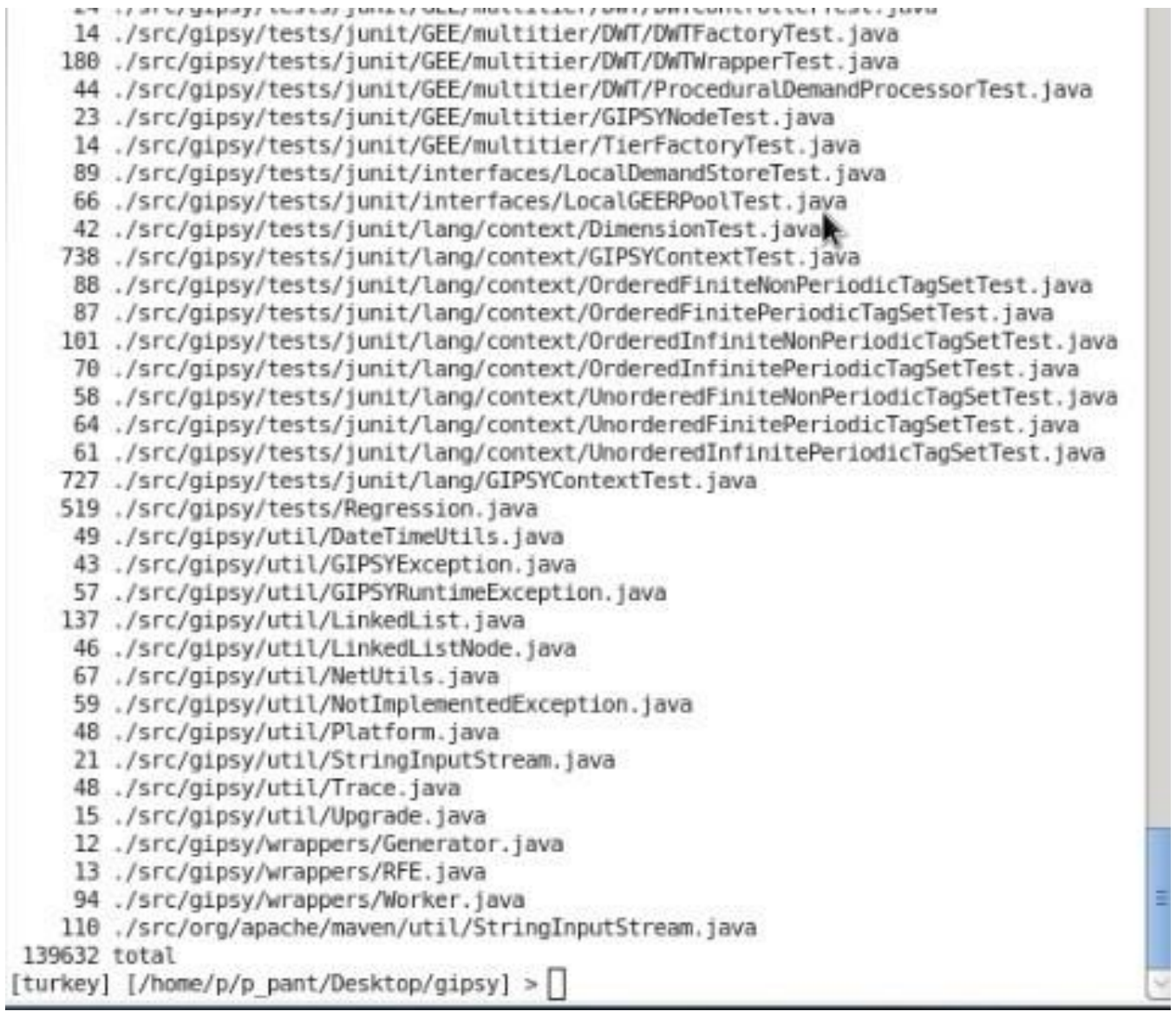

*Figure 60 Total Line of code GIPSY (Linux)*

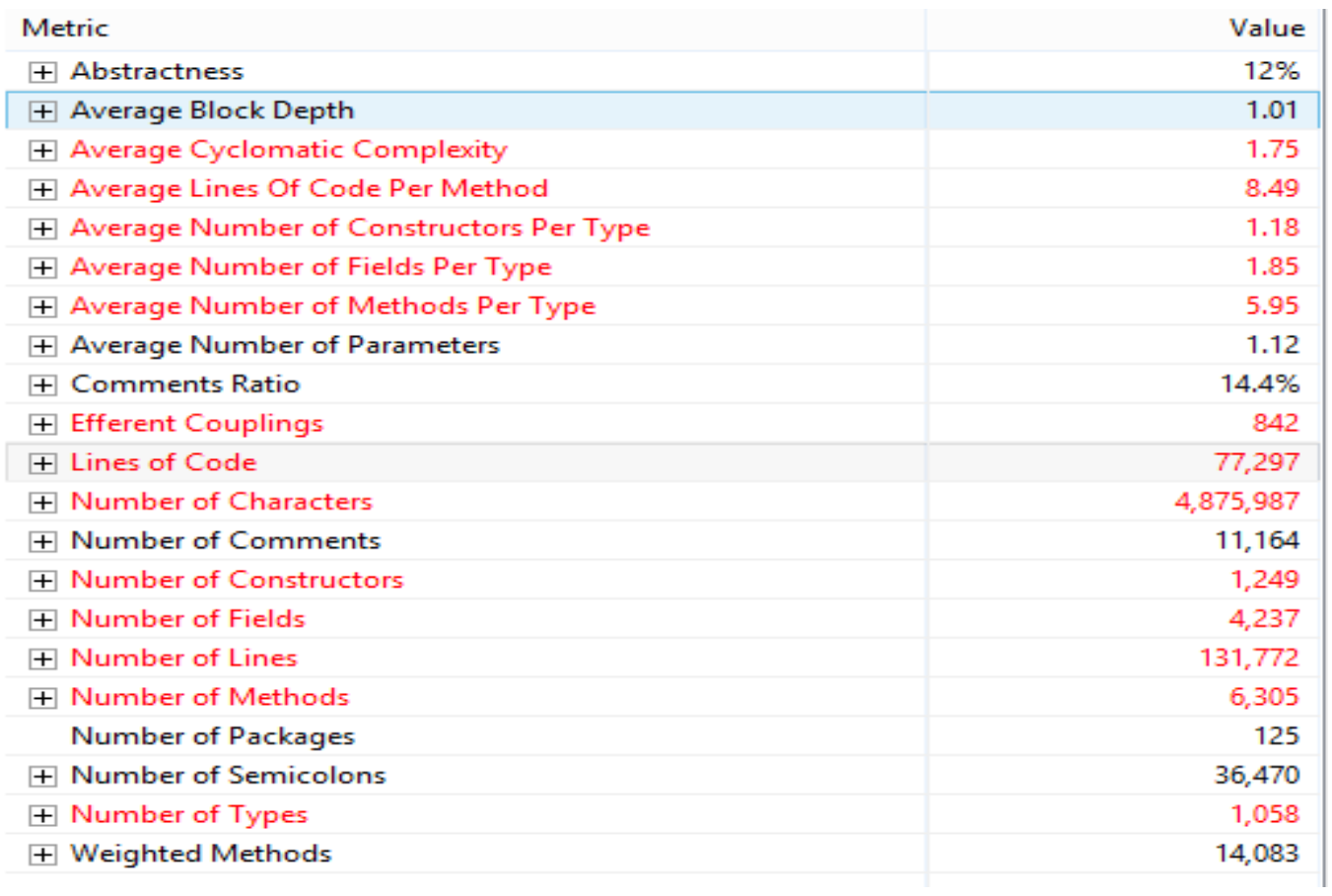

| Metric | Value |
|---|---|
| Abstractness | 12% |
| Average Block Depth | 1.01 |
| Average Cyclomatic Complexity | 1.75 |
| Average Lines Of Code Per Method | 8.49 |
| Average Number of Constructors Per Type | 1.18 |
| Average Number of Fields Per Type | 1.85 |
| Average Number of Methods Per Type | 5.95 |
| Average Number of Parameters | 1.12 |
| Comments Ratio | 14.4% |
| Efferent Couplings | 842 |
| Lines of Code | 77,297 |
| Number of Characters | 4,875,987 |
| Number of Comments | 11,164 |
| Number of Constructors | 1,249 |
| Number of Fields | 4,237 |
| Number of Lines | 131,772 |
| Number of Methods | 6,305 |
| Number of Packages | 125 |
| Number of Semicolons | 36,470 |
| Number of Types | 1,058 |
| Weighted Methods | 14,083 |

*Figure 61 DMARF CODE PRO RESULT*

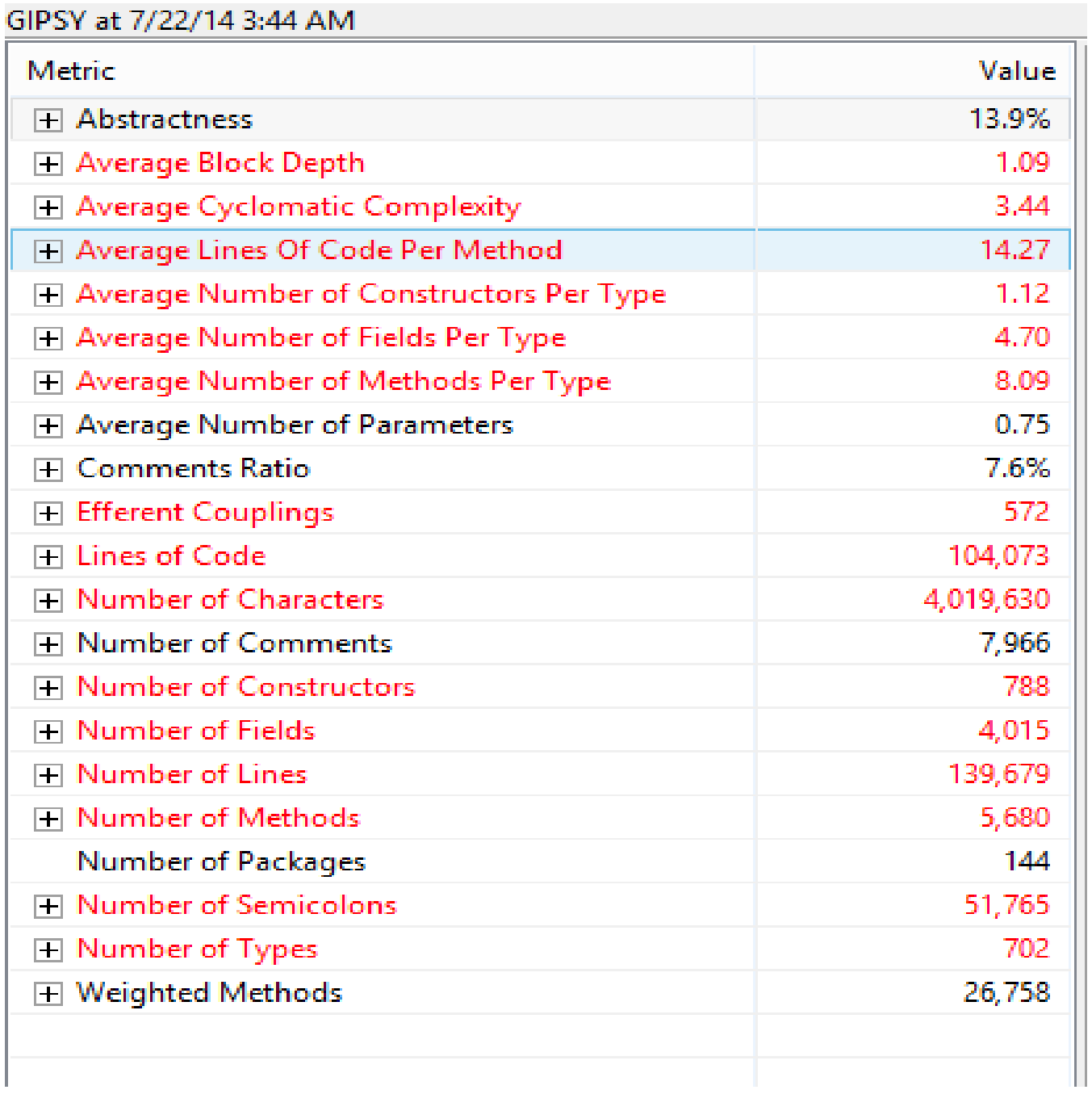

GIPSY at 7/22/14 3:44 AM

| Metric | Value |
| --- | --- |
| Abstractness | 13.9% |
| Average Block Depth | 1.09 |
| Average Cyclomatic Complexity | 3.44 |
| Average Lines Of Code Per Method | 14.27 |
| Average Number of Constructors Per Type | 1.12 |
| Average Number of Fields Per Type | 4.70 |
| Average Number of Methods Per Type | 8.09 |
| Average Number of Parameters | 0.75 |
| Comments Ratio | 7.6% |
| Efferent Couplings | 572 |
| Lines of Code | 104,073 |
| Number of Characters | 4,019,630 |
| Number of Comments | 7,966 |
| Number of Constructors | 788 |
| Number of Fields | 4,015 |
| Number of Lines | 139,679 |
| Number of Methods | 5,680 |
| Number of Packages | 144 |
| Number of Semicolons | 51,765 |
| Number of Types | 702 |
| Weighted Methods | 26,758 |

*Figure 62 GIPSY CODEPRO RESULT*

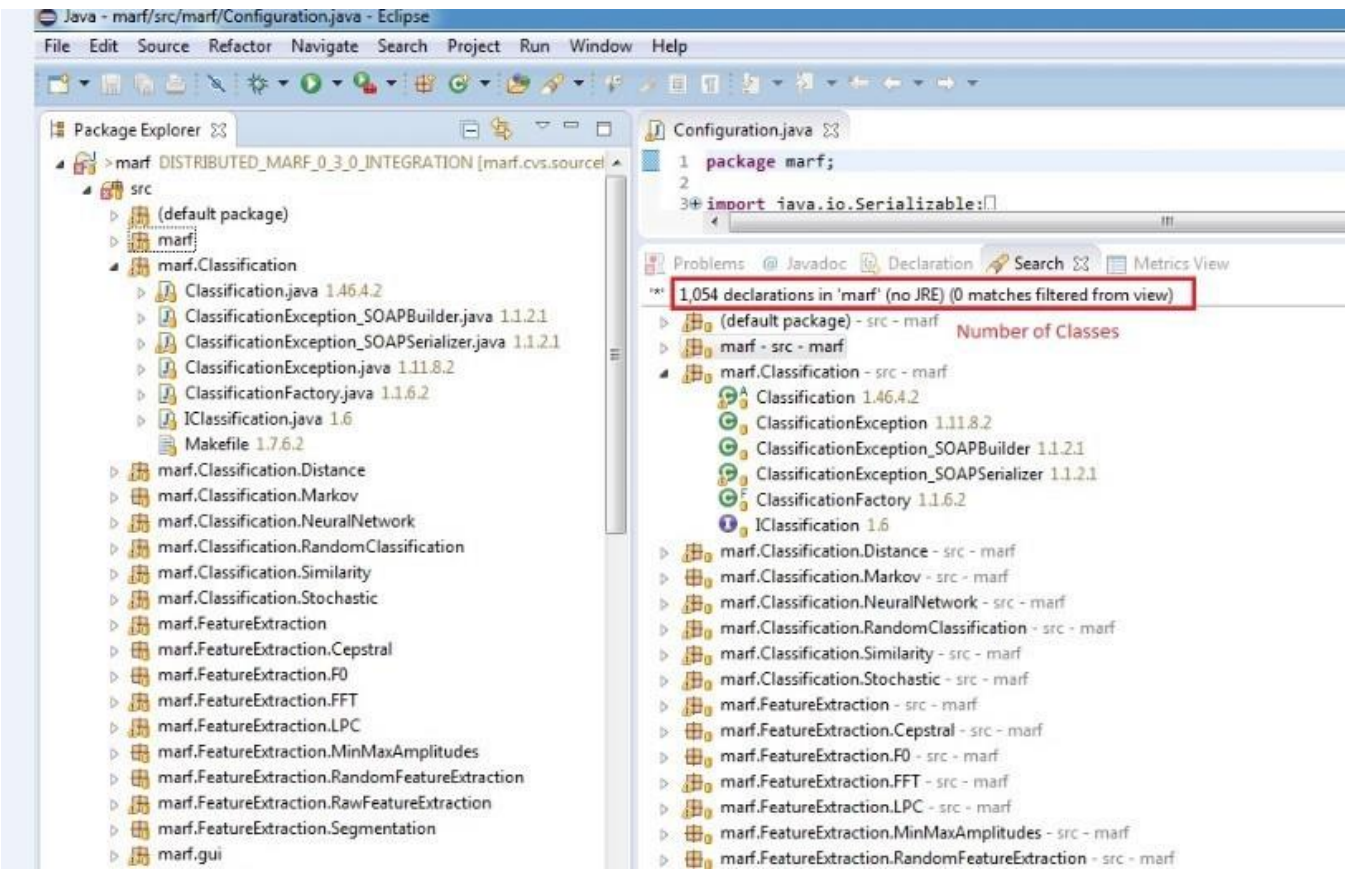

*Figure 63 DMARF_no_of_classes*

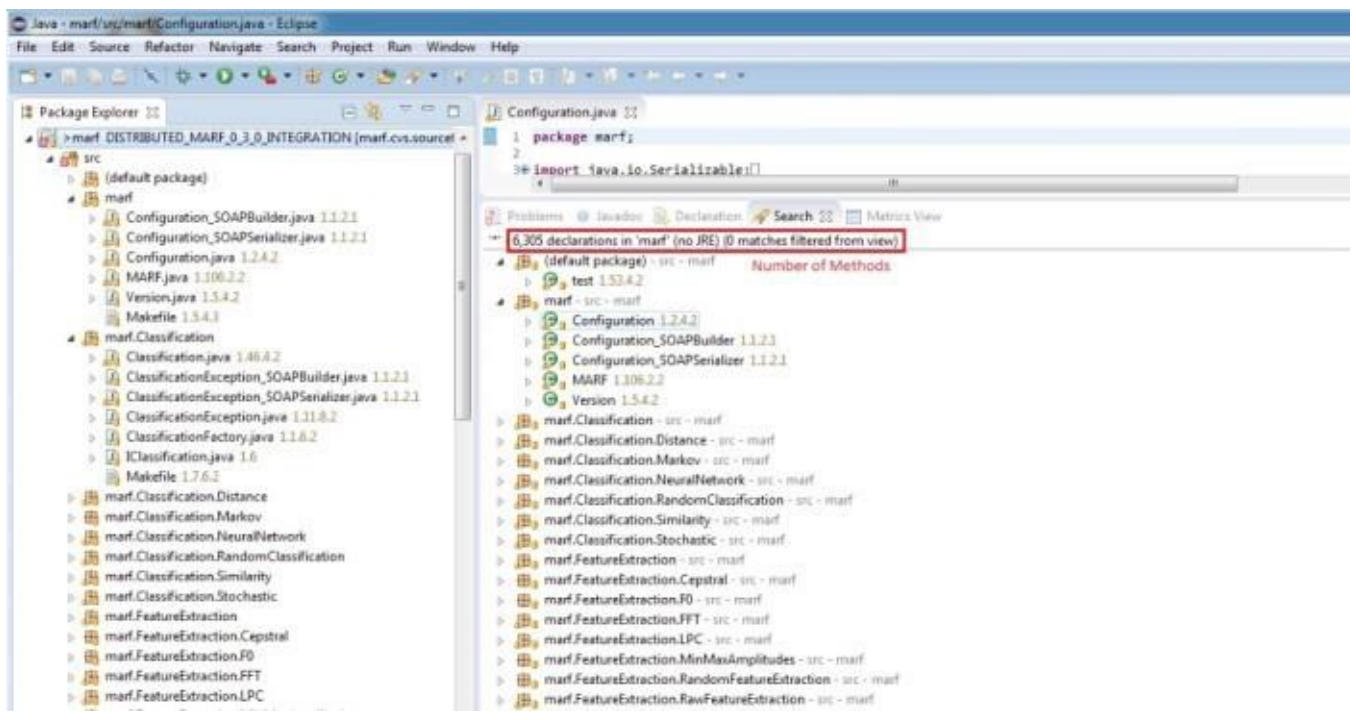

*Figure 64 Dmarf_no_of_Methods*

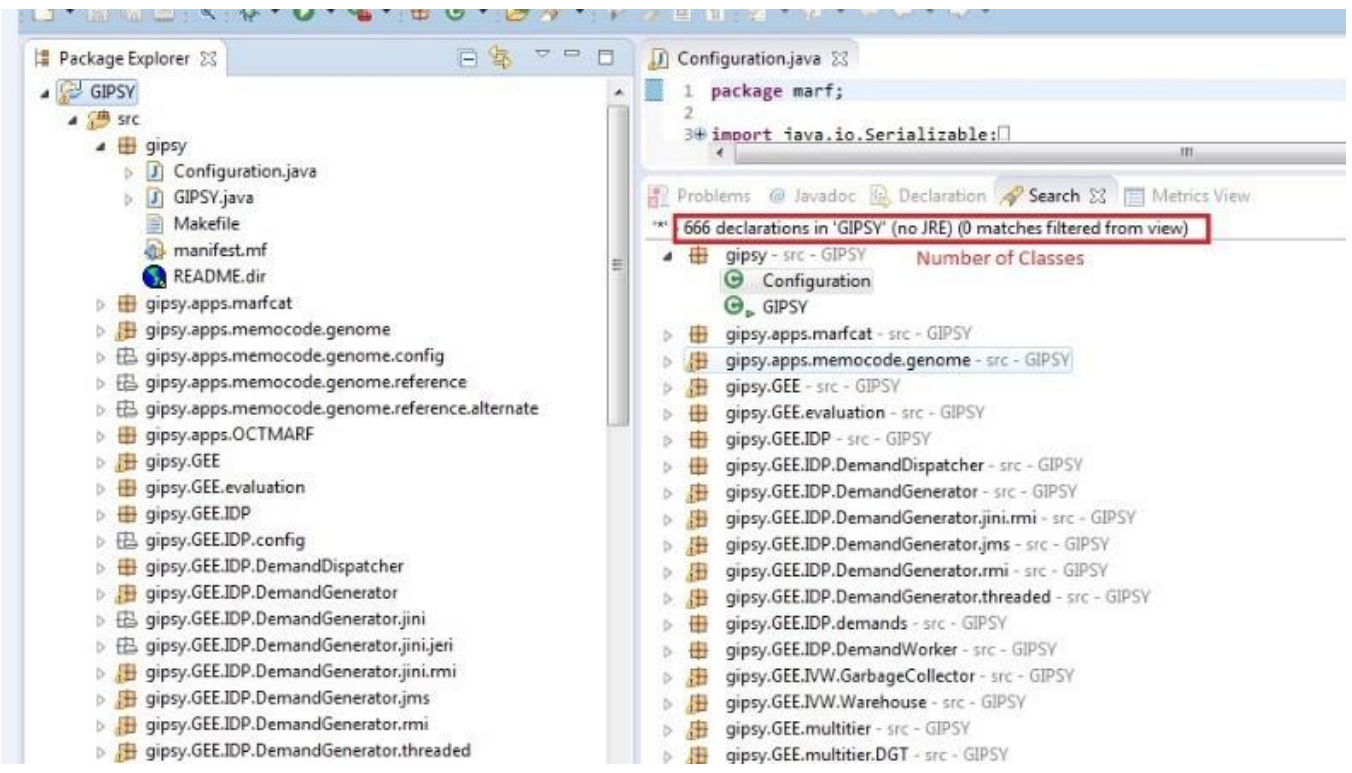

*Figure 65 GIPSY_no_of_classes*

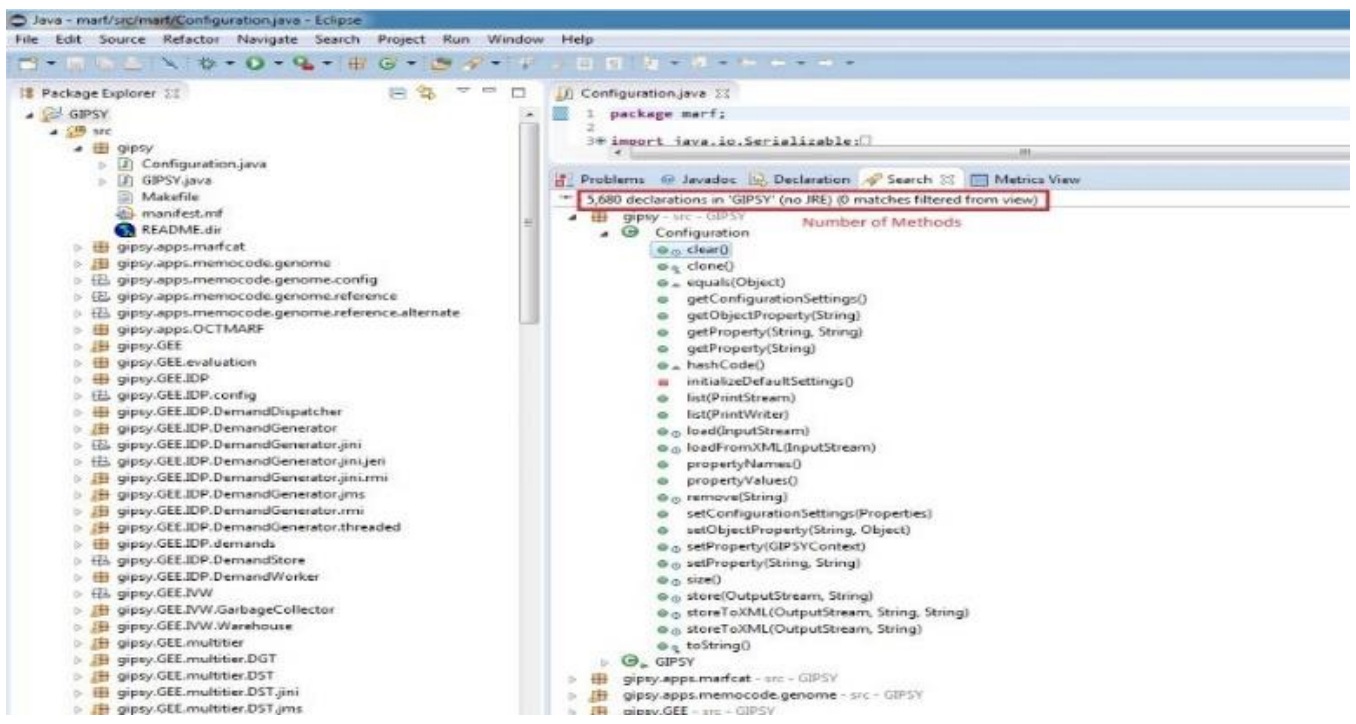

*Figure 66 GIPSY_no_of_Methods*

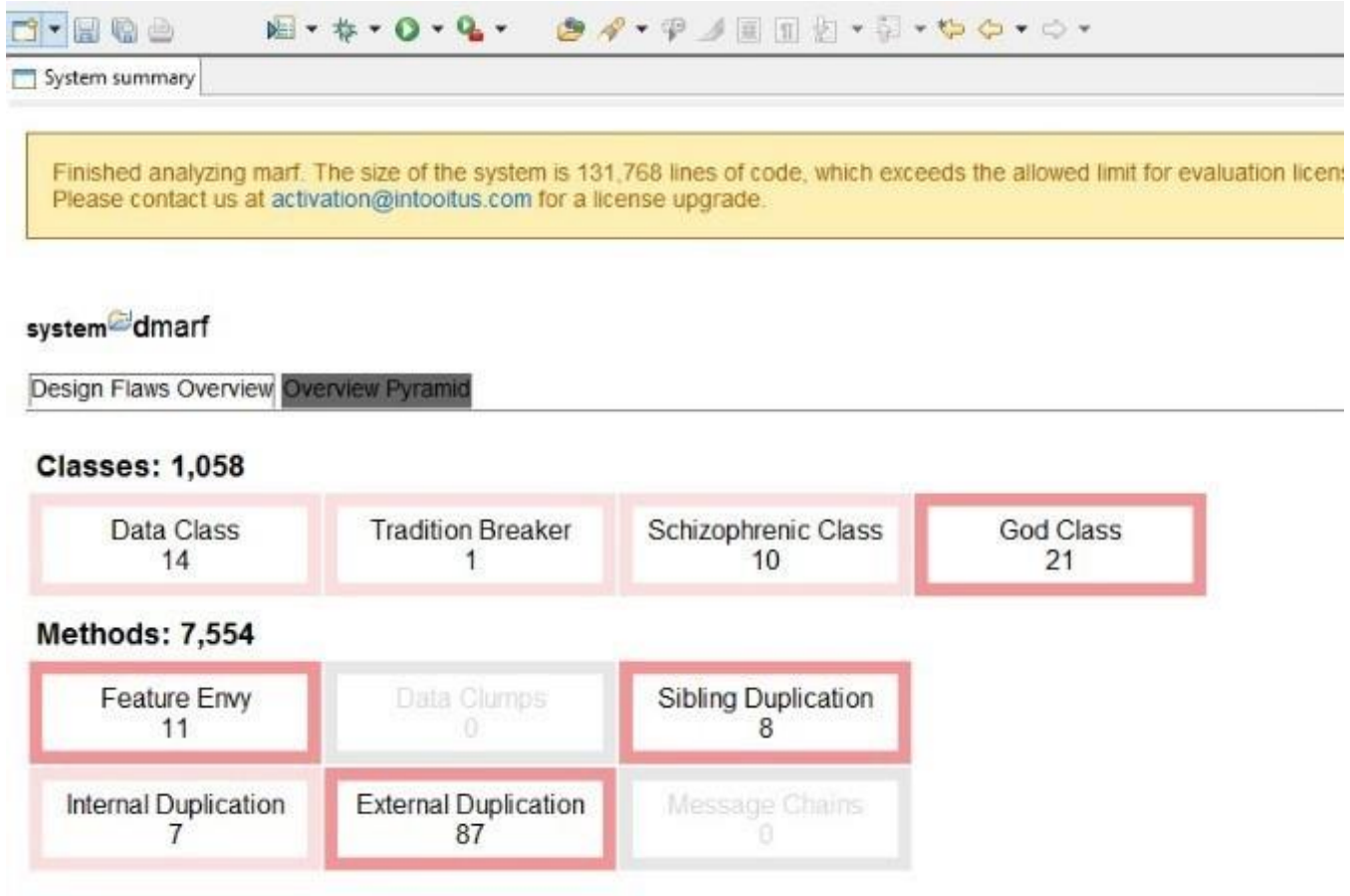

*Figure 67 DMARF InCode Analysis*

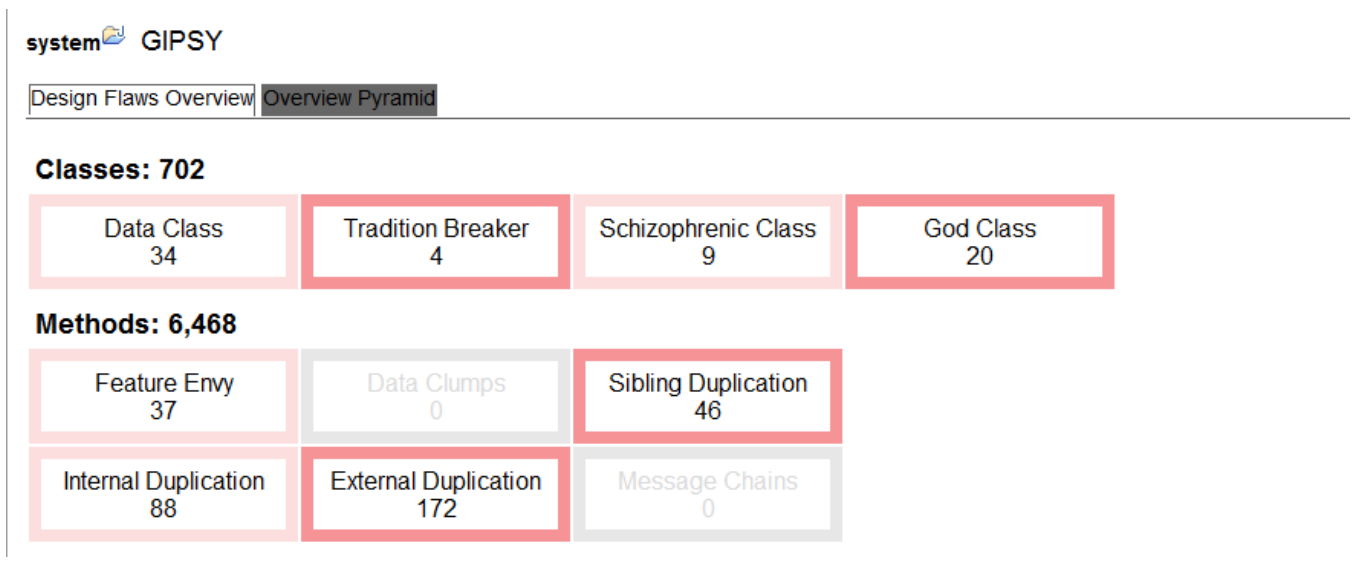

*Figure 68 Gipsy InCode Result*

## REFERENCES

[1] M. Young, The Technical Writer's Handbook. Mill Valley, CA: University Science, 1989. [1] Emil Vassev and Serguei A. Mokhov. Self-optimization property in autonomic specication of Dis-tributed MARF with ASSL. In Boris Shishkov, Jose Cordeiro, and Alpesh Ranchordas, editors, Proceedings of ICSOFT'09, volume 1, pages 331-335, Sofia, Bulgaria, July 2009.

[2] Serguei A. Mokhov, Emil Vassev, Joey Paquet, and Mourad Debbabi. Towards a self-forensics property in the ASSL toolset. In Proceedings of the Third C* Conference on Computer Science and Software Engineering (C3S2E'10), pages 108{113, New York, NY, USA, May 2010.

[3] Serguei A. Mokhov. Towards security hardening of scientic distributed demand-driven and pipelined computing systems. In Proceedings of the 7th International Symposium on Parallel and Distributed Computing (ISPDC'08), pages 375{382. IEEE Computer Society, July 2008.

[4] Serguei A. Mokhov, Lee Wei Huynh, and Jian Li. Managing distributed MARF with SNMP. Concordia Institute for Information Systems Engineering, Concordia University, Montreal, Canada, April 2007.M. Young, The Technical Writer's Handbook. Mill Valley, CA: University Science, 1989.

[5] Serguei A. Mokhov and Emil Vassev. Autonomic specication of self- protection for Distributed MARF with ASSL. In Proceedings of C3S2E'09, pages 175{183, New York, NY, USA, May 2009.

[6] Sleiman Rabah, Serguei A. Mokhov, and Joey Paquet. An interactive graph-based automation assistant: A case study to manage the GIPSY's distributed multi-tier run-time system. In Ching Y. Suen, Amir Aghdam, Minyi Guo, Jiman Hong, and Esmaeil Nadimi, editors, Proceedings of the ACM Research in Adaptive and Convergent Systems (RACS 2013), pages 387-394.

[7] Emil Vassev and Joey Paquet. Towards autonomic GIPSY. In Proceedings of the Fifth IEEE Workshop on Engineering of Autonomic and Autonomous Systems (EASE 2008), pages 25-34.

[8] Serguei A. Mokhov and Emil Vassev. Self-forensics through case studies of small to medium software systems. In Proceedings of IMF'09, pages 128-141

[9] Bin Han, Serguei A. Mokhov, and Joey Paquet. Advances in the design and implementation of a multi-tier architecture in the GIPSY environment with Java. In Proceedings of the 8th IEEE/ACIS International Conference on Software Engineering Research, Management and Applications (SERA 2010), pages 259-266.

[10] Joey Paquet. Towards autonomic GIPSY. In Proceedings of the Fifth IEEE Workshop on Joey Paquet and Peter G. Kropf. The GIPSY architecture. In Peter G. Kropf, Gilbert Babin, John Plaice, and Herwig Unger, editors, Proceedings of Distributed Computing on the Web, volume 1830 of Lecture Notes in Computer Science, pages 144-153.

[11] S. A. Mokhov and R. Jayakumar. Distributed modular audio recognition framework (DMARF) and its applications over web services.

[12] R. Murch. Autonomic Computing: On Demand Series. IBM Press, Prentice Hall, 2004.

[13] E. I. Vassev. Towards a Framework for Specification and Code Generation of Autonomic Systems. PhD thesis, Department of Computer Science and Software Engineering, Concordia University, Montreal, Canada ,2008.

[14] E. Vassev and J. Paquet. ASSL – Autonomic System Specification Language. In Proceedings if the 31st IEEE / NASA Software Engineering Workshop (SEW-31), pages 300–309, Baltimore, MD, USA, Mar. 2007. NASA/IEEE.

[15] S. A. Mokhov, "The role of self-forensics in vehicle crash investigations and event reconstruction," [online], May 2009, http://arxiv.org/abs/0905.2449.

[16] "Towards improving validation, verification, crash investigations, and event reconstruction of flight-critical systems with selfforensics," [online], Jun. 2009, http://arxiv.org/abs/0905.2449

[17] C. Gundabattula and V. G. Vaidya, "Building a state tracing Linux kernel," in Proceedings of the IT Incident Management and IT Forensics (IMF'08), O. Globel, S. Frings, D. Gunther, J. Nedon, and D. Schadt, Eds., Mannheim, Germany, Sep. 2008, pp. 173–196, LNI140.

[18] S. A. Mokhov. The role of self-forensics modeling for vehicle crash investigations and event reconstruction simulation. In Proceedings of HSC'09. SCS, Oct. 2009. To appear, online a http://arxiv.org/abs/0905.2449.

[19] S. A. Mokhov. Towards improving validation, verification, crash investigations, and event reconstruction of flight-critical systems with self-forensics. [Online], June 2009. A white paper submitted in response to NASA's RFI NNH09ZEA001L, http://arxiv.org/abs/0906.1845.

[20] S. A. Mokhov and E. Vassev. Self-forensics through case studies of small to medium software systems. In Proceedings of IMF'09, pages 128-141. IEEE Computer Society, Sept. 2009.

[21] G. Palmer (Editor). A road map for digital forensic research, report from first digital forensic research workshop (DFRWS). Technical report, DFRWS, 2001

[22] S. A. Mokhov and E. Vassev. Self-forensics through case studies of small to medium software systems. In Proceedings of IMF'09, pages 128-141. IEEE Computer Society, Sept. 2009.

[23] E. Vassev and J. Paquet. Towards autonomic GIPSY. In Proceedings of the Fifth IEEE Workshop on Engineering of Autonomic and Autonomous Systems (EASE 2008), pages 25-34. IEEE Computer Society, 2008.

[24] Yi Ji, Serguei A. Mokhov, and Joey Paquet. Unifying and refactoring DMF to support concurrent Jini and JMS DMS in GIPSY.

[25] Joey Paquet. Distributed eductive execution of hybrid intensional programs.

[26] Bin Han, Serguei A. Mokhov, and Joey Paquet. Advances in the design and implementation of a multi-tier architecture in the GIPSY environment with Java.

[27] Emil Vassev and Joey Paquet. Towards autonomic GIPSY.

[28] J. Paquet, "Distributed eductive execution of hybrid intensional programs," in Proceedings of the 33rd Annual IEEE International Computer Software and Applications Conference (COMPSAC'09). Seattle, Washington, USA: IEEE Computer Society, Jul. 2009, pp. 218– 224.

[29] Sun Microsystems, "Java Message Service (JMS)," [online], Sep. 2007, http://java.sun.com/products/jms/.

[30] Bin Han, Serguei A. Mokhov, and Joey Paquet. Advances in the design and implementation of a multi-tier architecture in the GIPSY environment with Java. In Proceedings of the 8th IEEE/ACIS International Conference on Software Engineering Research, Management and Applications (SERA2010), pages 259–266. IEEE Computer Society, May 2010. ISBN 978-0-7695-4075-7. doi: 10.1109/SERA.2010.40. Online at http://arxiv.org/abs/0906.4837.

[31] The Modular Audio Recognition Framework (MARF) and its Applications: Scientific and Software Engineering, http://arxiv.org/abs/0905.1235

[32] S. A. Mokhov and R. Jayakumar, “Distributed modular audio recognition framework (DMARF) and its applications over web services,” in Proceedings of TeNe’08. Springer, 2008, to appear.

[33] [S. A. Mokhov, “On design and implementation of distributed modular audio recognition framework: Requirements and specification design document,” Department of Computer Science and Software Engineering, Concordia University, Montreal, Canada, Aug. 2006, project report, http://marf.sf.net, last viewed December 2008.

[34] “Towards security hardening of scientific distributed demand driven and pipelined computing systems,” in Proceedings of the 7th International Symposium on Parallel and Distributed Computing (ISPDC’ 08). Krakow, Poland: IEEE Computer Society Press, Jul. 2008, to appear, http://ispdc2008.ipipan.waw.pl/.

[35] S. A. Mokhov, L. W. Huynh, and J. Li, “Managing distributed MARF’s nodes with SNMP,” in Proceedings of PDPTA’2008. Las Vegas, USA: CSREA Press, Aug. 2008, to appear.

[36] A. Wollrath and J. Waldo, “Java RMI tutorial,” Sun Microsystems, Inc., 1995–2005, http://java.sun.com/docs/books/tutorial/rmi/index.html.

[37] Sun Microsystems, “Java IDL,” Sun Microsystems, Inc., 2004, http:// java.sun.com/j2se/1.5.0/docs/guide/idl/index.html.

[38] “The java web services tutorial (for JavaWeb Services Developer’s Pack, v2.0),” Sun Microsystems, Inc., Feb. 2006, http://java.sun.com/ webservices/docs/2.0/tutorial/doc/index.html.

[39] Towards a framework for the general intensional programming compiler in the GIPSY,

http://dl.acm.org/citation.cfm?id=1028731.

[40] https://www.princeton.edu/~achaney/tmve/wiki100k/docs/Zipf_s_law.html

[41] http://nlp.stanford.edu/software/lex-parser.shtml

[42] S. A. Mokhov. Towards syntax and semantics of hierarchical contexts in multimedia processing applications using MARFL. In Proceedings of the 32nd Annual IEEE International Computer Software and Applications Conference (COMPSAC), pages 1288–1294, Turku, Finland, July 2008.IEEE Computer Society.